# Frequency Standard Contributions to Limitations on the Signal-to-Noise Ratio in Very Long Baseline Interferometric (VLBI) Observations

**E.A. Burt[1], T.A. Ely[1], G.C. Bower[2], T.J.W. Lazio[1], M. Anderson[1], and S. Hernandez[3]**

[1]Jet Propulsion Laboratory, California Institute of Technology, Pasadena, CA. [2]Academia Sinica Institute of Astronomy and Astrophysics: Hilo, HI, US. [3]Space Systems Development, Blue Origin, Kent, WA.

Corresponding author: Eric Burt (eric.a.burt@jpl.nasa.gov)

**Key Points:**

- The coherence function C(T) is not a sufficient metric for determining the optimal frequency standard for use in high-frequency space VLBI.
- A new metric giving frequency standard limitations to VLBI visibility S/N is derived.
- Monte Carlo simulations of this metric for real atomic clocks reveal few to be adequate. Promising clock technologies are identified.

## Abstract

Since its observation in 2019, the first image of a super-massive black hole using Very Long Baseline Interferometry (VLBI) with an Earth-scale baseline has generated much scientific and public interest. Work is now underway to extend the baseline into space to obtain higher image resolution. Operating one or more VLBI nodes in space will require the use of frequency standards that are space-qualified, greatly reducing the number of options available. The coherence function $C(T)$ is the metric usually used to determine the viability of a frequency standard. Here we show that $C(T)$ is a useful but not sufficient metric for gauging frequency standard performance in VLBI and instead derive an expression for the clock-limited VLBI visibility S/N. We evaluate this expression for real frequency standards and find only the Ultra-Stable Oscillator (USO) and hydrogen maser to be viable for upcoming high-frequency VLBI with the USO only useful for very limited integration times (30s at 90 GHz, 10s at 230 GHz, 5s at 345 GHz, and not viable at 630 GHz). The maser extends these, but may have prohibitive size for a space mission. We also evaluate emerging frequency standard technologies and find the optical local oscillator portion of optical clocks to be very promising (conservatively >100s at 90 GHz, 60s at 230 GHz, 40s at 345 GHz, and 22s at 630 GHz) when accounting for both performance and potential operation in space.

## Plain Language Summary

The next generation of instruments used to image super massive black holes will operate at least partially in space. A key enabling component of these instruments will be atomic clocks that can operate in space as well. In this article we derive a method for evaluating atomic clocks for this specific purpose and use it to evaluate both current and emerging clock technologies.

## 1 Introduction

Very Long Baseline Interferometry (VLBI) has been employed by the Event Horizon Telescope (EHT) to image super-massive black holes (SMBH) at the center of galaxies by using an aperture the size of the earth and an observation frequency of 230 GHz (Akiyama et al., 2024). Even larger apertures will yield resolution high enough to enable scientific returns such as better resolution of the first photon ring around SMBH's (Broderick et al., 2022), and better understanding of SMBH jet formation (Blandford et al., 2019).

The resolution of a VLBI measurement is approximately $R \approx \lambda/D$ where $\lambda$ is the observing wavelength and $D$ is the effective diameter of the synthesized telescope. In principle, to get higher resolution either the wavelength can be reduced or the baseline increased. However to maximize future scientific return, both are being pursued simultaneously: operating one of the VLBI nodes in space and observing at greater than 230 GHz. See Raymond et al. (2024) for recent ground observations at 345 GHz. Space VLBI has already been demonstrated at lower frequencies to achieve higher resolution. To set the scale, $\lambda/D \approx 25\ \mu as$ for the EHT while a value of less than $10\ \mu as$ is needed for photon ring science (Johnson et al., 2024). With the earth-sized baseline used by the EHT this resolution is not reached at 345 GHz ($16\ \mu as$), but can be achieved by adding an antenna in a 36,000 km geo-synchronous orbit ($5\ \mu as$).

As VLBI frequencies become higher, constraints on phase coherence between antennas will become more important. One of the main results of this work is to show that the coherence function $C(T)$ (Doeleman et al., 2011; M. Rioja et al., 2012; A. Rogers & Moran Jr, 1981) defined in section 2.4, which is often used to characterize decohering effects, gives an incomplete picture. $C(T)$ where $T$ is the integration time over which coherence is measured, varies from 1, perfect coherence, to 0, completely incoherent. We will show that commonly accepted $C(T)$ values of 0.8 or 0.9 can in some cases be insufficient to unambiguously detect the VLBI visibility fringe. An alternative metric that gives the frequency standard contribution to VLBI visibility S/N will be shown to give more reliable results, thereby enabling a better determination of what frequency standards can support high frequency VLBI in space. In this work, we will only concern ourselves with decohering effects due to the frequency standards, but also will derive a requirement on limitations to VLBI visibility S/N in the case where the S/N due to non-clock effects is also known. The frequency standard limited S/N metric will be shown to give a fairly precise statement of what frequency standards will be viable for new high-frequency VLBI measurements. To simplify reference, the main theoretical results derived in equations (40), (42), and Appendix C, are summarized in Appendix G, table A2.

While resolution is only one component of image synthesis, it is useful to compare the resolution of these various space-based instruments. In 1986, the Tracking and Data Relay Satellite System (TDRSS) was used to perform the first demonstration of VLBI in space (Levy et al., 1989). Observations were made at 15 GHz with the TDRSS satellite in a geosynchronous orbit at 36,000 km and reported a resolution of ~340 $\mu\mathrm{as}$. The VSOP mission (also called HALCA) was the first space satellite dedicated to VLBI (Hirabayashi et al., 2001; Hirayashi, 1991). VSOP was designed to operate at frequencies up to 22 GHz, but post-launch only demonstrated 5 GHz. The VSOP satellite apogee of 21,400 km results in a resolution of 580 $\mu\mathrm{as}$ though improved resolution was reported using a weighting algorithm (Lovell, 2004). Both the TDRSS experiment and VSOP used radiofrequency communications from the ground to establish a frequency and time standard for the space receiver so as to perform down-conversion and time stamping for later intercomparison. The Radioastron mission (Kardashev et al., 2017) was the first space VLBI to launch a high-performance frequency standard on board the satellite, in this case a hydrogen maser. (Technically, a clock consists of a frequency standard plus a mechanism for counting oscillations to produce a time standard. However, it is common to use the terms frequency standard and clock interchangeably unless the distinction is important.) Operating at 22 GHz with a satellite apogee of 128,000 km, Radioastron achieved a resolution of 21 $\mu\mathrm{as}$. For reference, the all-terrestrial Event Horizon Telescope achieved a resolution of 25 $\mu\mathrm{as}$ by operating in a smaller aperture but at the much higher frequency of 230 GHz. A resolution significantly lower will be required to resolve detailed features of SMBH's. Operating at an even higher frequency presents extreme challenges both in data recording methods and as we shall see in the choice of frequency standard. These initial space VLBI results are summarized in Table 1.

**Table 1.** A summary of space VLBI missions to date along with the terrestrial EHT performance for reference. Also shown for comparison is the radius of the M87 SMBH photon ring, which is of great scientific interest.

| **Mission** | **Baseline (km)** | **Observation frequency (GHz)** | $\boldsymbol{\lambda/D}$ $(\boldsymbol{\mu as})$ | **Resolution Reported** $(\boldsymbol{\mu as})$ |
|---|---|---|---|---|
| TDRSS | GEO: 36,000 | 15 | 115 | ~340 |
| VSOP | 21,400 | 22 | 580 | ~300 |
| Radioastron | 128,000[a] | 22 | 21 | 21 |
| EHT (terrestrial reference) | 10,000 | 230[b] | 25 | 25 |
| M87* photon ring (reference) | - | - | 39 | - |

[a]The Radioastron satellite had a maximum perigee of 360,000 km, but this was not used together with the highest frequency of 22-25 GHz. The best resolution was obtained with the stated earth-space baseline of 128,000 km.

[b] This has now been extended to 345 GHz with $\lambda/D = 19$ $\mu$as—see Raymond et al. (2024).

Various space VLBI proposals exist to increase the frequency and synthesized aperture, combining a space node with an existing terrestrial VLBI network. For example, see Tsuboi (2008), Johnson et al. (2024), and Fromm et al. (2021)—also Gurvits (2020) lays out some of the technical challenges of future space-based VLBI. All have in common the constraints on instrumentation imposed by the rigors of the space environment, the launch to get there, as well as limitations on size, weight, and power (SWaP). All high-performance VLBI measurements depend on having a precise frequency standard located at each node (Doeleman, 2009). For high-performance terrestrial VLBI these have been almost universally active hydrogen masers (R. F. Vessot, 1990). Masers do not currently limit SMBH measurements (Akiyama et al., 2019) though as we will see, they will begin to do so if higher frequencies are used. The maser is robust, mature, and commercially available, however work to space-qualify them has not achieved the same performance in space as that on the ground (Goujon et al., 2010; Kardashev et al., 2017; R. Vessot & Levine, 1979).

There are two types of hydrogen maser atomic clock: passive and active. In an active maser, the hydrogen atoms themselves generate the maser output frequency, while in a passive maser, the atoms only serve as a reference for a local oscillator. Passive masers, which are used on the European Galileo system, can be made significantly smaller than their active counterparts, but their performance is 1-2 orders of magnitude worse and not sufficient for high-performance VLBI. In this work, the term “maser” will be used to refer to the active type. Any other type will be qualified differently. The active maser SWaP is high: approximately 120 L × 60 kg × ~100 W (Vremya space hydrogen maser, 2011), which may not be feasible for a near-term space VLBI node.

The choice of other highly mature space clocks is limited. The currently available highest performing space-qualified frequency standards have > 10× degraded performance from the maser on key time scales of interest to VLBI. These include the USO (Ultra Stable Oscillator (USO), 2025), the Rubidium Atomic Frequency Standard (RAFS) used by the Global Positioning System (GPS) (see the reference, Teledyne, n.d.), the aforementioned space hydrogen masers, the *passive* hydrogen maser used by the Galileo positioning system (Droz et al., 2009), the Deep Space Atomic Clock (DSAC) (Burt et al., 2021), and a laser-cooled rubidium beam clock (Liu et al., 2018). Note that it is conventional to use the term USO to refer to a quartz oscillator. Here we follow this convention and explicitly identify other oscillators by using different names.

In addition to this lower level of performance, SWaP will be a concern for any space antenna, necessitating a further down-selection of viable instrumentation. To facilitate the complex frequency standard trade space, it will be necessary to have a requirement on what frequency standard performance is really needed. Clock uncertainty will impact VLBI through variations in the clock phase. The phase of the frequency standard at VLBI antenna *i,* can be written as $\varphi_i(t) = 2\pi\nu_{LO}\big(t + x_i(t)\big)$ where

$$x_i(t) = x_{i_0} + y_{i_0}t + 1/2\, D_i t^2 + \psi_i(t) \tag{1}$$

is the instantaneous phase noise of the clock at time $t$ (Allan, 1987). In this expression, $x_{i_0}$ is an initial phase offset in seconds, $y_{i_0}$ is an initial fractional frequency offset, $D_i$ is a fractional frequency drift, and $\psi_i(t)$ is the random phase noise. Throughout this work the term Local Oscillator (LO) will often be used to refer to the VLBI station reference clock and here, $\nu_{LO}$ refers to its frequency coherently multiplied up to the desired observation frequency. Without loss of generality, we have set the initial phase of the oscillators to zero because, using phase calibration tones, these initial phase offsets can be calibrated out of the signal (Sasao & Fletcher, 2008). For high performance frequency standards, amplitude noise is usually negligible compared to phase noise, so is not included here. Clock decoherence is a measure of how the phase difference between the two clocks on a single baseline will increase over time due to each of the terms shown in equation (1). As the phase difference becomes a significant portion of a radian, the interferometer fringes will lose contrast. Most treatments of the effect of clock variations on the VLBI visibility make the quasi-static approximation (Sasao & Fletcher, 2008; Thompson et al., 2017), which states that time variations in each of the terms in equation (1) are either slow on timescales of interest or small. For example, hydrogen maser fractional frequency drift is usually $D_i = 1 \times 10^{-15}/\mathrm{day}$. This component contributes to a time uncertainty for a 10 second integration of only about $1 \times 10^{-18}$ seconds and a corresponding phase uncertainty at 90 GHz of only about a micro-radian. The stochastic term $\psi_i(t)$ can vary rapidly, but the assumption that the amplitude of these variations was small was usually a good approximation. It was certainly true for hydrogen masers at lower frequencies, but it does not hold for other space clocks, and in the domain of much higher observation frequencies that we wish to consider it does not hold for the hydrogen maser either. For instance, all hydrogen maser noise taken together typically has a time uncertainty of $1.5 \times 10^{-13}$ seconds at 10 seconds of averaging time. At 630 GHz, this corresponds to almost a radian of error.

Considering the limited selection of space-qualified clocks, the highly mature, rubidium space atomic clock used in the Global Positioning System (Teledyne, n.d.) would be an obvious candidate for any space mission requiring an atomic clock due to its good stability and low SWaP. However, at 345 GHz and 10 seconds of integration time the phase uncertainty of this clock rises to almost $2\pi\nu_{LO}\psi_i(t) = 14$ radians, which would make the interferometer fringe undetectable (note that this can be partially compensated by fringe fitting, which we will address).

To quantify the impact of frequency standard noise on VLBI measurements, the coherence function $C(T)$ defined in section 2.4 is often used (Doeleman et al., 2011; M. Rioja et al., 2012; A. Rogers & Moran Jr, 1981), since it can indicate how the visibility signal will decrease with time as a function of decoherence. However, ultimately any VLBI measurement will place a requirement on the visibility Signal to Noise Ratio (S/N) and a knowledge of signal decoherence alone is not sufficient. The complex VLBI visibility is defined as $\mathcal{V} = \mathcal{V}_0 e^{i\varphi(t)}$ where the phase term contains both deterministic and stochastic components. The stochastic components consist of, among other things, noise from the antenna, the receiver, the atmosphere, and the frequency standard.

The usefulness of $C(T)$ as a metric is clear: it is a monotonically decreasing function from 1 to 0 that as a function of time gives a quantitative expression for how the visibility signal will decrease due to clock (or other) decoherence. Values of $C(T)$ in the range of 0.8 or 0.9 might seem adequate because they will leave the bulk of the signal unchanged, but to facilitate choosing a specific clock, a more precise requirement is needed because this range is too broad to downselect among several viable space clock candidates. In addition, $C(T)$ only shows how the visibility signal varies with time and not the visibility signal to noise ratio (S/N). For instance, we will show that at the averaging time for which the visibility S/N due to clock noise falls below the level required for unambiguous fringe detection, $C(T)$ can be higher than $0.97$. In other words, a clock coherence of 0.9 that otherwise might have been considered quite good, may not result in adequate visibility S/N.

The primary results of this paper will be to evaluate the impact of clock noise on VLBI visibility S/N, derive a metric that allows a quantitative comparison between clocks to be made for general combinations of pure noise types, show how the equations can be solved for some special cases, and give a method for solving them numerically in the general case. These derivations will not make the quasi-static approximation for clock noise, nor will they assume that clock noise is strictly stationary over the time scales of interest. The results will be applied to several existing and potentially future space-qualified clocks. The main goal of the paper is to show the limitations of currently available clocks in the context of future high-performance VLBI, and to determine what are the noise characteristics of future clocks that could most benefit VLBI.

In section 2, we will derive a requirement on VLBI visibility S/N limitations caused by the frequency standard and define various metrics for characterizing clock noise. In section 3, we will examine each of the ways that clock noise can impact the visibility and derive an expression for the visibility that includes all clock noise contributions. From this we will derive a general expression for the visibility S/N. We then define the clock-induced S/N limit metric, which gives the maximum S/N achievable with only clock noise present (an upper bound for any real measurement). We solve this exactly in the special case of clock white frequency noise. Most clocks are represented by a combination of noise sources for which the visibility S/N cannot be solved exactly so we close this section with a description of the numerical approach that we use for other types of noise and show S/N limits due to frequency standard noise in the case of white phase, flicker frequency, and random walk frequency noise. In section 4, we apply the results of section 3 to real clocks of interest to VLBI. With an ultimate goal of gaining insight into space VLBI, we will apply the results to existing and emerging space clocks as well as clock combinations likely to be employed in space VLBI. In section 5, we compare the S/N limitations derived here to the coherence time and comment on the Frequency to Phase Transfer technique (Zhao et al., 2017) used to reduce the impact of clock noise. In section 6, we will give recommendations for what clocks are most viable for future VLBI, in particular space VLBI.

## 2 Preliminaries

### 2.1 Requirement on the VLBI Visibility Signal to Noise Ratio

VLBI consists of a network of multiple antenna pairs, each referred to as a baseline. As the number and size of the different baselines is increased, the resulting image becomes more complete with higher resolution. Associated with each antenna baseline is a "visibility" formed by correlating the signals arriving at the two antennas. As such, each baseline can be thought of sampling one spatial frequency of the source corresponding to the baseline length. When many baselines are mathematically combined, the result is the Fourier Transform of the image. For each baseline, the basic operation of VLBI is to correlate the two signals. It is usually assumed that different points in the source are uncorrelated so that the time-averaged product of these is zero. The correlation consists of shifting the two signals relative to each other in time to determine when their product is maximized and the true time offset $\tau$ between them. Thus, the correlator output for the visibility associated with a single baseline is of the form

$$\mathcal{V} = \langle v_1(t) v_2(t+\tau) \rangle \tag{2}$$

where $v_1(t)$ is the signal received from antenna 1 in the baseline at time $t$, $v_2(t+\tau)$ is the signal from antenna 2 at time $t+\tau$ with $\tau$ adjusted to maximize $\mathcal{V}$. In general, this is a complex quantity with an amplitude and a phase that can be written as

$$\mathcal{V} = \mathcal{V}_0 e^{i\varphi(t)} \tag{3}$$

where $\varphi(t)$ has contributions from both deterministic and stochastic sources, including the local frequency standards used. The signal to noise ratio (S/N) for $\mathcal{V}$ at a given time is then defined by the ratio of the expected value for $\mathcal{V}$ at that time divided by its standard deviation at that time,

$$\mathcal{R} = \frac{|\overline{\mathcal{V}}|}{\sigma_{\mathcal{V}}} = \frac{|E[\mathcal{V}]|}{\sqrt{E[\mathcal{V}^2] - E[\mathcal{V}]^2}} \tag{4}$$

where $E[\cdot]$ is the (ensemble) expectation operator, $\overline{\mathcal{V}}$ is the mean value of $\mathcal{V}$, and $\sigma_{\mathcal{V}}$ is the standard deviation. Usually, noise in the visibility phase is more important than visibility amplitude noise (Perley, 1999), so we will assume that the amplitude $\mathcal{V}_0$ is constant. Once we have derived an expression for $\mathcal{V}$, and therefore $\mathcal{R}$, we want to know the requirement on $\mathcal{R}$. While this ultimately may be determined by a specific observation, a minimum useful single baseline visibility S/N is taken here to be 6.5 to reduce the probability of signal misidentification to less than 0.1% (Thompson et al., 2017). We would like the frequency standard to contribute to the overall S/N at the 10% level or less, so that it does not play a significant role in the overall noise budget. Given that noise from different sources will add in quadrature, for the clock contribution to the overall S/N to be less than 10%, the clock induced S/N limit ($\mathcal{R}_c$) should be at least 3× times the desired overall value, or about 20. That is, if the signal isn't changed, a 3× higher S/N means 3× lower noise. If the amplitude of clock noise is 3× lower than other noise sources, clock noise will contribute at the 10% level to the overall noise when quadrature summed with other noise sources. Thus, we will take the working requirement for clock performance to be

$$\mathcal{R}_c > 3 \times 6.5 \cong 20 \tag{5}$$

In the following sections, when we derive the S/N limits induced by various frequency standards, we will compare the resulting curves to this value of 20 to determine if the frequency standard is able to meet this threshold and at what integration times.

2.2 Including Other Noise Sources

The focus of this paper will be S/N limits imposed by clocks. The previous section gives an argument for why $\mathcal{R}_c \geq 20$ is a good starting point for a requirement on the clock contribution to a single baseline visibility S/N. This value of 20 assumes that the VLBI system is being operated at its limit such that the S/N due to non-clock sources is already close to 6.5 and in order to work, the contribution of the frequency standard to the total S/N should be 10%. We will show this threshold in graphs that follow to represent a conservative target that will work in most cases. However, if the S/N due to other noise sources is known, we can easily be more precise on the clock requirement. In appendix A, we derive the following expression that allows us to place a constraint on the visibility S/N due to the clocks used ($\mathcal{R}_c$) as a function of the target S/N ($\mathcal{R}_t$) and the S/N from other sources ($\mathcal{R}_n$)

$$\mathcal{R}_c \geq \mathcal{R}_t \sqrt{\frac{1}{1-\left(\frac{\mathcal{R}_t}{\mathcal{R}_n}\right)}} \tag{6}$$

For instance, assuming a requirement that $\mathcal{R}_t = 6.5$, if $\mathcal{R}_n = 7$ then we must have $\mathcal{R}_c \geq 17.5$. More generally, Figure 1 gives $\mathcal{R}_c$ as a function of $\mathcal{R}_n$ assuming that $\mathcal{R}_t = 6.5$. We see from the graph that the requirement of $\mathcal{R}_c \geq 20$ will effectively remove the clock noise from the overall visibility error budget in most situations.

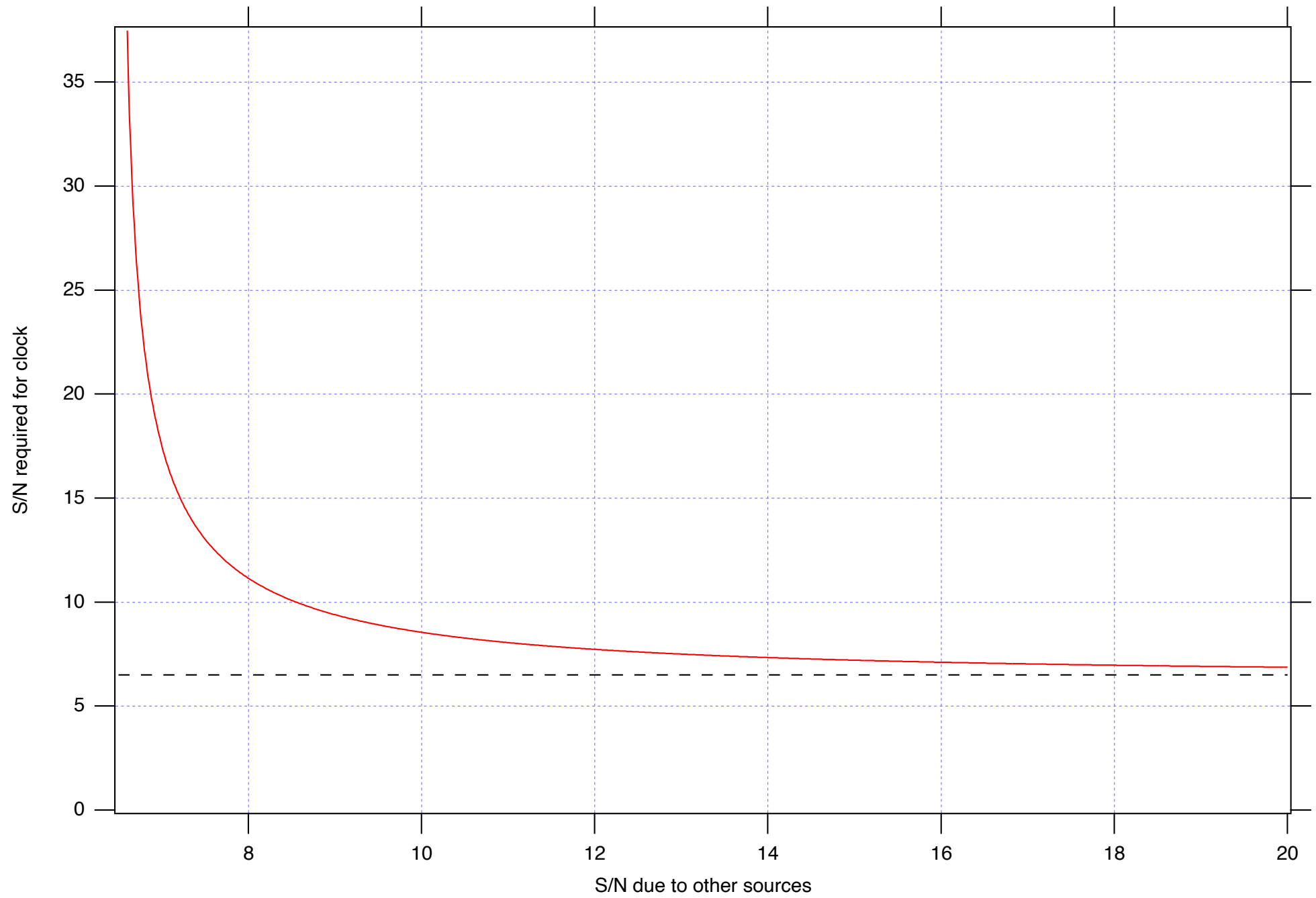

**Figure 1.** The visibility S/N required of the clocks as a function of the S/N due to all other noise sources assuming a total S/N requirement of 6.5 (red). The value of 6.5 is also shown (dashed black).

2.3 Frequency Standard Noise Characteristics

The noise of all frequency standards can be characterized as a combination of several pure noise types (rarely just one). The most common are white frequency noise, white phase noise (frequency noise in which the derived phase is white rather than the frequency itself), flicker frequency noise, and random walk frequency noise. The fractional frequency of a clock is represented by $y(t)$ and its associated phase is represented by $x(t)$ where $x(t) = \int_0^t y(t')dt'$. Frequency standard noise is usually characterized by the two-sample Allan Deviation of frequency at an averaging time $\tau$ (Allan, 1987) defined by,

$$\sigma_y(\tau) = \sqrt{\left\langle \frac{1}{2} (\bar{y}_{k-1} - \bar{y}_k)^2 \right\rangle} \tag{7}$$

where,

$$\bar{y}_k = \frac{x(t_k + \tau) - x(t_k)}{\tau} \tag{8}$$

The Allan Deviation for each pure noise type behaves differently with $\tau$: for white frequency noise $\sigma_y(t) \propto \tau^{-1/2}$, for frequency noise whose associated phase is white (white phase noise) $\sigma_y(t) \propto \tau^{-1}$, for flicker frequency noise $\sigma_y(t)$ is independent of $\tau$, and for random walk frequency noise $\sigma_y(t) \propto \tau^{+1/2}$. Note that drift is also a characteristic that we will include but is not stochastic. For drift $\sigma_y(t) \propto \tau^{+1}$. While real frequency standards can consist of a complex combination of these pure noise types, typically a particular noise type dominates the others over a given range of averaging times. One of the features of the Allan Deviation is that when displayed on a log-log plot the power of $\tau$ dependence becomes the slope of $\sigma_y(t)$ making it easy to identify a noise type at a given $\tau$.

Another useful representation of the Allan deviation of an atomic frequency standard in the special case of white frequency noise is,

$$\sigma_y(\tau) = \frac{1}{\pi(S/N_c)Q}\sqrt{\frac{T_c}{\tau}} \tag{9}$$

Here, $S/N_c$ refers to the signal to noise with which the atomic clock transition can be detected (not to be confused with VLBI visibility *S/N*), $Q = f/\Delta f$ where $f$ is the atomic clock frequency and $\Delta f$ is the full width at half maximum (FWHM) of the atomic transition line, and $T_c$ is the clock cycle time—the time required to make a single measurement of the atomic transition. From this expression we see that a higher $Q$ will result in better clock stability. For instance, in the current generation of highest performing atomic clocks, $f$ can be on the order of $10^{15}$ Hz and $\Delta f$, usually limited by the LO linewidth, can be less than 1 Hz such that the $Q$, one metric for potential clock performance, can be $10^{15}$ or greater (Margolis, 2010). By comparison, the $Q$ of a maser is about $10^9$.

The Allan Deviation is a time domain metric and is typically used to characterize clocks on time scales of 1 second or longer, thus corresponding to low frequencies. To characterize clocks for shorter times corresponding to higher frequencies, the power spectral density of frequency $S_y(f)$ or the power spectral density of phase $S_\varphi(f)$ are used (Stein, 1985). These two metrics are related (Barnes, 1971), but it is only possible to unambiguously convert the power spectral density into the Allan deviation, not the inverse. When displayed on a log-log plot, power spectral densities also have characteristic slopes. It is important not to confuse the term "phase noise", which refers to one of the frequency domain metrics, with "white phase noise" that refers to a certain noise type irrespective of the metric. Table 2 summarizes the noise characteristics.

**Table 2.** Frequency standard noise characteristics. For each of the metrics the proportionality to either frequency or averaging time is given (drift is included even though strictly speaking it is not noise).

| **Noise type** | $S_y(f)$ | $S_\varphi(f)$ | $\sigma_y(\tau)$ |
|---|---|---|---|
| White phase | $f^2$ | constant | $\tau^{-1}$ |
| White frequency | constant | $f^{-2}$ | $\tau^{-1/2}$ |
| Flicker frequency | $f^{-1}$ | $f^{-3}$ | constant |
| Random walk frequency | $f^{-2}$ | $f^{-4}$ | $\tau^{1/2}$ |
| Drift | $f^{-3}$ | $f^{-5}$ | $\tau^{1}$ |

2.4 Coherence Function C(T)

For VLBI to function the two signals collected on each baseline must remain coherent with each other over the integration time of the measurement. In practice, this is achieved by referencing these signals to an ultra-stable frequency standard located at each site. The two main causes of decoherence are limitations in the noise characteristics of these frequency standards and Earth's atmosphere. A good comparison of these two sources can be found in (Doeleman, 2011). The metric most often used to determine decoherence effects in the VLBI visibility is the coherence function $C(T)$ defined as (A. Rogers & J. Moran Jr, 1981; Thompson et al., 2017)

$$C(T) = \left| \frac{1}{T} \int_0^T e^{i\varphi(t)} dt \right| \tag{10}$$

For any given realization of phase noise, calculating this function will itself yield a random variable; therefore, we are often more interested in the statistics of this function that can, depending on the noise type, also be time varying functions. That is, we seek the following expected mean value

$$\bar{C}(T) = E[C(T)] \tag{11}$$

and its root mean square (RMS) value,

$$C_{RMS}(T) = \sqrt{E[C(T)C^*(T)]} \tag{12}$$

The result in equation (12) is typically used to determine the coherence time $T_c$ via comparing to a specified threshold value.

The coherence function and its associated statistics are monotonically decreasing with values between 1 and 0 indicting the relative magnitude of visibility phase errors due to a given noise source. As we shall see, the visibility magnitude is closely related to $C(T)$, so it also shows how the visibility degrades as a function of averaging time from decoherence. $\bar{C}(T)$ and $C_{RMS}(T)$ can be calculated exactly for white phase and white frequency noise (A. Rogers & J. Moran, 1981) and evaluated numerically for other noise types. It is often assumed that values for $C_{RMS}(T)$ for the reference clock at the VLBI antenna of above ~ 0.92, or about 1 radian of error in the white frequency noise case, are adequate. However, as we will see, this value depends on the noise type and in some cases will not yield sufficient performance. While $C_{RMS}(T)$ can be a useful metric for estimating clock impact on VLBI measurements, it suffers from four deficiencies: 1) it does not give the S/N ratio for a visibility—it is effectively only the numerator in that ratio and doesn't completely account for the noise contribution; 2) the threshold for coherence time is usually taken to be the time $T$ when the value of $C_{RMS}(T)$ corresponds to approximately 1 radian of phase error, but this is a somewhat arbitrary – why not 1.2 radians or 0.8 radians; 3) the threshold value of $C_{RMS}(T)$ required depends on the noise type and since each frequency standard usually has its own unique mix of noise types, the threshold for coherence will be different for every frequency standard; and 4) it doesn't account for other non-clock noise sources. For the latter, the actual $C_{RMS}(T)$ required for the clock will depend on the magnitude of those other sources. As other noise sources increase, the requirement on frequency standard coherence must also change so that the combination remains acceptable. In principle, this could be addressed with $C_{RMS}(T)$, but usually isn't. This is a problem for any metric and for the S/N metric we propose, we addressed this in section 2.2. Later in section 5.1 we will give a quantitative comparison between $C_{RMS}(T)$ and the S/N metric that illustrates these issues.

### 2.5 Relevant Time Scales

Historically, VLBI averaging times could vary from fractions of a second for strong sources to many thousands of seconds for weak sources or small antennas. In this paper, we are interested in VLBI with the highest level of performance and will require observing at very high frequencies (> 100 GHz). At these observation frequencies, VLBI averaging (or integration) time is often limited by atmospheric decoherence (Pesce et al., 2024; A. E. Rogers, 1988) to about 10 or several 10's of seconds. This is still the case with space VLBI that have terrestrial nodes in the network. Eventually space VLBI may exist with all antennas in space, thereby eliminating atmospheric effects. In this case, much longer averaging times may become possible even at high frequencies and other noise sources, such as those due to the frequency standards, will limit these averaging times. In this paper, we are primarily interested in the next generation of VLBI with one antenna in space, but will consider all-space VLBI as well, so that the averaging time scales of interest will range from several hundred milli-seconds to 100 s.

Another time scale of interest is the accumulation period ($T_a$) (Sasao & Fletcher, 2008). This is the period over which initial averaging is performed so as to reduce system noise (eg., receiver noise) to the point where a reliable fringe search can be performed. $T_a$ is usually much smaller than the total integration or averaging time. For instance, EHT measurements used $T_a$ = 0.4 seconds and an integration time of 10–15 seconds (Akiyama et al., 2019).

## 3 Clock Contributions to Visibility Noise

### 3.1 Derivation of Clock Contributions to the Visibility Signal to Noise Ratio

We start by deriving an expression for the VLBI visibility, including all contributions of clock noise and then derive a limit on the visibility S/N imposed by the clocks. We will then apply this to a hypothetical clock having pure white frequency noise, solving this special case both exactly and numerically. This is done in part to build intuition and to validate the numerical approach. We will also solve the equations numerically for other pure noise types typically found in clocks. In the next section, we will take on real clocks that have a mixture of pure noise types.

#### 3.1.1 Received Signal

Let the geometry of two antennae be as shown in Figure 2 with a baseline $\boldsymbol{D_\lambda}$ in units of $\lambda$, a vector $\boldsymbol{s_0}$ from the center of the baseline to the phase center of the source, a vector $\boldsymbol{\sigma}$ from the source phase center to a point of interest in the source such that $\boldsymbol{s} = \boldsymbol{s_0} + \boldsymbol{\sigma}$ (note we have assumed the source is inertially fixed with no proper motion), and a differential portion of the source around $\boldsymbol{s}$ proscribed by the angle $d\Omega$.

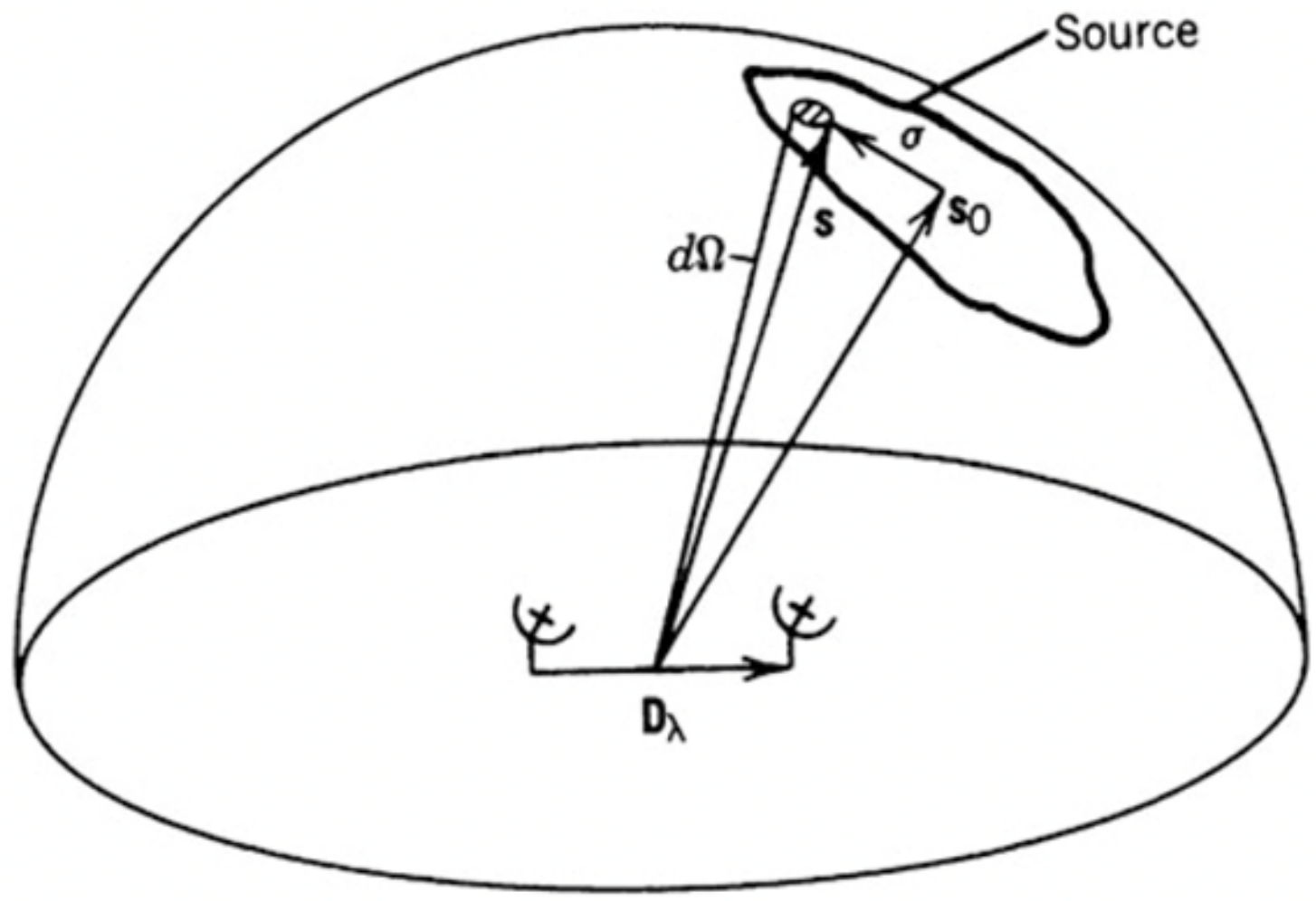

**Figure 2.** Schematic representation of two VLBI antennas and their relationship to a source (Thompson et al., 2017, figure 3.1). The vector between the two antennas is $\boldsymbol{D}_\lambda$ measured in wavelengths or $\boldsymbol{D}$ measured in meters. Note that both $\boldsymbol{s}$ and $\boldsymbol{s_0}$ as depicted in the figure are unit vectors and will be represented as $\hat{\boldsymbol{s}}$ and $\hat{\boldsymbol{s}}_\mathbf{0}$ in the text to avoid confusion. The source is projected onto the spherical surface determined by $\hat{\boldsymbol{s}}_\mathbf{0}$. The vector $\boldsymbol{\sigma} = \hat{\boldsymbol{s}} - \hat{\boldsymbol{s}}_\mathbf{0}$ points to a location in the projected source.

Given a source signal at frequency $\nu$ ("sky frequency"), the voltage signal arriving at antenna 1 is $v_1(t)$ and at antenna 2 is $v_2(t)$ from the same source point but differing by the geometric delay $\tau_g(t)$ (note that this is a time varying function) and can be expressed as follows

$$v_1(t) = \int_{-\infty}^{+\infty} \int_S v_0(\hat{\boldsymbol{s}}) e^{2\pi i \nu t} d\Omega d\nu \tag{13}$$

and

$$v_2(t) = \int_{-\infty}^{+\infty} \int_S v_0(\hat{\boldsymbol{s}}) e^{2\pi i (\nu t - \nu \boldsymbol{D}\cdot\hat{\mathbf{s}}/\boldsymbol{c})} d\Omega d\nu = \int_{-\infty}^{+\infty} \int_S v_0(\hat{\boldsymbol{s}}) e^{2\pi i \left(\nu t - \nu \tau_g(t)\right)} d\Omega d\nu \tag{14}$$

where $v_0(\hat{\boldsymbol{s}})$ is the received signal Fourier amplitude. Note that, $c\tau_g(t) = \boldsymbol{D}(t) \cdot \hat{\boldsymbol{s}} = \boldsymbol{D}(t) \cdot (\hat{\boldsymbol{s}}_\mathbf{0} + \boldsymbol{\sigma}) = \boldsymbol{D}(t) \cdot \hat{\boldsymbol{s}}_\mathbf{0} + \boldsymbol{D}(t) \cdot \boldsymbol{\sigma} = c\big(\tau_0(t) + \tau_\sigma(t)\big)$ where we define the delay to the source phase center $\tau_0(t)$ with

$$\tau_0(t) = \boldsymbol{D}(t) \cdot \hat{\boldsymbol{s}}_\mathbf{0}/c \tag{15}$$

and

$$\tau_\sigma(t) = \boldsymbol{D}(t) \cdot \boldsymbol{\sigma}/c \tag{16}$$

These two quantities will be useful later in separating out components that change significantly with time and those that do not. The expressions in equations (13) and (14) will be evaluated over some bandwidth $\Delta\nu$ once the signal is filtered, but at this point the frequency range is from $-\infty$ to $+\infty$. The region in the sky subtended by the source is projected onto a subsurface $S$ of a sphere determined by the unit vector $\hat{\boldsymbol{s}}_\mathbf{0}$. Throughout our derivation, we will assume that the source is incoherent: signals from different points in the source have no well-defined or correlated phase relationship. We assume that the respective signal amplitudes $v_{i_0}(\hat{\boldsymbol{s}}) \equiv v_0(\hat{\boldsymbol{s}})$ are the same for the two signals since the propagation paths will be very similar. Note that each antenna will normally have an antenna pattern $A_i(\boldsymbol{\sigma})$ giving the effective area of the antenna. For this derivation we absorb $A_i(\boldsymbol{\sigma})$ into $v_{i_0}(\hat{\boldsymbol{s}})$. Finally, we also assume that the signal amplitudes are not frequency dependent and are constant across the detection band.

In the receiver at each VLBI station, these signals are separately downconverted by mixing with the local oscillator (LO) at frequency $\nu_{LO}$, band-pass filtered (BPF), and then digitally sampled. The LO consists of a synthesizer at frequency $\nu_{LO}$ phase locked to the output of a frequency standard usually at 10 or 100 MHz such that $\nu_{LO}$ has all of the frequency stability characteristics of the underlying frequency standard. Figure 3 shows a schematic of the signal processing that takes place at each station.

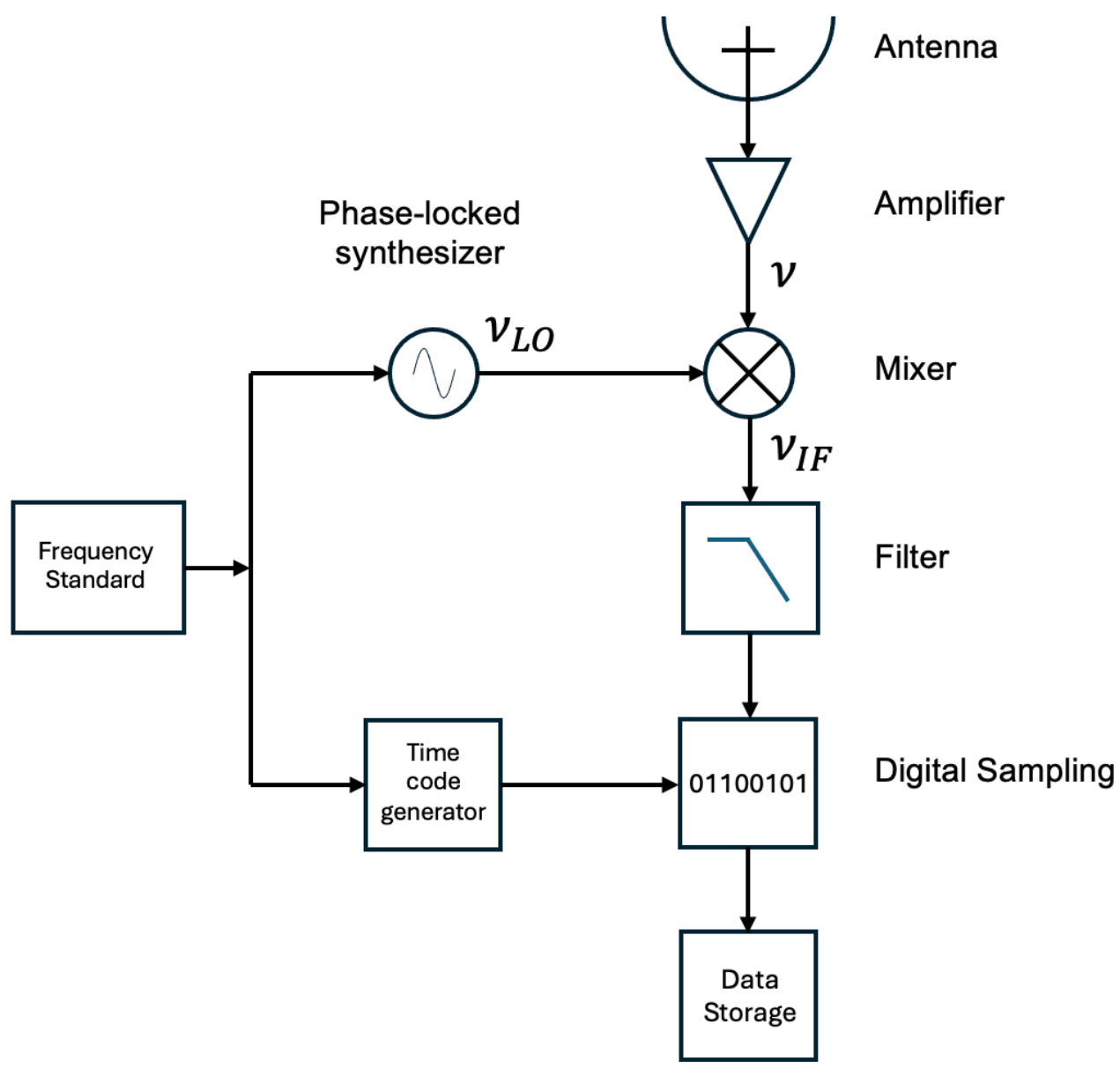

**Figure 3**. Schematic representation of the signal processing at one station. The signal at sky frequency $\nu$ is received by the antenna and amplified. This is input to a mixer to downcovert the signal to $\nu_{IF} = \nu - \nu_{LO}$ where the other input to the mixer $\nu_{LO}$ is the output of a synthesizer phase locked to the station frequency standard. The output of the mixer $\nu_{IF}$ is input to a filter and then to a digital sampler, which is also referenced to the frequency standard. Finally, the output of the sampler is stored on media that can be transported to the central correlator.

For high-frequency (> 100 GHz) VLBI, usually only a high-performance frequency standard, such as a hydrogen maser, will suffice as the LO. (Normally, only weak/distant sources will be observed, and long interrogation times will be required to obtain sufficient signal. If the antennas have large collection areas shorter integration times may be possible, thereby enabling the use of lower performing LO’s.) The hydrogen maser frequency standard noise can be approximated by white phase noise over short averaging times. As we shall see, this gives it a distinct advantage over other potential LO’s. Once the signals are processed at the individual antenna stations, they are transported to a central site for correlation. There are several different types of correlators. Here, we are only interested in the impact of frequency standard noise on the final VLBI visibility with the final result not dependent on the type of correlator, so we treat the correlation step in a general fashion without differentiating which correlator type.

3.1.2 Heterodyne Downconversion

The first step in the VLBI process that is affected by clock noise is the down conversion by the LO using a heterodyne mixer to an intermediate frequency (IF). The result of this process yields two sidebands, one at $\nu_{IF} = \nu - \nu_{LO}$ (upper) and $\nu_{IF} = \nu_{LO} - \nu$ (lower) where $\nu_{IF}$ is the variable intermediate frequency in the band $\Delta\nu$ (to be determined by the filters that are applied in the amplification stage directly after mixing) centered at $\nu_0$ . We focus on the upper sideband and absorb the LO amplitude into the $v_0$ amplitude since it will not play a part in what follows. The resulting signals after the mixer are (the superscript "h" indicates "heterodyne mixing")

$$v_1^h(t) = \int_{-\infty}^{+\infty} \int_S v_0(\hat{\boldsymbol{s}}) e^{2\pi i(\nu_{IF}t - \nu_{LO}x_1(t))} d\Omega d\nu_{IF} \tag{17}$$

and

$$v_2^h(t) = \int_{-\infty}^{+\infty} \int_S v_0(\hat{\boldsymbol{s}}) e^{2\pi i\left(\nu_{IF}t - \nu\tau_g(t) - \nu_{LO}x_2(t)\right)} d\Omega d\nu_{IF} \tag{18}$$

where the $x_i(t)$ are defined in equation (1).

3.1.3 Band Pass Filtering

Assuming filter response functions, $h_1(t)$ and $h_2(t)$, we first transform these and the voltage signals into the frequency domain resulting in $V_1^h(\nu_{IF})$, $V_2^h(\nu_{IF})$, $H_1(\nu_{IF})$, and $H_2(\nu_{IF})$ where, using the convolution theorem, the filter impact is simply the product of these

$$\begin{aligned} V_1^{hf}(\nu_{IF}^{'}) &= V_1^h(\nu_{IF}^{'})H_1(\nu_{IF}^{'}) \\ &= \left[\int_{-\infty}^{+\infty}\int_{-\infty}^{+\infty}\int_S v_0(\hat{\boldsymbol{s}}) e^{2\pi i(\nu_{IF}t - \nu_{LO}x_1(t))} d\Omega d\nu_{IF} e^{2\pi i\nu_{IF}^{'}t} dt\right] H_1(\nu_{IF}^{'}) \\ &= V_{0,1}\left[\int_{-\infty}^{+\infty}\int_{-\infty}^{+\infty} e^{2\pi i(\nu_{IF}t - \nu_{LO}x_1(t))} d\nu_{IF} e^{2\pi i\nu_{IF}^{'}t} dt\right] H_1(\nu_{IF}^{'}) \end{aligned} \tag{19}$$

where $V_{0,1} = \int_S v_0(\hat{\boldsymbol{s}})\, d\Omega$ groups the spatially varying terms, which can then be moved outside of the $d\nu_{IF}$ and $dt$ integrals. The superscript "hf" stands for "heterodyne mixed" and "filtered". In principle, $v_0(\hat{\boldsymbol{s}})$ is a variable time series and so time dependent (the time dependence is via $\boldsymbol{\sigma}$); however, as previously mentioned, visibility amplitude noise is considered insignificant compared to visibility phase noise (Perley, 1999), and we assume that the voltage amplitude varies slowly on the integration time scale. Performing the same operation on the second signal $v_2^h(t)$, using equations (15) and (16), expanding the sky frequency $\nu$ with $\nu = \nu_{LO} + \nu_{IF}$ and the delay $\tau_g(t)$ with $\tau_g(t) = \tau_0(t) + \tau_\sigma(t)$, leads to

$$
\begin{aligned}
V_2^{hf}(\nu_{IF}^{'}) &= V_2^{h}(\nu_{IF}^{'})H_2(\nu_{IF}^{'}) \\
&= \left[\int_{-\infty}^{+\infty}\int_{-\infty}^{+\infty}\int_{S} v_0(\hat{\boldsymbol{s}})e^{2\pi i\left(\nu_{IF}t-\nu\tau_g(t)-\nu_{LO}x_2(t)\right)}d\Omega d\nu_{IF}e^{-2\pi i\nu_{IF}^{'}t}dt\right]H_2(\nu_{IF}^{'}) \\
&= \left[\int_{-\infty}^{+\infty}\int_{-\infty}^{+\infty}\int_{S} v_0(\hat{\boldsymbol{s}})e^{2\pi i\left(\nu_{IF}t-(\nu_{LO}+\nu_{IF})(\tau_\sigma(t)+\tau_0(t))-\nu_{LO}x_2(t)\right)}d\nu_{IF}e^{-2\pi i\nu_{IF}^{'}t}dt\right]H_2(\nu_{IF}^{'})d\Omega \\
&= \left[\int_{S} v_0(\hat{\boldsymbol{s}})e^{-2\pi i(\nu_{LO}+\nu_0)\tau_\sigma(t)}\,d\Omega\right. \\
&\qquad \left.\times\int_{-\infty}^{+\infty}\int_{-\infty}^{+\infty} e^{2\pi i(\nu_{IF}t-\nu\tau_0(t)-\nu_{LO}x_2(t))}d\nu_{IF}e^{-2\pi i\nu_{IF}^{'}t}dt\right]H_2(\nu_{IF}^{'})
\end{aligned}
\tag{20}
$$

In equation (20) we are able to take the term $e^{-2\pi i(\nu_{LO}+\nu_{IF})\tau_\sigma(t)}$ out of the $d\nu_{IF}$ integral because over the bandwidth $\Delta\nu$ it has very little variation and it can be closely approximated by $e^{-2\pi i(\nu_{LO}+\nu_0)\tau_\sigma(t)}$ where $\nu_0$ is the center of the integration bandwidth. This is a statement that for the source, the quantity $\nu_{IF}\boldsymbol{D}\cdot\boldsymbol{\sigma}/\boldsymbol{c} \ll 1$. This is not true in general but is true for compact objects such as the SMBH's of interest here, specifically, for M87* and SagA*. Virtually all other super massive black holes will have a smaller ratio $\boldsymbol{D}\cdot\boldsymbol{\sigma}/\boldsymbol{c}$ so for the detection of any SMBH (and likely any black hole) the inequality holds as well. (The SMBH at the center of M87 is one of the largest known and is one of the closest. Any other SMBH at the center of a different galaxy will likely be further away and have a diameter approximately equal to or less than that of M87* so that the maximum extent of $\sigma_{other} \ll \sigma_{M87*}$. Though not of interest here, the argument is likely to hold for any black hole since stellar class black holes within the reach of observational methods may be much closer, but their diameter is proportionally even smaller. Increases in the IF frequency by up to a factor of 10 also leave the inequality intact, but beyond that the inequality must be re-verified.) By making this approximation we can remove this term from the $dt$ integral as well and group all of the spatially varying components together defining

$$
V_{0,2} = \int_{S} v_0(\hat{\boldsymbol{s}})e^{-2\pi i(\nu_{LO}+\nu_0)\tau_\sigma(t)}\,d\Omega \tag{21}
$$

such that

$$
V_2^{hf}(\nu_{IF}') = V_{0,2}\left[\int_{-\infty}^{+\infty}\int_{-\infty}^{+\infty} e^{2\pi i(\nu_{IF}t-\nu\tau_0(t)-\nu_{LO}x_2(t))}d\nu_{IF}e^{-2\pi i\nu_{IF}'t}dt\right]H_2(\nu_{IF}') \tag{22}
$$

We then transform back to the time domain, (note that there are two different $\nu_{IF}$ integrals and so two different integration variables $\nu_{IF}$ and $\nu_{IF}'$)

$$
\begin{aligned}
&v_1^{hf}(t') \\
&= V_{0,1}\int_{-\infty}^{\infty}\left[\int_{-\infty}^{+\infty}\int_{-\infty}^{+\infty} e^{2\pi i(\nu_{IF}t-\nu_{LO}x_1(t))}d\nu_{IF}e^{2\pi i\nu_{IF}'t}dt\right]H_1(\nu_{IF}')e^{-2\pi i\nu_{IF}'t'}d\nu_{IF}'
\end{aligned}
\tag{23}
$$

and

$$
\begin{aligned}
&v_2^{hf}(t') \\
&= V_{0,2}\int_{-\infty}^{+\infty}\left[\int_{-\infty}^{+\infty}\int_{-\infty}^{+\infty} e^{2\pi i(\nu_{IF}t-\nu\tau_0(t)-\nu_{LO}x_2(t))}d\nu_{IF}e^{-2\pi i\nu_{IF}'t}dt\right]H_2(\nu_{IF}')e^{2\pi i\nu_{IF}'t'}d\nu_{IF}'
\end{aligned}
\tag{24}
$$

#### 3.1.4 Digital Sampling

The next step that is affected by clock noise is the sampling process in which the respective signals are time tagged using the LO for later correlation. Each signal is digitally sampled at twice the IF frequency (Nyquist sampling), with sampling times at the two receivers being $t_{s_1} = t - x_1(t)$ and $t_{s_2} = t + \Delta t_s - x_2(t + \Delta t_s)$, where $\Delta t_s$ allows for a time difference introduced by the receiver electronics. The sampling time at each receiver also introduces noise due to the local oscillator at each antenna. Since $\Delta t_s$ will be nanoseconds or less and the clock noise is not expected to change on this time scale, we will take $\Delta t_s \approx 0$ and $t_{s_i} = t - x_i(t)$. The noise occurs at the IF frequency. The superscript "hfs" adds "sampled" to "heterodyne mixed and filtered." Note that sampling can introduce noise from other sources as well, but here we are only interested in the effect from the clock. In both equations (23) and (24) $t$ becomes $t - x_i(t)$ so the term $\nu_{IF}t - \nu_{LO}x_i(t)$ becomes $\nu_{IF}\big(t - x_i(t)\big) - \nu_{LO}x_i(t) = \nu_{IF}t - \big(\nu_{IF}x_i(t) + \nu_{LO}x_i(t)\big) = \nu_{IF}t - \nu x_i(t)$ where $\nu = \nu_{LO} + \nu_{IF}$. Using these simplifications the resulting sampled signals are expressed as

$$v_1^{hfs}(t') = V_{0,1} \int_{-\infty}^{\infty} \left[ \int_{-\infty}^{+\infty} \int_{-\infty}^{+\infty} e^{2\pi i (\nu_{IF}t - \nu x_1(t))} d\nu_{IF} e^{-2\pi i \nu'_{IF} t} dt \right] H_1(\nu'_{IF}) e^{2\pi i \nu'_{IF} t'} d\nu'_{IF} \quad (25)$$

and

$$v_2^{hfs}(t') = V_{0,2} \int_{-\infty}^{+\infty} \left[ \int_{-\infty}^{+\infty} \int_{-\infty}^{+\infty} e^{2\pi i (\nu_{IF}t - \nu\tau_0(t) - \nu x_2(t))} d\nu_{IF} e^{-2\pi i \nu'_{IF} t} dt \right] H_2(\nu'_{IF}) e^{2\pi i \nu'_{IF} t'} d\nu'_{IF} \quad (26)$$

The noise on the phase contributed by the LO at the IF frequency is lower than that at the LO frequency by the ratio $\nu_{IF}/\nu_{LO}$ and in certain cases can be ignored, but we retain it here for generality (for example, in very broad band detection $\nu_{IF}/\nu_{LO}$ would approach 1 and this contribution would not be negligeable).

#### 3.1.5 Delay Tracking and Fringe Stopping

Prior to correlation, the time series from the second antenna $v_2^{hfs}(t)$ is compensated for the geometric delay, called *delay tracking* and *fringe stopping,* using current best estimates (i.e., expected mean value) of the delay $\bar{\tau}_0 = \overline{\boldsymbol{D}} \cdot \hat{\boldsymbol{s}}_0/c$. The overbar '‾' is used to denote the (ensemble) expectation of the random variable defined as

$$\bar{\tau}_0(t) = E[\tau_0(t)] \quad (27)$$

The delay tracking advances $v_2^{hfs}(t)$ forward in phase at the IF rate $\nu_{IF}$ and the fringe stopping advances the phase forward at the LO rate $\nu_{LO}$ with the combined rate at $\nu$ using $\nu = \nu_{LO} + \nu_{IF}$. These delay and stopping processes yield similar effects on $v_2^{hfs}(t)$ as did sampling, in that, they effectively shift the signal in time, the resulting effects on $v_2^{hfs}(t)$ yield

$$v_2^{hfsd}(t') \\ = V_{0,2} \int_{-\infty}^{+\infty} \left[ \int_{-\infty}^{+\infty} \int_{-\infty}^{+\infty} e^{2\pi i (\nu_{IF} t - \nu \Delta\tau(t) - \nu x_2(t))} d\nu_{IF} e^{-2\pi i \nu'_{IF} t} dt \right] H_2(\nu'_{IF}) e^{2\pi i \nu'_{IF} t'} d\nu'_{IF} \tag{28}$$

where the superscript has added "d" for "delayed" and the error in the delay compensation is defined using $\Delta\tau(t) \triangleq \tau_0(t) - \bar{\tau}_0(t) = (\boldsymbol{D}(t) \cdot \hat{\boldsymbol{s}}_0 - \overline{\boldsymbol{D}}(t) \cdot \hat{\boldsymbol{s}}_0)/c$ (note that we have dropped the subscript zero on $\Delta\tau$ to make the notation more compact).

Technically, advancing the signal by $\bar{\tau}_0$ has shifted the clock error process from $x_2(t)$ to $x_2(t + \bar{\tau}_0)$; however, since our focus is on the effect of clock statistical properties on the VLBI processing the particulars of $x_2(t)$ versus $x_2(t + \bar{\tau}_0)$ are not relevant; thus, we simplify the development by replacing $x_2(t + \tilde{\tau}_g)$ with $x_2(t)$. Also, note that we do not make the quasi-static approximation here as is typically done. As previously mentioned, this approximation states that the clock phase varies slowly on the time scale of the measurement. This is true for the deterministic drift of the phase, but not of the noise on the phase. Thus, for the purpose of evaluating the effect of clock noise on the overall VLBI signal, the $x_i(t)$ terms cannot be treated as approximately constant and cannot be removed from the time integral.

We now have the signals processed through the individual receivers and turn to the central correlation that results in the visibility.

#### 3.1.6 Correlation

Correlating the two preceding voltage signals yields the measured visibility, which contains the ideal visibility as well as decorrelating effects of the oscillators and baseline errors. Assessing the impact of the oscillators on the measured visibility is the goal of the present work. In practice, the measured output of the correlator $\mathcal{V}_m(t_k; T)$ (where the subscript '*m*' represents measured value) is the averaged product of the two signals (expressed at the same effective time) over some integration time $T$ (shown as a parameter) and can be expressed as the real value of the following complex-valued expression

$$\mathcal{V}_m(t_k; T) = \mathbb{R}\left\{ \left\langle v_1^{hfs}(t_k) v_2^{hfsd*}(t_k) \right\rangle_T \right\} = \mathbb{R}\left\{ \frac{1}{T} \int_{t_k}^{t_k+T} v_1^{hfs}(s) v_2^{hfsd*}(s)\, ds \right\} \tag{29}$$

where we have introduced the discrete time $t_k$ and index $k$ to acknowledge that a typical observation often concatenates a set of $\{\mathcal{V}_m(t_k; T)\}$ over a period time to form a test function that is used in fringe fitting to find a correlation peak. To simplify the expressions that follow, we will drop the $\mathbb{R}\{\quad\}$ notation and implicitly assume the real valued part of the complex expressions is the relavant result. Because equation (29) is an integral over a finite time, the result is a random variable with its own statistics. Furthermore, since oscillator and baseline errors are in general nonstationary, the result of equation (29) is also a random process. As a result, we will need the expected value and associated variance for the result in equation (29). That is, the mean value of the measured $\mathcal{V}_m(t_k; T)$ is defined as

$$\overline{\mathcal{V}}_m(t_k; T) \triangleq E\left[ \left\langle v_1^{hfs}(t_k) v_2^{hfsd*}(t_k) \right\rangle_T \right] \tag{30}$$

where we again use the overbar '$\overline{\quad}$' to denote the (ensemble) expectation. Now, if all error sources were stationary processes with zero mean and time-independent variance, the expectation itself could be expressed as a further time average; however,this condition does not hold for many of the noise sources considered here and we will not make this assumption. (In the case of all stationary processes, the expectation would be redundant and not present in the analysis. That is, the time average operation $\left\langle v_1^{hfs}(t')v_2^{hfsd^*}(t')\right\rangle_T$ yields a value that is approximately constant across time, and the expectation of a constant returns itself; therefore, the additional expectation operation could be eliminated from the analysis. In our case with nonstationary noise sources, the finite time average operation yields another random variable and the expectation operation is needed to determine the mean value of this average.) Substituting equations (25) and (28) into (30) yields

$$\begin{aligned}\overline{\mathcal{V}}_m(t_k;T) &= E\left[\left\langle v_1^{hfs}(t_k)v_2^{hfsd^*}(t_k)\right\rangle_T\right] \\ &= E\left[\frac{1}{T}\int_{t_k}^{t_k+T}\left(V_{0,1}\int_{-\infty}^{\infty}\left[\int_{-\infty}^{+\infty}\int_{-\infty}^{+\infty}e^{2\pi i(\nu_{IF}t-\nu x_1(t))}d\nu_{IF}e^{-2\pi i\nu'_{IF}t}dt\right]H_1(\nu'_{IF})e^{2\pi i\nu'_{IF}s}d\nu'_{IF}\right)\right. \\ &\left.\times\left(V_{0,2}\int_{-\infty}^{+\infty}\left[\int_{-\infty}^{+\infty}\int_{-\infty}^{+\infty}e^{-2\pi i\left(f_{IF}r-\nu(\Delta\tau(r)+x_2(r))\right)}d\nu_{IF}e^{2\pi if'_{IF}r}dr\right]H_2^*(f'_{IF})e^{-2\pi if'_{IF}s}df'_{IF}\right)ds\right]\end{aligned} \tag{31}$$

where we have introduced dummy variables to aid in distinguishing the different independent variables in the seven different integrals. Rearranging equation (31) leads to the following intermediate expression that we will operate on

$$\begin{aligned}\overline{\mathcal{V}}_m(t_k;T) = E\Bigg[\frac{1}{T}\int_{t_k}^{t_k+T}V_{0,1}V_{0,2}^*\int_{-\infty}^{\infty}\int_{-\infty}^{\infty}\int_{-\infty}^{\infty}\int_{-\infty}^{\infty}\int_{-\infty}^{\infty}\int_{-\infty}^{\infty}e^{2\pi i\nu\left(x_2(r)-x_1(t)+\Delta\tau(r)\right)} \\ \times e^{2\pi i(\nu_{IF}t-f_{IF}r)}e^{2\pi i(f'_{IF}r-\nu'_{IF}t)}e^{2\pi i(\nu'_{IF}-f'_{IF})s} \\ \times H_1(\nu'_{IF})H_2^*(f'_{IF})d\nu_{IF}df_{IF}dtdrd\nu'_{IF}df'_{IF}ds\Bigg]\end{aligned} \tag{32}$$

Before proceeding with the operations in equation (32) we note that, unlike the clock noise or baseline errors, the source intensity in the received signals are stationary with zero mean (i.e, $E[v_0(\hat{\boldsymbol{s}})]=0$) and have constant variances. Furthermore, the intensity is incoherent spatially, in time, and in frequency. These observations lead to useful simplifications of equation (32) that make it analytically tractable in certain cases. First, we define the ideal visibility, consistent with (Thompson et al., 2017), as

$$\mathcal{V}_0 \triangleq \int_S I(\hat{\boldsymbol{s}})e^{2\pi i\nu\tau_\sigma}\,d\Omega = \int_{-\infty}^{\infty}\int_{-\infty}^{\infty}I(l,m)e^{2\pi i\nu\left(ul+vm+w\left(\sqrt{1-l^2-m^2}-1\right)\right)}\frac{dldm}{\sqrt{1-l^2-m^2}} \tag{33}$$

where, to emphasize the spatial dependence of equation (33), we have made explicit the relationship to the $(u,v,w)$ coordinate system ($w$ is aligned along $\hat{\boldsymbol{s}}_{\mathbf{0}}$ and $(u,v)$ define the plane that is orthogonal) and the associated direction cosines $(l,m)$ measured with respect to $u$ and $v$, respectively. Using the incoherence properties of $v_0(\hat{\boldsymbol{s}})$ we can express $E[v_0(\hat{\boldsymbol{s}})v_0^*(\hat{\boldsymbol{s}}')]$ as

$$E[v_0(\hat{\boldsymbol{s}})v_0^*(\hat{\boldsymbol{s}}')] = I(\hat{\boldsymbol{s}})\delta(\hat{\boldsymbol{s}}-\hat{\boldsymbol{s}}')\delta(\nu_{IF}-\nu'_{IF})\delta(\nu_{IF}-f_{IF})\delta(\nu'_{IF}-f'_{IF})\delta(t-r)\delta(t-s) \tag{34}$$

Applying equation (34) to $E[V_{0,1}V_{0,2}^*]$ we arrive at the following useful expression

$$\begin{aligned} E[V_{0,1}V_{0,2}^*] &= E\left[\int_S\int_S v_0(\hat{\boldsymbol{s}})v_0^*(\hat{\boldsymbol{s}}')e^{2\pi i\nu\tau_{\sigma'}}\, d\Omega d\Omega'\right] \\ &= \int_S\int_S E[v_0(\hat{\boldsymbol{s}})v_0^*(\hat{\boldsymbol{s}}')]e^{2\pi i\nu\tau_{\sigma'}}\, d\Omega d\Omega' \\ &= \int_S\int_S I(\hat{\boldsymbol{s}})e^{2\pi i\nu\tau_{\sigma'}}\, d\Omega d\Omega'\; \delta(\hat{\boldsymbol{s}}-\hat{\boldsymbol{s}}')\delta(\nu_{IF}-f_{IF})\delta(\nu_{IF}'-f_{IF}')\delta(t-r)\delta(\nu_{IF}-\nu_{IF}')\delta(t \\ &\quad - s) \\ &= \int_S I(\hat{\boldsymbol{s}})e^{2\pi i\nu\tau_{\sigma'}}\, d\Omega\delta(\nu_{IF}-f_{IF})\delta(\nu_{IF}'-f_{IF}')\delta(t-r)\delta(\nu_{IF}-\nu_{IF}')\delta(t-s) \\ &= \mathcal{V}_0\delta(\nu_{IF}-f_{IF})\delta(\nu_{IF}-f_{IF})\delta(\nu_{IF}'-f_{IF}')\delta(t-r)\delta(\nu_{IF}-\nu_{IF}')\delta(t-s) \end{aligned} \tag{35}$$

Turning back to equation (32), we use Fubini's theorem to move the expectation inside of the integrals and apply the independence of the separate noise/error processes from each other to yield

$$\begin{aligned} \overline{\mathcal{V}}_m(t_k;T) &= \frac{1}{T}\int_{t_k}^{t_k+T} E[V_{0,1}V_{0,2}^*]\int_{-\infty}^{\infty}\int_{-\infty}^{\infty}\int_{-\infty}^{\infty}\int_{-\infty}^{\infty}\int_{-\infty}^{\infty}\int_{-\infty}^{\infty} E\left[e^{2\pi i\nu(x_2(r)-x_1(t)+\Delta\tau(r))}\right] \\ &\quad \times e^{2\pi i(\nu_{IF}t-f_{IF}r)}e^{2\pi i(f_{IF}'r-\nu_{IF}'t)}e^{2\pi i(\nu_{IF}'-f_{IF}')s}H_1(\nu_{IF}')H_2^*(f_{IF}')d\nu_{IF}df_{IF}dtdrd\nu_{IF}'df_{IF}'ds \end{aligned} \tag{36}$$

Subsitituting equation (35) into equation (36) results in

$$\begin{aligned} \overline{\mathcal{V}}_m(t_k;T) = \frac{\mathcal{V}_0}{T}\int_{t'}^{t'+T} &\int_{-\infty}^{\infty}\int_{-\infty}^{\infty}\int_{-\infty}^{\infty}\int_{-\infty}^{\infty}\int_{-\infty}^{\infty}\int_{-\infty}^{\infty} E\left[e^{2\pi i\nu(x_2(r)-x_1(t)+\Delta\tau(r))}\right] \\ &\times e^{2\pi i(\nu_{IF}t-f_{IF}r)}e^{2\pi i(f_{IF}'r-\nu_{IF}'t)}e^{2\pi i(\nu_{IF}'-f_{IF}')s} \\ &\times H_1(\nu_{IF}')H_2^*(f_{IF}')\delta(\nu_{IF}-f_{IF})\delta(\nu_{IF}'-f_{IF}')\delta(t-r)\delta(\nu_{IF}-\nu_{IF}')\delta(t-s) \\ &\times d\nu_{IF}df_{IF}dtdrd\nu_{IF}'df_{IF}'ds \end{aligned} \tag{37}$$

where we note that $\mathcal{V}_0$ has been moved outside of the time integral because it can be approximated as a constant. Applying the delta functions produces the following expression for the expected value of the measured visbility as output by the correlator

$$\overline{\mathcal{V}}_m(t_k;T) = \frac{\mathcal{V}_0}{T}\int_{t_k}^{t_k+T}\int_{-\infty}^{\infty} E\left[e^{2\pi i\nu(x_2(s)-x_1(s))}\right]E\left[e^{2\pi i\nu\Delta\tau(s)}\right]H_1(\nu_{IF})H_2^*(\nu_{IF})d\nu_{IF}ds \tag{38}$$

Since our present analysis is focused on the effects of clock noise (i.e., $\psi_i(t)$ in equation (1)) on the measured visibility we will assume the baseline correction $\bar{\tau}_0(t)$ is perfect and that the deterministic clock error components are also perfect (i.e., and $\{x_{i_0}, y_{i_0}, D_i\} = 0$). That is, the following can be assumed

$$\begin{aligned} &\Delta\tau(t) = 0 \text{ if } \bar{\tau}_0(t) \text{ is exact} \\ &x_i(t) = \psi_i(t) \text{ if } \{x_{i_0}, y_{i_0}, D_i\} \triangleq 0 \end{aligned} \tag{39}$$

Leading to our final measured visibility expression for analyzing clock noise effects

$$\boxed{\overline{\mathcal{V}}_m(t_k;T) = \frac{\mathcal{V}_0}{T}\int_{t_k}^{t_k+T}\int_{-\infty}^{\infty} E[e^{i\Phi(s,\nu)}]H_1(\nu_{IF})H_2^*(\nu_{IF})d\nu_{IF}ds} \tag{40}$$

where $\Phi(t,\nu) \triangleq 2\pi\nu\big(\psi_2(t)-\psi_1(t)\big)$.

The quantity $E\left[e^{i\Phi(t,\nu)}\right]$ contains all of the clock noise and is central to calculating the impact of this source of noise on VLBI S/N. From appendix B equation (A19) we derive an expression for this expectation in the special case of white frequency noise

$$E\left[e^{i\Phi(t,\nu)}\right] = e^{-4\pi^2\nu^2\sigma_y^2(1)t} \tag{41}$$

$(\nu = \nu_{LO} + \nu_{IF})$, which for a given clock with Allan Variance at 1 second $\sigma_y^2(1)$ can be evaluated and substituted into equation (40) to obtain the measured visibility $\mathcal{V}_m$. Note the explicit dependence on $t$. For each value of $t$, $E\left[e^{\mathrm{i}\Phi(t,\nu)}\right]$ and the integral in equation (40) can have a different value. Equation (41) is only valid for white frequency noise (and random walk in the derived phase). For other noise types, $E\left[e^{i\Phi(t,\nu)}\right]$ will be evaluated numerically. The VLBI visibility S/N, given in equation (4), can be expressed more explicitly as a function of $\overline{\mathcal{V}}_m(t_k;T)$ and its associated variance $E[\mathcal{V}_m(t_k;T)^2]$ as follows

$$\boxed{\mathcal{R} = \frac{\left|\overline{\mathcal{V}}_m(t_k;T)\right|}{\sigma_{\mathcal{V}_m}} = \frac{\left|\overline{\mathcal{V}}_m(t_k;T)\right|}{\sqrt{E[\mathcal{V}_m(t_k;T)^2] - \overline{\mathcal{V}}_m(t_k;T)^2}}} \tag{42}$$

When calculated for clock noise only (no other noise source), the result in equation (42) will represent a metric for the "limitation on S/N imposed by the clocks" and is denoted $\mathcal{R}_c$, which will be an upper bound on the S/N possible once other noise sources are added in. Note that the ideal visibility (no noise), $\mathcal{V}_0$ is in common to the numerator and denominator of $\mathcal{R}_c$ and will cancel out leaving a source-independent result, as expected for multiplicative noise.

Equations (40) – (42) are the primary theoretical results of this paper. In some cases, they can be solved exactly, but in most cases, they must be solved numerically. In the next section we will solve them exactly in the special case of white frequency noise and then evaluate them numerically as a way of partially validating the numerical approach. The method used to numerically evaluate the S/N for a frequency standard with an arbitrary Allan deviation is as follows.

Assuming an integration time $T$,

1. Generate two time series of frequency offsets representing the clocks at each end of a baseline. Each time series should have an Allan deviation $\sigma_{yi}(\tau)$—this could be a pure noise type or a mixture of different noise types. In general $\sigma_{yi}(\tau)$ can be different for different clocks, but can be the same if the same clock type is used at each station.
2. Integrate each frequency time series to get two phase time series $\varphi_1(t)$ and $\varphi_2(t)$.
3. Form a phase difference time series, $\Phi(t) = \varphi_2(t) - \varphi_1(t)$.
4. If the accumulation period is enabled, time average $\Phi(t)$ over the accumulation period $T_a$ to get a new time series at $T_a$ time steps $\Phi(t_k)$ where $t_k$ identifies the end time of each accumulation period in the sequence that makes up the integration time $T$.
5. If fringe rate correction is enabled, simulate it here: Fit this $T_a$-averaged time series to a straight line with slope $m(T)$ and remove $m(T)$ from $\Phi(t_k)$ to get a corrected phase $\Phi_c(t_k)$. (In real VLBI this can't be done since the visibility amplitude $\mathcal{V}_0$ is not known – only the product of $\mathcal{V}_0$ and a phase term. Here, for convenience, we have set $\mathcal{V}_0 = 1$ so that we can invert $e^{i\Phi_c(t)}$ to get $\Phi_c(t)$ and do the fringe rate fitting directly. The two approaches will give the same result as long as $\mathcal{R}_n$ is sufficiently large so as to allow unambiguous detection of the fringe, which is the situation of interest here.)

6. Take $\mathbb{R}[e^{i\Phi_c(t_k)}] = \cos[\Phi_c(t_k)]$.
7. If fringe rate correction is enabled, form the visibility, $\mathcal{V}(T) = \langle\cos[\Phi_c(t_k)]\rangle_T$, at integration time $T$ and its square, otherwise $\mathcal{V}(T) = \langle\cos[\Phi(t)]\rangle_T$ (and its square).
8. Repeat steps 1-7 $N_{ave}$ times to obtain the expectations, $E[\mathcal{V}(T)]$ and $E[\mathcal{V}^2(T)]$.
9. Calculate the standard deviation of $\mathcal{V}(T)$, $\sigma_{\mathcal{V}} = \sqrt{E[\mathcal{V}^2(T)] - (E[\mathcal{V}(T)])^2}$.
10. Calculate the clock-limited S/N at integration time $T$, $\mathcal{R}_c(T) = E[\mathcal{V}(T)]/\sigma_{\mathcal{V}}$.
11. Repeat steps 1-10 for each integration time $T$ of interest.

This procedure for numerically calculating $\mathcal{R}_c$ is the next main result of the paper. It is valid for any mixture of noise types thereby enabling the analysis of realistic clocks, which typically have noise characteristics not well-represented by white frequency noise alone.

### 3.2 A "Toy" Clock Model: the Signal to Noise Ratio for White Frequency Noise

In this section we consider the exact expression for $\mathcal{R}_c$ in the special case of a "toy" hypothetical clock with pure white frequency noise that we solve both exactly and numerically. The focus here will not be on real clocks but on illustrating the S/N behavior in a simple case. Real clocks have a combination of noise types that we will treat in subsequent sections. Often the complex noise characteristics of clocks are most apparent when all time scales are included but many can be represented by white frequency noise over certain averaging times; thus, the solution in the white frequency case can serve as a useful approximation. It is one of the few cases that can be solved exactly and, in addition to being a useful approximation to more complex cases, it will serve to validate the numerical approach that must be used for these.

The standard metric for clock performance is the Allan deviation described in section 2.3. We consider a clock with Allan deviation $1 \times 10^{-13}/\sqrt{T}$, which is similar to a hydrogen maser (masers average down faster at short times from about 1 to 10 seconds but are typically noisier than this at 1 second). The frequency difference between two clocks with this performance level will have an Allan deviation $\sqrt{2}$ higher or $\sqrt{2} \times 10^{-13}/\sqrt{T}$. Figure 4 shows the Allan deviation and RMS coherence function $C_{RMS}(T)$ introduced in section 2.4. We calculate $C_{RMS}(T)$ to second order in the case of white frequency noise using (Rogers, A., & Moran Jr, J., 1981). We also use 230 GHz as representative of current high performance VLBI. Later we will consider other observation frequencies. Both the Allan deviation and the coherence function shown in figure 4 are for the difference of the two clocks most relevant to the VLBI correlation process. One radian of phase error in the clocks at the averaging time of interest is often considered adequate for VLBI and for white frequency noise corresponds to $C_{RMS}(T) \approx 0.92$, which here occurs at about 33 seconds of averaging time.

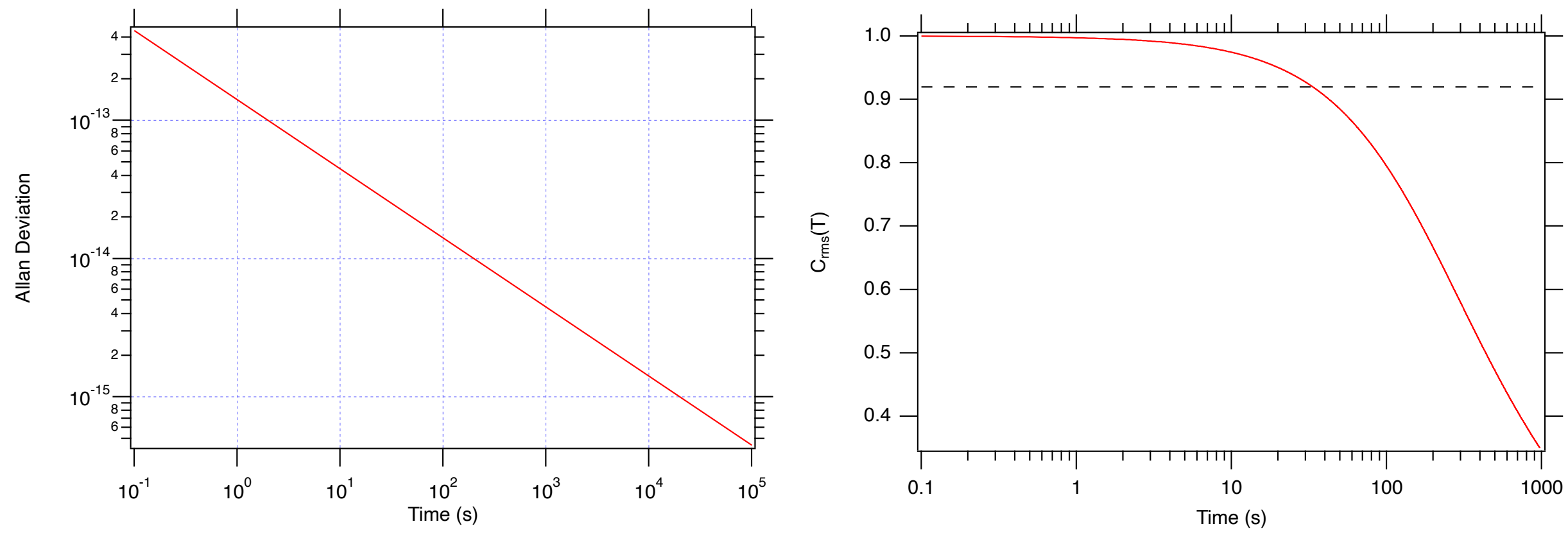

**Figure 4.** On the left is the Allan deviation of fractional frequency variations for the difference of two hypothetical clocks each with stability $1 \times 10^{-13}/\sqrt{\tau}$. On the right is the coherence function $C_{RMS}(T)$ for the difference between two clocks with this performance level. The black dashed line indicates the value of 0.92, which corresponds to about 1 radian of phase error for this noise type.

From appendix B, equation (A19) and the previous section, an important intermediate result is for the expectation of the clock noise in the carrier signal

$$E\left[e^{i\Phi(t,\nu)}\right] = E\left[e^{i\left(\varphi_2(t,\nu)-\varphi_1(t,\nu)\right)}\right] = e^{-4\pi^2\nu^2\sigma_{y1}^2 t}$$

where $\nu = \nu_{LO} + \nu_{IF}$ is the VLBI observation frequency.The solution for $\mathcal{R}_c$ in the white frequency noise case is derived in appendix C. We need to calculate the expected measured visibility $\overline{\mathcal{V}}_m(T)$ given by equations (A27) and (A31) in that appendix with the results

$$\overline{\mathcal{V}}_m(T) = \Delta\nu \mathcal{V}_0 H_0^2 \frac{1 - e^{-\eta T}}{\eta T} \tag{43}$$

$$E[\mathcal{V}_m(T)^2] = (\Delta\nu \mathcal{V}_0 H_0^2)^2 \frac{e^{-4\eta T} + 8e^{-\eta T} + 12\eta T - 9}{12(\eta T)^2} \tag{44}$$

where $\eta \triangleq 4\pi^2\nu^2\sigma_{y1}^2 \approx 4\pi^2(\nu_{LO} + \nu_0)^2\sigma_{y1}^2$ and $\Delta\nu$ is the VLBI bandwidth. Note that $\eta$ has units of $1/seconds$ so that $\eta T$ is unitless (see Appendix B).

Finally, from appendix C equation (A32), the clock noise induced signal to noise ratio $\mathcal{R}_c$ in the special case of white frequency noise is given by

$$\mathcal{R}_c = \frac{\sqrt{12}(1 - e^{-\eta T})}{\sqrt{e^{-4T\eta} - 12e^{-2T\eta} + 32e^{-T\eta} + 12T\eta - 21}} \tag{45}$$

This expression for $\mathcal{R}_c$ gives an approximation to the limit on S/N due to the clocks at each antenna in the case of clock white frequency noise and infinite detection bandwidth.

In real VLBI measurements, the signal often has a power spectrum (measured in Jy) orders of magnitude lower than the thermal noise of the receiver. Since thermal noise is white, it averages down both with time and with bandwidth. To analyze data, a VLBI measurement incorporates a bandwidth and an “Accumulation Period” ($T_a$) over which averaging is performed. The value of $T_a$ is usually much shorter than the full integration time $T$ but is set as long as possible so as to reduce the thermal noise without the risk of accumulating more than a radian of phase and incurring a cycle slip as with the example already given for EHT measurements where $T_a = 0.4$ seconds, while $T =$10–15 seconds. For each baseline, the correlated data is first averaged over $T_a$ to produce a sub-series of values. The rate change of the time delay between the two stations ("fringe rate") and the slope of the frequency dependence ("group delay") are both varied to form a two-dimensional search for the correlated fringe peak—a process referred to as "fringe fitting". Fringe fitting for the fringe rate is equivalent to removing a slope from the visibility phase over the integration time.

In the case of clock noise that has a white phase characteristic, this process will not improve the clock noise contributions to limitations in S/N. In fact, it may make it worse: a slope calculated for pure white noise is itself a stochastic variable. Subtracting or adding such a slope to the data can add noise. In the case of white frequency clock noise, the phase is random walk, whose variance increases with time, and removing a slope can reduce that variance even though there is no *deterministic* slope present. This may be true of other noise types whose variances are time-dependent as well. However, in cases where there is no deterministic drift due to the clock, removing a slope has some risk. Never-the-less, the fringe fitting process must be performed in order to optimize signal detection for other reasons and the fringe fitting process will not be able to distinguish between an actual deterministic rate due to, for instance, a clock frequency offset and a stochastic rate due to the time-dependence of the variance.

A slope is removed in the visibility phase to simulate solving for the fringe rate in the numerical solutions. However, in the exact solution, removal of a phase slope is perfect thereby giving an infinite S/N, so is not included here.

We will now solve the white frequency noise case numerically and compare it to the analytic result in equation (45) as a partial validation of the numerical approach, which must be used in more complex cases. For comparison to the exact solution, we will perform the numerical solution with fringe rate fitting disabled using the algorithm described in the previous section.

For the comparison of the numerical and exact solutions in the white frequency noise case, we look at the VLBI frequency 230 GHz. As shown in Figure 5a the numerical solutions diverge slightly from the exact solution at the shortest times simulated due to the finite time step used: as the time step is reduced (effectively higher bandwidth), agreement improves and the graph shows that a time step of 10× smaller than the smallest time scale of interest should be used. However, all solutions are independent of the time step for integration times greater than a second, which is of most interest to high frequency VLBI. Simulation results had little sensitivity to varying the longest averaging time from 1000 to 10,000 seconds. In the simulations that follow, a step size of 0.1 seconds was chosen as a good compromise between higher resolution and the ability to run the simulation out to integration times of interest in a reasonable amount of time. The graphs therefore plot integration times from 1 to 100 s. This represents a narrow bandwidth but is not too different from what is done in a real VLBI measurement where averaging is performed over the accumulation period of approximately 1 second mentioned previously. In the special case considered here, the exact solution shows that the clock-limited S/N has a constant slope in a log plot. In the simulation we need only sample a sub-band to determine what that slope is. We then expand the band to verify that the solution doesn't change. For other noise types and mixtures the S/N may not have a constant log-log slope on the time frame of interest. In these cases, we will use the same technique of expanding the time range in the simulation to make sure that the result doesn't change. This approach is justified for clock noise because most clock noise types fall off rapidly at short times and high frequency (see Table 2). The exception is white phase noise, but with this noise type averaging is a valid method for reducing noise.

As described above, a real VLBI measurement will average the raw data over an accumulation period $T_a$, which is much less than the integration time $T$ in order to reduce white thermal noise from the receiver. In addition, it is necessary to perform a 2-dimensional fringe search, which determines the optimal group delay ($\partial\mathcal{V}/\partial\omega$) and fringe rate ($\partial\mathcal{V}/\partial t$). Figure 5b shows the numerical solution that now accounts for $T_a$ and the fringe rate as well as the original solution with neither of these (an exact solution accounting for the fringe rate is not included for the reason mentioned above). It is clear that in the white frequency clock noise case, there is significant advantage to fringe fitting due to the random walk characteristic of the associated phase and its time-dependent variance. The fringe fitting process (slope removal) prevents the phase from running away, thus reducing the variance and increasing the S/N. As noted above, this process may not improve the results for certain noise types. However, for most clock noise characteristics considered here this potential degradation in S/N is small and there is no disadvantage to fringe fitting. As mentioned, for real VLBI it must be done anyway for other reasons, so we must perform it in the simulation in order to adequately represent real measurements.

The corrected solution approaches the uncorrected solution at short simulation times because there the phase has not yet deviated significantly from the single point noise level and the correction has little effect. This reflects the fact that there is no advantage to removing a visibility slope in the data until the total phase has diverged beyond the noise in adjacent phase values.

Note that even with fringe rate fitting turned on (Figure 5b), the clock-limited S/N quickly drops below 20. This however does not rule out the use of such a frequency standard at longer integration times. It simply means that as integration time increases, the incidence rate of data for which the fringe cannot be resolved goes up. Statistically there will still be times when the clock noise will not be a limitation. We will return to this when we look at Ultra-Stable Oscillators and the Hydrogen Maser in the next sections.

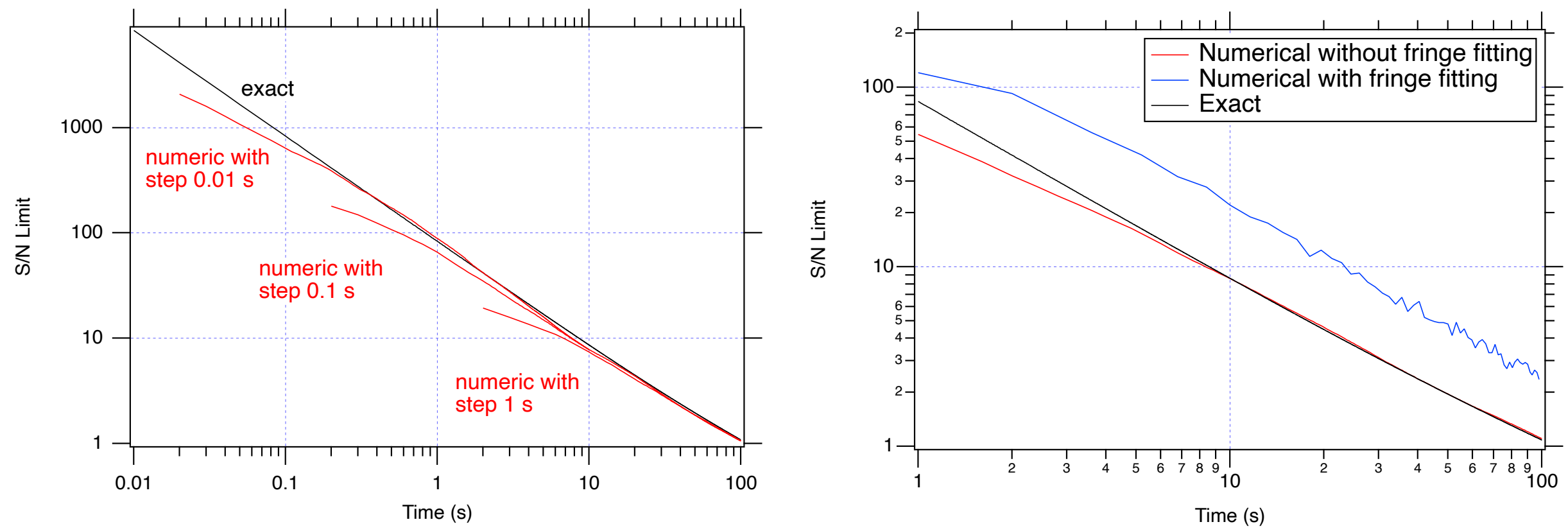

**Figure 5. a.** On the left, for a VLBI frequency of 230 GHz, a comparison between the exact (black trace) and numerical expressions (red traces) for the Clock-limited S/N in the case of white frequency noise where each clock has an ADEV of $1 \times 10^{-13}/\sqrt{\tau}$ and no accumulation period averaging is performed. Results for numerical time steps of 0.01 s, 0.1 s, 1 second are superimposed on the exact solution giving good agreement when the time step is 10× the shortest time scale of interest. **b.** On the right is shown the same parameters but with the $T_a$ = 0.4 s. In solid blue is a numerical simulation of the S/N limit with fringe fitting incorporated and for comparison the numerical solution with no fringe fitting (red) and the exact solution with no fringe fitting (black).

We now extend the white frequency noise case to other VLBI frequencies of interest. Figure 6a shows the comparison between exact and numeric solutions for the VLBI frequencies 90, 230, 345, and 630 GHz using a time step of 0.01 s, no accumulation period, and fringe fitting off. Again, we see good agreement between the simulation and the exact solution. Figure 6b shows simulation results with $T_a$ = 0.4 seconds and fringe fitting turned on. In this case the results show that fringe fitting improves the usable integration time by about a factor of 2.

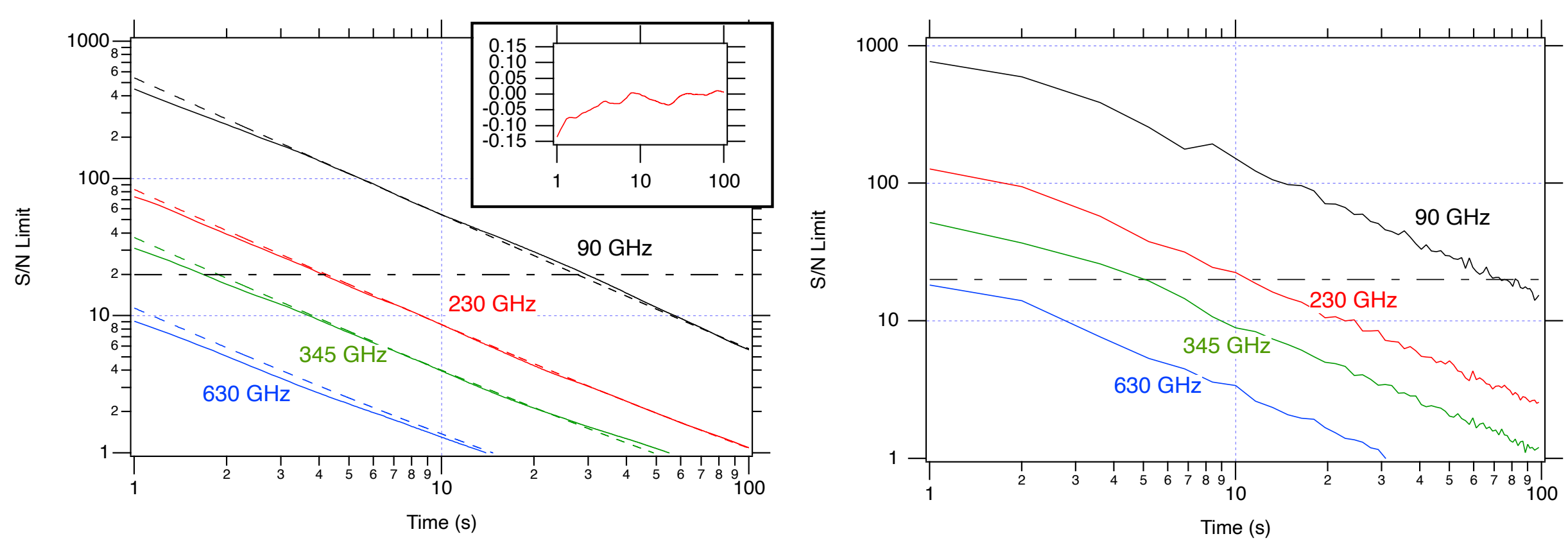

**Figure 6. a.** On the left is a graph extending the comparison between exact (dashed) and numeric (solid) solutions in the case of the clock white frequency noise of $1 \times 10^{-13}/\sqrt{\tau}$ for the VLBI frequencies 90 (black), 230 (red), 345 (green) and 630 (blue) GHz. The inset shows the fractional variation between the exact and numeric solution in the representative case of 230 GHz indicating agreement within 10% for integration times of interest. **b.** On the right is a graph

showing simulation results for the same clock stability but now with $T_a$ = 0.4 seconds and with fringe fitting enabled. In both graphs the threshold S/N of 20 is the alternating dashed black line.

With the understanding that the numerical solution should only be used for averaging times 10× the time step (0.1 s), the agreement between the exact and numerical solutions is at the 10% level or better (see the inset in Figure 6a, thereby giving confidence in the numerical approach.

It was stated above that a commonly accepted value for the coherence of 0.92 might be adequate for VLBI in the case of white frequency noise because it corresponds to about 1 radian of phase error. Note that in the case of this hypothetical clock pair, while the coherence function is above the candidate threshold of 0.92 for up to 33 seconds when observing at 230 GHz (see Figure 4), the S/N limit imposed by these clocks is already below 3 for uncorrected and below 7 for corrected data at this time (Figures 5b and 6b). A S/N of 7 would normally be too low to unambiguously detect the interferometric fringe because this is already at the lower limit and other noise sources have not been included (see section 2.1 above). *This is a key finding of this work: the coherence function for clock noise may not be an adequate metric for determining the clock performance required by high-frequency VLBI. In section 2.1 we suggested a threshold on the S/N limit imposed by the clock of 20 as an alternative. Here, for the case of 230 GHz, the corrected clock-limited S/N goes below this threshold at about 11 s*.

As shown in Figure 6b, in the case of white frequency noise, fringe fitting improves the situation for all frequencies considered. For instance, at 90 GHz, the time at which the S/N falls below the threshold of 20 improves by about a factor of 2. However, Figure 6 also shows that even with fringe fitting, white frequency noise at this level, which would generally be considered excellent relative to most clocks currently capable of operating continuously in the field and in particular space clocks, may not be sufficient to support high performance VLBI. For actual high-frequency VLBI, astonomers typically use the hydrogen maser, which will be discussed in a subsequent section.

Figure 7a shows this comparison applied to the visibility amplitude of 0.1 Jy reported in the first EHT results for the Submillimeter Telescope (SMT)—IRAM 30 m telescope (PV) baseline (Akiyama et al., 2024). To focus on the effect of clock noise we further assume a hypothetical baseline that does not move in time. To simulate the clock noise contribution we assume a clock with an Allan deviation of $1 \times 10^{-13}/\sqrt{\tau}$ at each site. This approximates the hydrogen maser that was actually used since even though it is slightly degraded from the maser for $\tau = 10 - 30s$, it is slightly better than the maser at $\tau = 1s$. In this figure the light gray trace gives a hypothetical time series of visibilities with fringe fitting disabled for this baseline assuming that the only noise source is due to the frequency standards at the 11 second averaging time predicted by the S/N requirement and corresponding to a coherence of 0.97. The spread in values shown in the light gray trace exceeds the spread of values shown for this baseline in Akiyama et al. (2024), but the black trace shows the same parameters with fringe rate correction enabled showing the expected improvement in variance and most measurements falling within the published spread. The S/N of the simulated uncorrected and corrected time series shown in (a) is in good agreement with the prediction shown in Figure 6. Figure 7b shows the same time series but now averaged at the 33 seconds indicated by the coherence of 0.92 that is generally accepted as being sufficient for white frequency noise (see Figure 4). Both the spread of uncorrected and corrected values is outside the spread of visibilities detected for this baseline by the EHT indicating that for this type of clock noise with and without fringe correction the S/N would not be sufficient to unambiguously detect the fringe at the integration time of 33s corresponding to the accepted coherence of 0.92 for this noise type.

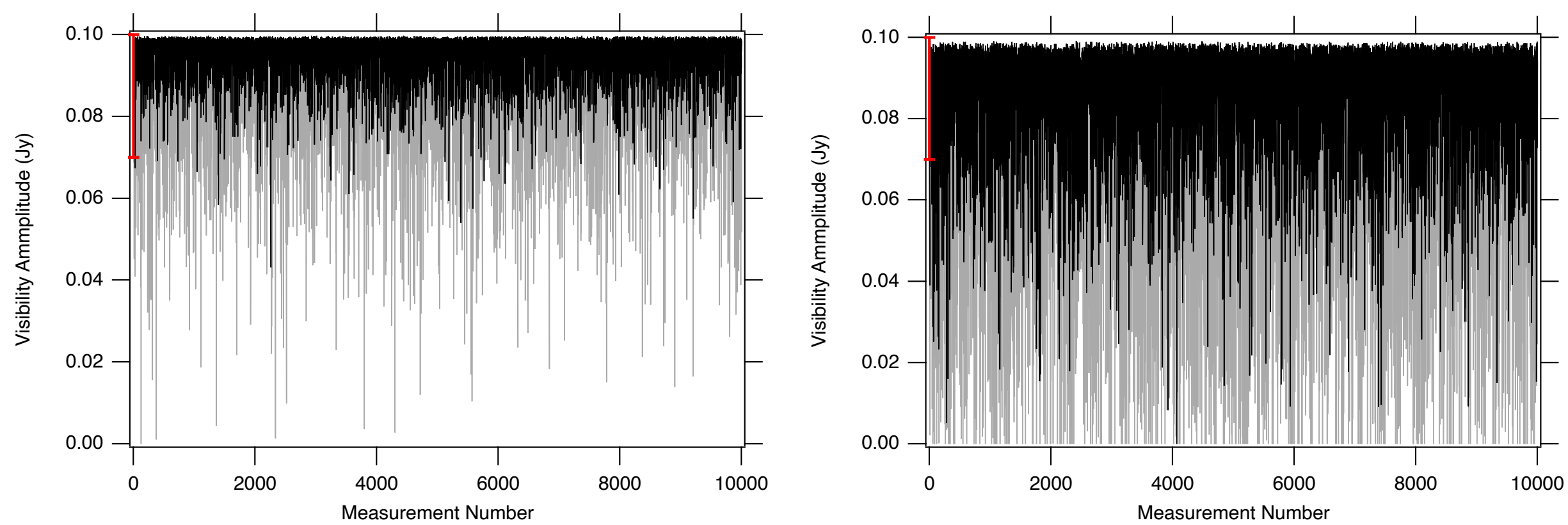

**Figure 7.** Simulated visibility for the SMT—PV baseline reported by the first EHT measurement at 230 GHz using a hypothetical clock at each antenna each with an Allan deviation of $1 \times 10^{-13}/\sqrt{\tau}$ and only clock noise present. In each graph the light gray trace is a time series of uncorrected visibility values while the black trace is the same time series corrected for the fringe rate over the given averaging time. The graph on the left is for an averaging time of 11 seconds and the graph on the right is for an averaging time of 33 seconds. In all cases an accumulation period of 0.4 seconds was used. The vertical red bar shows the approximate spread of values in the SMT-PV baseline.

The increase in the noise level for longer averaging times may seem counter-intuitive, but is just an example of the fact that it is the noise of the clock phase that is most important rather than the noise of its frequency. While the frequency noise is white and averages down with the square root of time, the associated phase is random walk and increases with the square root of time (see section 2.2).

### 3.3 A "Toy" Clock Model: the Signal to Noise Ratio for Other Pure Noise Types

Noise in real atomic clocks usually consists of a mix of several pure noise types. The most common among these are white frequency, white phase (where the phase of the clock has a white noise characteristic instead of the frequency), flicker frequency and random walk frequency noise (Stein, 1985). Usually, a passive clock, which consists of a local oscillator (LO) steered to an atomic reference, will consist of white phase noise that averages down linearly with time, reflecting the LO noise characteristics for times shorter than the clock control loop, and then white frequency noise that averages down with the square root of time for longer times until it reaches a flicker frequency "noise floor". Beyond this, further averaging does not result in improved stability. The noise may also transition to random walk frequency noise at this point, which will degrade frequency stability as the square root of further averaging time, or to drift, which degrades frequency stability linearly with time. Active atomic clocks, such as the hydrogen maser, are able to use a phase locked loop to steer the LO and will exhibit white phase noise on longer time scales than passive clocks. As we shall see, a white phase noise characteristic is extremely beneficial to VLBI. The lock loop used to steer the LO will also introduce additional complexity in the noise characteristics and is likely to do so on the short time scales of 1-100 seconds most interesting to high-frequency VLBI. In this section, we will show the limitations on S/N due to these three additional pure noise types and compare the results to those obtained in the previous section for white frequency noise.

Figure 8 shows the Allan deviation for a single hypothetical clock for the 4 noise types considered here where each has a stability of $1 \times 10^{-13}$ at 1 second. Also shown is the corresponding coherence function at 230 GHz for a pair of these clocks.

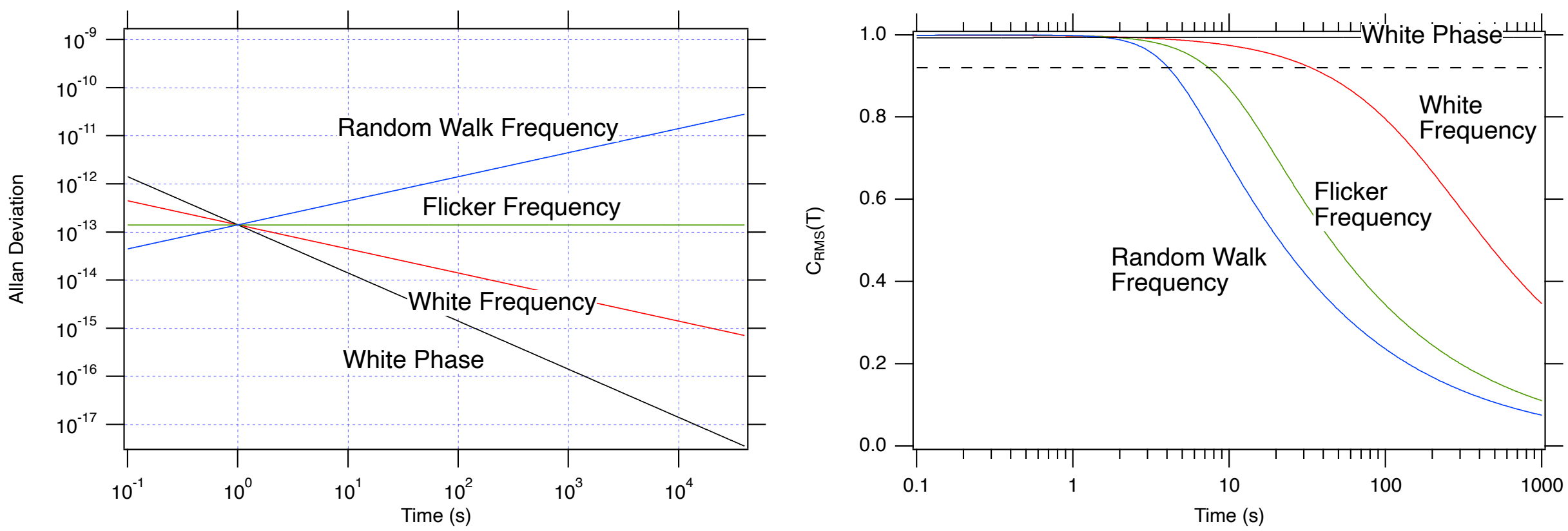

**Figure 8.** On the left is shown the Allan deviation for random walk frequency (blue), flicker frequency (green), white frequency (red—repeated here for comparison) and white phase (black). For each noise type the stability is $1 \times 10^{-13}$ at 1 second. On the right is shown the corresponding coherence function (same color for each noise type) evaluated at an observation frequency of 230 GHz. As in the previous section, the white frequency noise threshold at 0.92 is shown, but this is only for comparison. The threshold is noise type dependent: it is 0.85 for white phase and slightly higher than 0.92 for flicker and random walk.

In Figure 9 we compare the impact of the four basic clock noise types considered in this paper on limits to VLBI visibility S/N (calculated numerically) in the special case of a 230 GHz VLBI frequency. In each case, each of the two clocks in a single VLBI baseline has an Allan deviation of $1 \times 10^{-13}$ at 1 second but the clock noise scales differently at other times depending on the noise type. As expected, the S/N limits agree near, but not exactly, at 1 second where the clock frequency noise is the same and deviate before and after. Only the S/N limit for a frequency standard with a phase that has a white noise characteristic *doesn't degrade* with time. This is because white phase noise is stationary (time-independent variance) and longer averaging times result in reduced uncertainty. Most clocks are characterized by other noise types with time-dependent (non stationary) variances. So, for most clocks, the S/N limit *degrades* with averaging time in contrast to other VLBI noise sources, which may improve with averaging time. It is important to average VLBI data as much as possible (the accumulation period) before attempting fringe fitting, but this aspect of clock noise will impose an optimal averaging time where the noise in the clock phase crosses over noise from other sources. Longer averaging will not improve the detected signal. We will return to this topic when we analyze VLBI S/N characteristics of clocks that are actually used in VLBI.

Figure 9a shows simulation results for each of the four noise types with fringe rate correction disabled. Figure 9b shows the same data with a 0.4 second accumulation period and fringe rate correction enabled. For each noise type, corrected S/N's are all improved but note that all curves, corrected and un-corrected except that of white phase noise, eventually have a slope similar to that of WFN. This is because once the noise in the phase $\varphi(t)$ has an amplitude larger than $\pi$ radians, $e^{i\varphi(t)}$ becomes bounded, regardless of noise type. It is interesting that the corrected data reaches this slope at a later time. This is because the correction in phase rate delays the time at which the phase reaches $\pi$ radians. The S/N limit for flicker frequency and random walk also suggest that a significantly shorter averaging time than is predicted by the coherence function alone (Figure 8b) will be necessary.

An important result of this simulation is that there is little improvement in the S/N limit with fringe fitting enabled in the case of white phase noise since the variance for this noise type doesn't increase with time.

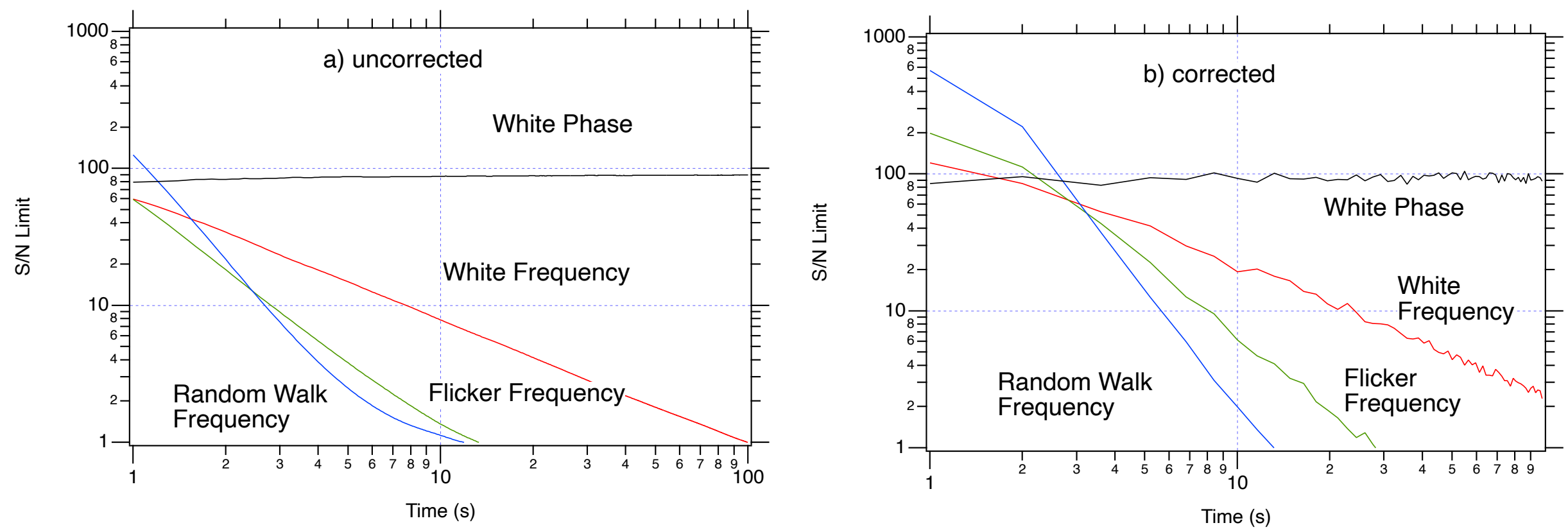

**Figure 9. a.** Comparison of clock-imposed limitations on S/N for 4 pure noise types and no fringe rate correction: white frequency (red), flicker frequency (green), random walk noise (blue) and white phase (black). In each case the hypothetical single-clock Allan deviation was taken to be $1 \times 10^{-13}$ at 1 second. **b.** Same simulation with an accumulation period of 0.4 seconds and fringe rate correction applied.

## 4 Application to Real Clocks

### 4.1 Signal to Noise Limits for Clocks Currently Used in Terrestrial VLBI

Having shown that the numeric and exact solutions for the clock-imposed limits on VLBI visibility S/N agree in the special case of white frequency noise, thereby giving confidence in the numerical approach, we now turn to calculating numerical solutions for clock-imposed limits on VLBI visibility S/N for various real clocks. In particular we are most interested in clocks that are currently used for VLBI and clocks that might be used in space VLBI. We start with clocks that are currently used in terrestrial VLBI. For all simulations that follow we will show only results that include fringe rate correction.

For terrestrial VLBI at high frequencies (>100 GHz), currently the only commercially available and reliable atomic clocks that provide sufficient performance are the hydrogen maser and, in some cases, Ultra-Stable (quartz) Oscillators (USOs) (see the reference, "Ultra Stable Oscillator (USO)") . Some experimental work has been carried out with high-performance (better than either the maser or the USO in the short term) cryogenic oscillators (Doeleman et al., 2011), but these are not commercially available and require a specialist to operate.

Figure 10 shows typical clock Allan deviations for the hydrogen maser and the USO (see appendix F for details). On short time scales, the maser Allan deviation is determined by its internal USO local oscillator so the Allan deviation for the maser and the USO agree for times shorter than 1 second. After that, a combination of flicker frequency noise, random walk frequency noise, and drift limit the USO stability. The white phase noise component of masers at short times less than 10 seconds is particularly useful for VLBI. As we will see, it enables these standards to have equal or better VLBI performance compared to that obtained using atomic frequency standards with better stability at 1 second but other noise types.

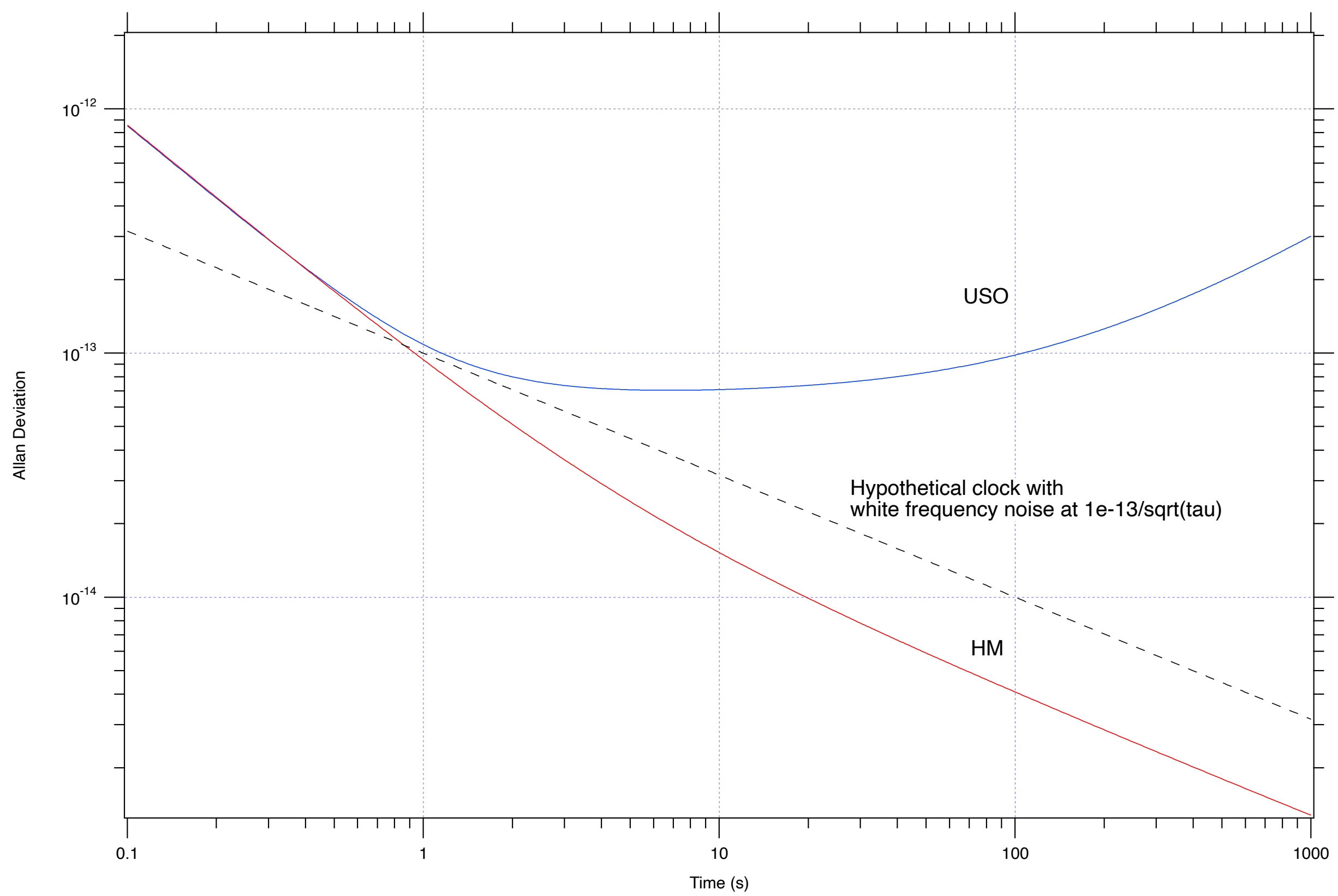

**Figure 10.** Single clock Allan Deviations for real clocks used in current high frequency terrestrial VLBI: hydrogen maser (red) and USO (blue). A hypothetical atomic clock with $1 \times 10^{-13}/\sqrt{\tau}$ stability (dashed black) is shown for reference.

For these two clocks, due to their complex combination of noise types, exact solutions for the S/N limit are not possible. We plot numerical solutions of equations (40) and (42) in Figure 11 for several VLBI frequencies of interest. The different noise types have clear signatures in these S/N limit graphs as well. As the maser noise transitions from white phase noise to white frequency noise, its associated S/N limit turns over and has the same slope as the hypothetical pure white frequency noise clock. But it is important to note that even though all three clocks have about the same *frequency* stability at 1 second, the *phase* of the white frequency noise clock is diverging with time while the phase of the maser is stable. It is this noise characteristic at short averaging times that make masers such good frequency standards for VLBI. The USO also has white phase noise at short times less than 1 second and initially looks similar to the maser, but its frequency noise is more complex, consisting of flicker frequency noise, random walk frequency noise and a much higher drift (typically $1 \times 10^{-15}$/s vs. $1 \times 10^{-20}$/s for the maser). Each of these additional noise sources, which come into play at about one 1 second, cause the phase to diverge and lead to rapidly degrading S/N performance.

In Figure 11, each graph has a horizontal line showing the S/N threshold of 20 described previously in section 2.1. Impressively, even at a frequency of 345 GHz, the USO is able to stay above this threshold for short integration times such as would be possible with large area VLBI telescopes and/or strong signals. However, space VLBI will likely have small area antennas and signal detection may require longer integration times and a better frequency standard.

Finally, none of these frequency standards are able to support a S/N greater than 20 at 630 GHz for any integration time indicating the need for new types of frequency standards for future VLBI networks that could be located entirely in space using higher frequencies. Such a network would be free from atmospheric decoherence effects thereby enabling longer integration times, but it would not be possible for the maser or the USO to support these.

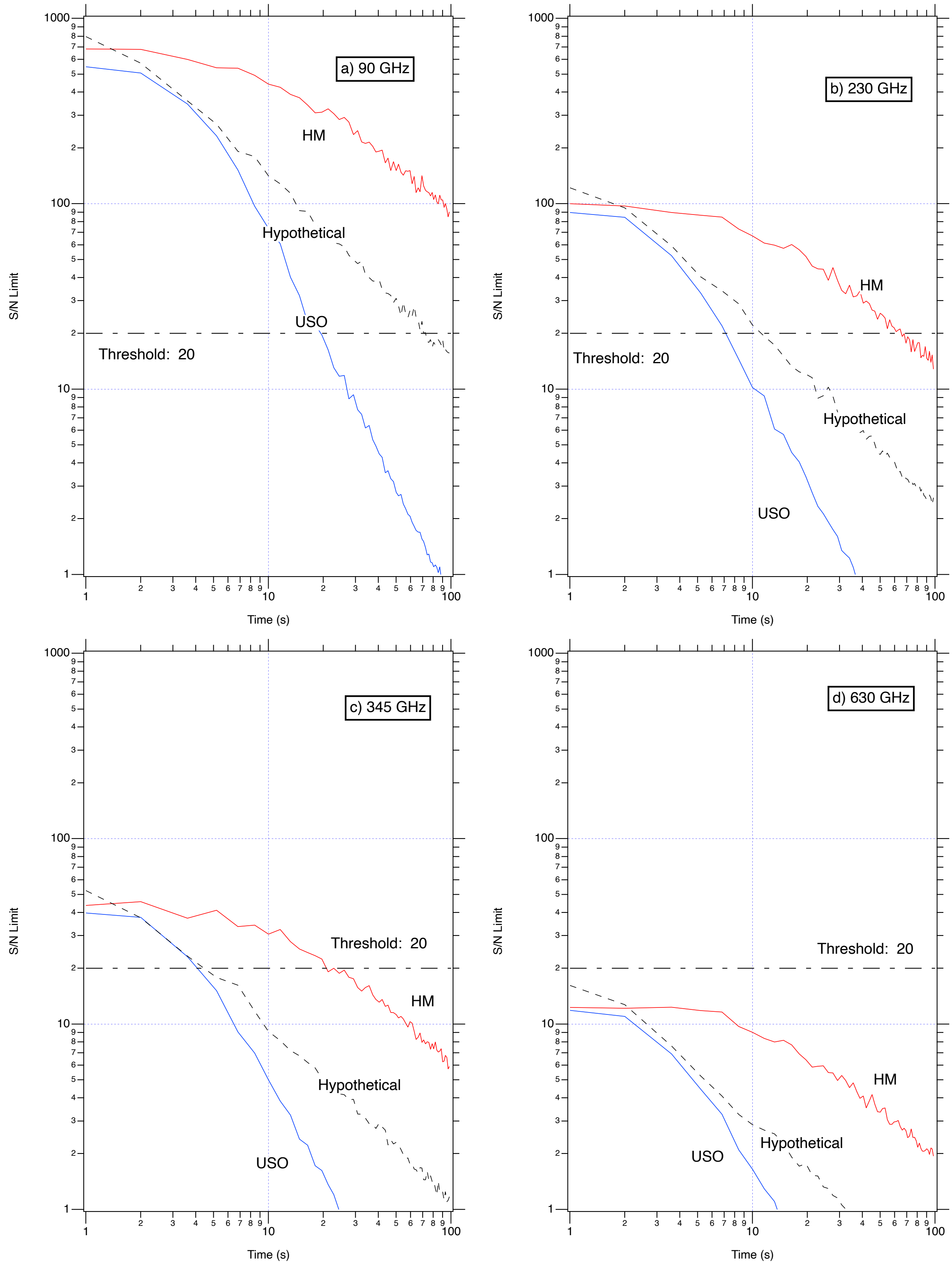

a) 90 GHz
HM
Hypothetical
USO
Threshold: 20
b) 230 GHz
HM
Hypothetical
USO
Threshold: 20
c) 345 GHz
HM
Hypothetical
USO
Threshold: 20
d) 630 GHz
HM
Hypothetical
USO
Threshold: 20
S/N Limit
Time (s)

**Figure 11.** Clock-limited Visibility S/N calculated numerically for: Hydrogen maser (red), a USO (blue) and a hypothetical clock with an Allan Deviation of $1 \times 10^{-13}/\sqrt{\tau}$ (dashed black) for reference. Each graph shows results for a different VLBI frequency: a) 90 GHz, b) 230 GHz, c) 345 GHz, and d) 630 GHz. In each graph the threshold value of 20 described in the text (broken dashed black) is shown for reference.

The turnover in the graphs shown in Figure 11 for both frequency standards reveals interesting aspects of the underlying noise processes. In each case, the noise type changes at about the same time. In the case of the maser, it transitions from white phase noise, for which the S/N will stay approximately constant with time, to white frequency noise, whose corresponding phase will be random walk and whose S/N will degrade with time. Similarly, the USO also starts out with white phase noise but instead transitions to flicker frequency noise and then to random walk and drift, all of which have a corresponding phase whose derived S/N degrades with time even faster than white frequency noise.

The graphs in Figure 11 illustrate the point made in the last section that most clock noise causes the VLBI visibility S/N to *degrade* with time and the optimal averaging time will occur at the cross over of S/N limits imposed by the frequency standard and those of potentially other sources. For instance, assume a VLBI measurement at 230 GHz with non-clock contributions to visibility S/N of 5 at an averaging time of 10 seconds and that this non-clock noise continues to improve for longer averaging times. If a space VLBI satellite used a USO as its on-board clock, it would only be able to take advantage of improved non-clock noise contributions at times less than 10 seconds since the clock contribution at this time is below 10 and continuing to degrade with time.

Recently the EHT consortium published experimental results for visibility S/N at 345 GHz over several baselines (Raymond et al., 2024). This gives a unique opportunity to validate the model presented here. In these real EHT measurements, a high-performance hydrogen maser was used at each site. In Figure 12 we show the clock S/N limit imposed by the maser at 345 GHz derived in Figure 11 and superimpose this with the real 345 GHz EHT visibility S/N data. The results are in good agreement with the model predictions. The model falls near the middle of the real spread, but in fact, the actual spread extends further on the lower end. This is because data is not recorded for S/N less than 8 where the fringe cannot be unambiguously detected. This puts the model prediction towards the higher side of the spread in the real data as expected since the real data will include other noise contributors such as the atmosphere.

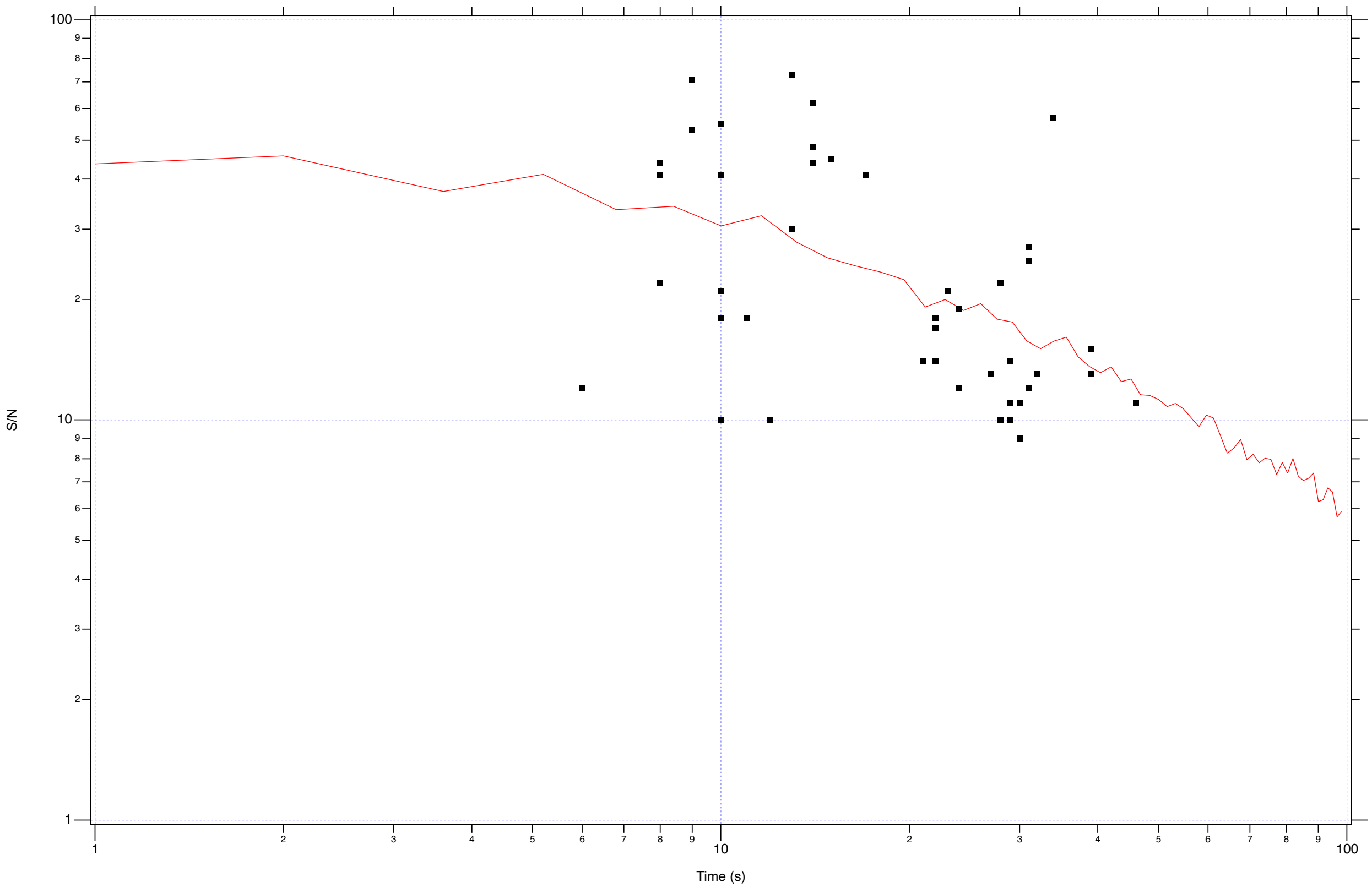

**Figure 12.** Modeled S/N limit imposed by high performance hydrogen masers at 345 GHz (red) and real visibility data at 345 GHz taken by the EHT consortium over several baselines including the SMA and ALMA (black dots).

### 4.2 Signal to Noise Limits for Currently Available Space Clocks

Space VLBI limits the choices of frequency standard to those that have been flight-qualified and may limit the frequency standard Size Weight and Power (SWaP). Since the hydrogen maser has been so successful in terrestrial VLBI, one might think that a hydrogen maser that has already flown in space (Kardashev et al., 2017; R. Vessot & Levine, 1979) would be the ideal choice. However, it has been challenging to achieve the same maser performance in space as on the ground. In addition, the hydrogen maser has a relatively high SWaP and may not be feasible for some missions. The introduction to this work described two types of masers: active and passive. High performance terrestrial VLBI uses active hydrogen masers almost exclusively, however we will look at both the active and passive masers since flight-qualified versions for both exist (the latter is the frequency standard used by the European Galileo navigation system.) Other space-qualified frequency standards include the Rubidium Atomic Frequency Standard (RAFS) used by GPS (see the reference, Teledyne, n.d.) and the USO already mentioned above.

Laser cooling and trapping of atoms and ions has revolutionized atomic clock technology (Brewer et al., 2019; Cutler et al., 1981; Hinkley et al., 2013; Prestage et al., 1991; Tjoelker et al., 1996). Recently a trapped ion atomic clock was demonstrated in space (Burt et al., 2021). While this technology is very promising for applications requiring long-term stability, relative to an active hydrogen maser its short-term stability has increased phase noise due to a relatively slow control loop used to steer its LO. This increased phase noise on short time scales on the order of the approximate control loop time of about 10 seconds makes this approach less ideal for VLBI. A laser-cooled clock was also recently flown in space (Liu et al., 2018). However, this clock does not perform as well in the short-term as the space-qualified active hydrogen maser and has an even higher SWaP.

As in the previous analysis, we will assume the same clock exists at both nodes of the visibility. This would likely be the case for VLBI with all nodes located in space. Near-term space VLBI will likely consist of a hybrid of terrestrial and space antennas. In this case, the two nodes of a visibility may have different clocks. We will take this case up in the next section. Figure 13 shows the Allan deviations of the space clocks considered here.

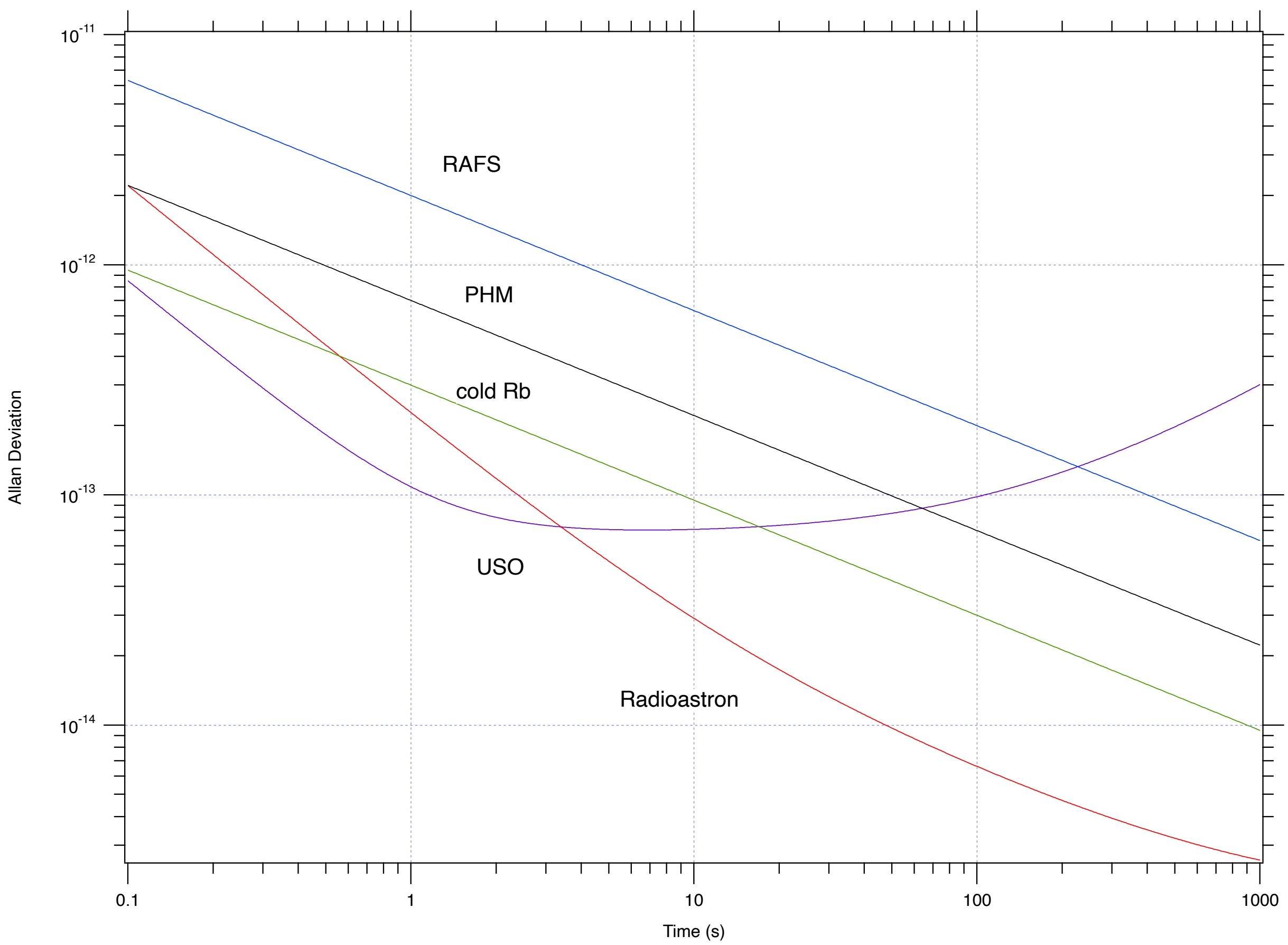

**Figure 13.** Allan deviation of currently operating high-performance space clocks: USO (purple), GPS Rubidium Atomic Frequency Standard or RAFS (blue), Galileo Passive Hydrogen Maser or PHM (black), Radioastron active hydrogen maser (red), and a laser-cooled rubidium clock or cold Rb (green).

Figure 14 shows the results of simulating clock-limited S/N from equations (40) and (42) for the currently operating high-performance space clocks. As before, each graph corresponds to a different VLBI frequency. It is important to note that while the USO has a very high drift rate, its VLBI S/N limit still exceeds all others except the Radioastron maser because its frequency is governed by white phase noise in the short term (Rubiola, E., Giordano, V. 2006), which is the aspect that is the most important for VLBI. Even at 230 GHz, it is the only clock that exceeds the threshold of 20, but only for the relatively short integration time of 7 s. While the Radioastron maser stability is determined by its internal USO at short times less than 1 second, this USO is not as stable as those that are currently available for space as stand-alone instruments. The USO result shown here is for the highest level of performance for currently available USO's. At 345 GHz the viable USO integration time is reduced to 4 seconds and at 630 GHz, no frequency standard is above the threshold. While the Radioastron maser has a lower S/N limit than the USO in the short term, thanks to the longer duration of its white phase characteristic, it has better VLBI S/N performance after 10 seconds and in the case of 230 GHz is still viable for up to 30 s. Another interesting observation is that even though the cold rubidium clock has superior performance to the USO on time scales greater than 10 s, the USO is far superior for VLBI applications: another example showing that short-term clock performance, and in particular white phase noise is more important than long-term performance for VLBI. In fact, the USO performance with fringe fitting enabled is still viable as the frequency standard at 345 GHz, although only marginally so with a limited averaging time, while the cold rubidium standard is only viable at 90 GHz. The viability of the USO can be further improved using Frequency To Phase Transfer (see section 5.2). Finally, the results show that the RAFS and PHM are not viable for any observation frequency.

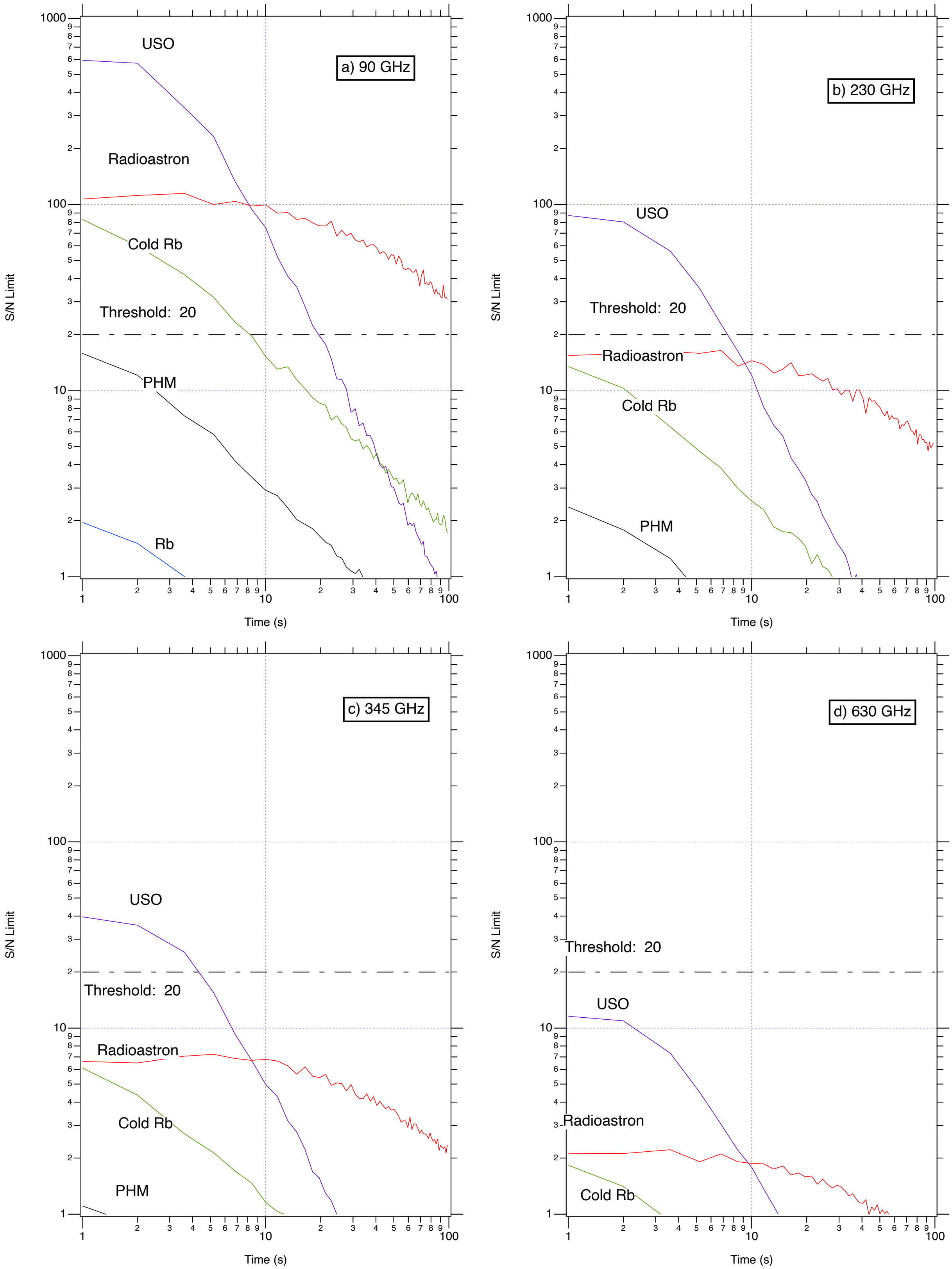
a) 90 GHz
USO
Radioastron
Cold Rb
Threshold: 20
PHM
Rb
b) 230 GHz
USO
Threshold: 20
Radioastron
Cold Rb
PHM
c) 345 GHz
USO
Threshold: 20
Radioastron
Cold Rb
PHM
d) 630 GHz
Threshold: 20
USO
Radioastron
Cold Rb
S/N Limit
Time (s)
1000
100
10
1

**Figure 14.** Clock-limited Visibility S/N calculated numerically for: the Radioastron hydrogen maser (red), a USO (purple), a laser-cooled rubidium space clock (green), the Galileo passive hydrogen maser (black), and the GPS rubidium atomic frequency standard (blue). Each graph shows results for a different VLBI frequency: a) 90 GHz, b) 230 GHz, c) 345 GHz, and d) 630 GHz. In each graph a threshold value of 20 (broken dashed black) is shown for reference.

### 4.3 Signal to Noise Limits for Hybrid Space-Terrestrial VLBI

Ultimately, all-space VLBI networks may exist, but initially space VLBI is more likely to combine a terrestrial network with one antenna in space (Kardashev et al., 2017). In this case all terrestrial nodes would probably have a high-performance ground-based hydrogen maser and only the space node would have a lower performing clock. In this section we will only consider the USO as the frequency standard in the space node. Of the currently operating space clocks considered here, the USO performs better than all others except the Radioastron maser. The latter has significantly higher SWaP and cost and is not commercially available.

Figure 15 shows the clock-induced S/N limit for an earth-space VLBI visibility in which the space node is referenced to a USO and the earth node is referenced to a hydrogen maser. The simulation was run with a 0.4 second accumulation period and fringe fitting enabled. The results are shown for several VLBI frequencies.

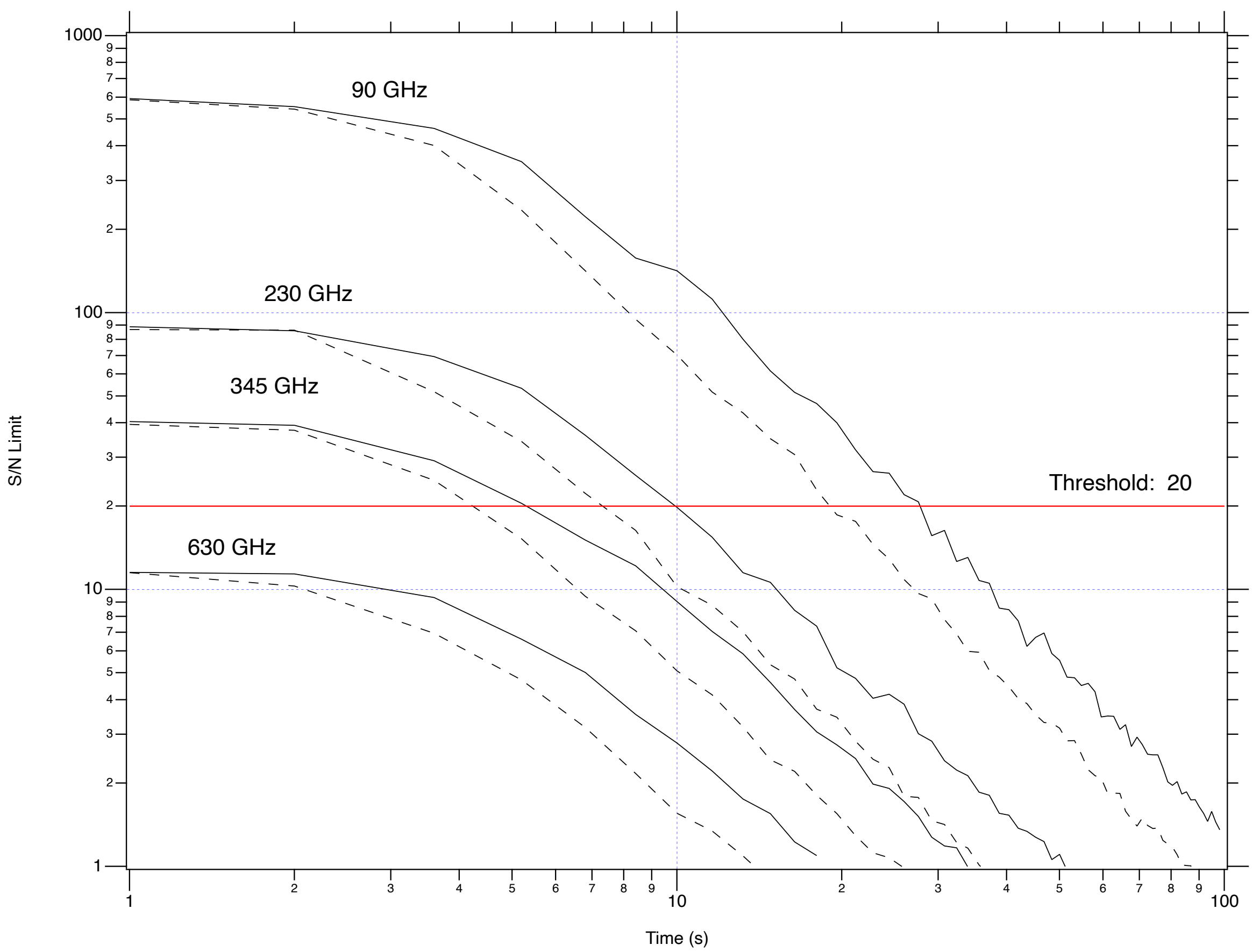

**Figure 15.** Clock-induced S/N limits for a visibility with a USO at one node and a maser at the other (solid black). At each frequency the case of two USOs is shown for comparison (the adjacent dashed black trace just below the associated solid black trace). The S/N threshold of 20 is also shown (red).

For each frequency, there is an approximate $\sqrt{2}$ improvement in the time at which the S/N crosses the threshold of 20. For instance, at 345 GHz, the maximum integration time for which the S/N limit stays above the threshold of 20 goes from about 4 seconds for two USO's to about 5.5 seconds for a maser and a USO. This analysis suggests that a USO reference in space combined with a maser reference on the ground may be viable, even at 345 GHz, as long as a total integration time of less than 10 seconds can be tolerated.

### 4.4 Signal to Noise Limits for Emerging Clocks of Interest to Space VLBI

Next, we examine promising high-performance clocks that may be available for operation in space in the near future. There are several new clocks that are either space-qualified or in the process. Here we restrict ourselves to those whose performance is better than the existing space clocks already considered. These include the ACES hydrogen maser (Goujon et al., 2010), which is already built and space-qualified and projected to have stability similar to a ground maser (and better than the Radioastron maser), an optical iodine atomic clock (Roslund, 2024), an Optical Local Oscillator (OLO) (see below) and a high performance optical clock (HPOC) (Margolis, 2010).

As mentioned in section 4.2 techniques to laser cool and trap atoms have had a profound impact on atomic clock performance. More recently optical technology has further revolutionized clock performance in both the short and long term (Margolis, 2010). Optical clocks have largely remained a laboratory instrument not yet mature enough for operation in space or in the field. However, in the last few years there have been many efforts to make them portable (Brand et al., 2019; Ohmae et al., 2021; Stuhler et al., 2021). One of the key technologies that made optical clocks possible is the optical frequency comb (Diddams, 2010), which enables direct conversion of optical frequencies into the microwave domain without adding noise. The LO for an optical clock is a stabilized laser, often with sub-Hz linewidth. When a frequency comb is locked to this laser, the optical stability can be down-converted to the microwave frequency spacing of the comb "teeth". A high-performance optical clock consists of a laser LO locked to an ultra-stable cavity in the short term and steered to an atomic reference in the long term where the atoms are electromagnetically trapped and cooled. The combination of a laser stabilized to a cavity and a comb forms the Optical Local Oscillator (OLO) used by high-performance optical clocks.

A simpler type of optical clock in which the laser is locked directly to the atomic reference and the atoms are not laser-cooled or trapped but simply contained within a cell, can be built into a smaller package. This minimal sapproach, which could be called a "warm-cell optical clock", trades some performance for a package that has a clearer path to a space-qualified instrument and is still capable of performance better than a maser in the short term. Interactions between the atoms and the walls of the vacuum chamber in this approach means that a strong drift component is introduced and long-term performance is degraded, but for high-frequency VLBI applications, long-term performance usually isn't important. A commercial version of this type of optical clock is now available (Roslund et al., 2023) making it immediately viable for terrestrial VLBI applications and potentially viable for space applications.

Optical clocks have not yet operated in space, partially due to the complexity of the laser and trapping systems. In addition to the warm cell optical clock, another viable approach to space qualifying this technology in the near term is to use the OLO by itself without the atomic reference. We will show that for VLBI applications, only the OLO portion of the optical clock is likely to be needed for the forseeable future (observation frequencies at or below the THz level and integration times less than 100 s). As with a USO, the OLO has a high drift rate that governs its stability in the long term, but long-term stability is not required for high-frequency VLBI, which is usually only sensitive to frequency standard performance from seconds to 10's of seconds. The OLO by itself has potentially several advantages over the full optical clock including: 1) lower SWaP, 2) it is likely to be space-qualified sooner, 3) higher reliability, and 4) longer operational life.

The Allan deviation for the ACES hydrogen maser, a warm cell optical clock based on iodine atoms, and an estimate for a space-qualified OLO are shown in Figure 16. Note that the instability shown for the space-qualified OLO is about 10× worse than for a typical ground-based OLO. This is conservative and it is likely that a space OLO with low SWaP and better performance will be possible as the technology matures. OLO performance below what is shown here in the short-term has already been demonstrated in a miniature cm-scale optical cavity (McLemore, 2022) though in that work performance for averaging times from 10 to 100 seconds is limited by a drift of about 1 Hz/second while a drift of about 200 mHz/second is needed to reach the desired OLO stability at 100 seconds. A drift of 25 mHz/second or better is obtained in larger cavities (Hirata, 2014), and it is likely that a drift of about 200 mHz/second needed to support the desired stability at 100 seconds would be obtainable with a larger cavity still acceptably small enough for space. Since the technology is at an early stage, it is likely to advance rapidly, so we add the better ground based OLO performance ("OLO-future"). We also add the full high-performance optical clock ("HPOC"), which will eventually be space-qualified. Both the optical iodine clock and OLO performance exceeds that of the maser on all but very long time scales that are not of interest to VLBI where short-term characteristics are more important. The ground based OLO performance is about 10× better than the anticipated performance of the first space-qualified version. It is assumed that a future HPOC would use an even more stable OLO (typical of ground performance) so its performance is slightly better than the future OLO even in the short term. In addition, once the HPOC closes the loop on steering this LO at about 1 second it continues to average down further.

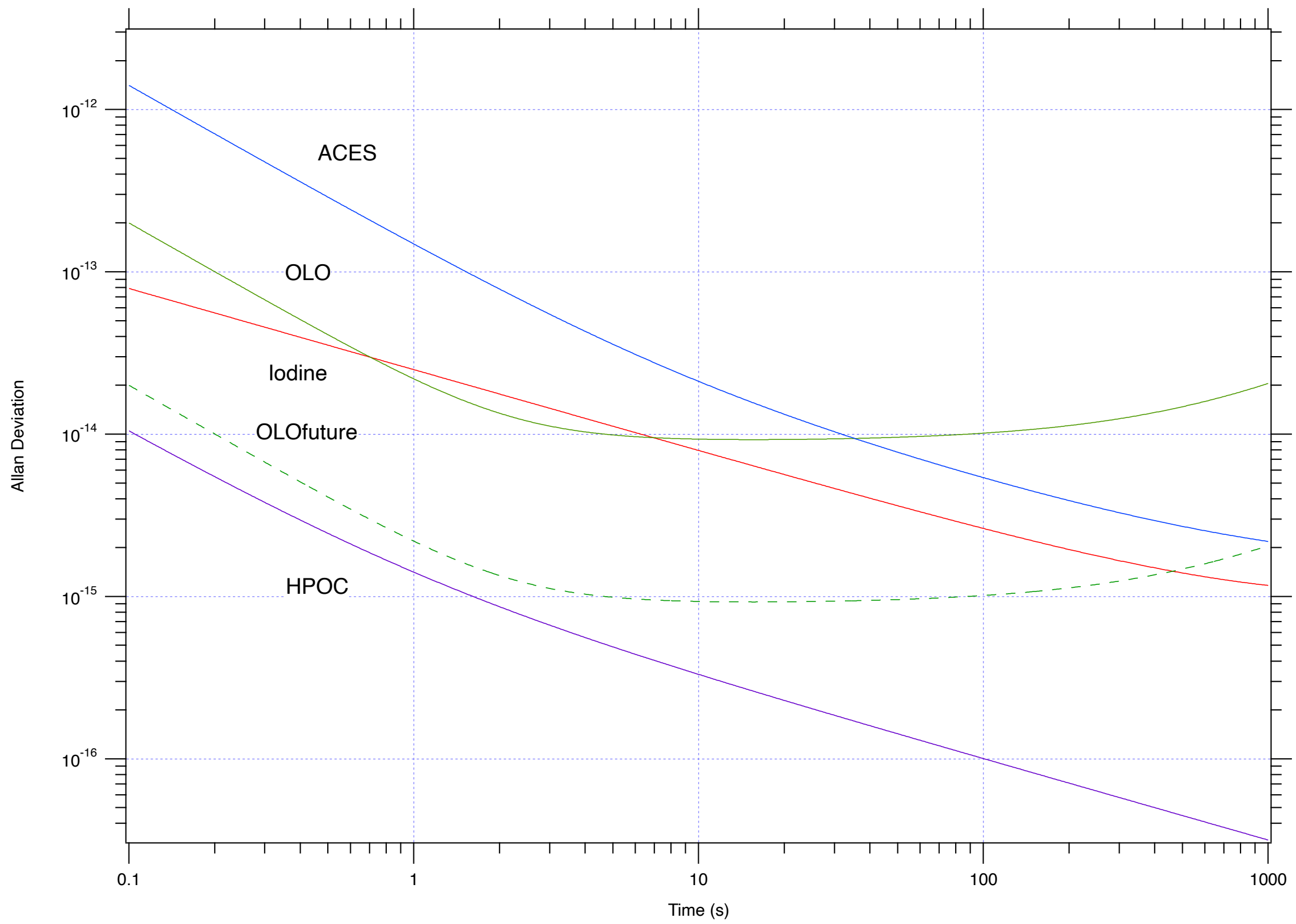

**Figure 16.** Allan deviation for emerging clocks of interest to VLBI: The space-qualified ACES hydrogen maser (blue), a "warm cell" optical iodine clock (red), the estimated performance of an initial space-qualified OLO (green), a future version of the space OLO with anticipated performance (dashed green), and a high-performance space-qualified optical clock - HPOC (purple).

As before we now calculate the clock-induced S/N limits for each of the VLBI frequencies 90, 230, 345, and 630 GHz with an accumulation period of 0.4 seconds and fringe fitting enabled, and show the results in Figure 17. Here, all of these standards are equal to or better than a maser, so unlike the case of the USO above, we assume that the ground component would also use the same clock. The OLO is significantly better than the ACES maser for all times less than several hundred seconds and even better than the iodine clock between 1 and 10 seconds. Space-qualified OLO performance will improve significantly and a conservative estimate of future OLO performance remains above the other two clock types for all averaging times of interest. In fact, even at 630 GHz, the future OLO S/N limit is still 5x above the threshold of 20 at an integration time of 100 s. Of course, the HPOC does even better, but the improvement is unlikely to be useful since other noise sources will almost certainly dominate. Based on this result we conclude that for future space VLBI at 630 GHz or even higher, a simple OLO would suffice as the clock—the atomic reference of the full optical atomic clock is not needed. This is a large simplification and a future OLO performing at this level could be space-qualified sooner than the full clock.

It is also notable that for the other three clocks, which could be space-qualified the soonest (the ACES maser already is), the S/N limit remains close to the threshold of 20 at 345 GHz for the maser and above 20 for over 20 seconds for the OLO and the iodine clock, even at 630 GHz (longer integration times will likely be limited by atmospheric decoherence (Raymond, et al., 2024), in the single space node topology). Along with the OLO, the iodine warm cell optical clock technology is at the beginning of its development cycle and like the OLO its performance level will likely improve over time. With no atmosphere to contend with, an all-space VLBI network could consider much longer integration times as long as the clocks could support this. As shown in Figure 17, all three of these clocks are viable as a starting point for such a network with the iodine clock and the OLO having a path to further improvements as needed (masers are very mature and unlikely to improve).

To summarize the situation with emerging clocks, considering SWaP and maturity in addition to performance, a space-qualified OLO or warm cell optical clock may be the optimal choice for a frequency standard in a hybrid earth-space VLBI even up to a VLBI frequency of 630 GHz. Both the OLO and iodine clock performance will continue to improve in the future giving a viable long-term path as well. The simplicity of cell clocks may make them more robust in space while the OLOs probably have more room to improve their performance. Both options are simpler and more mature than a full optical clock whose long-term stability is not necessary for high-frequency VLBI.

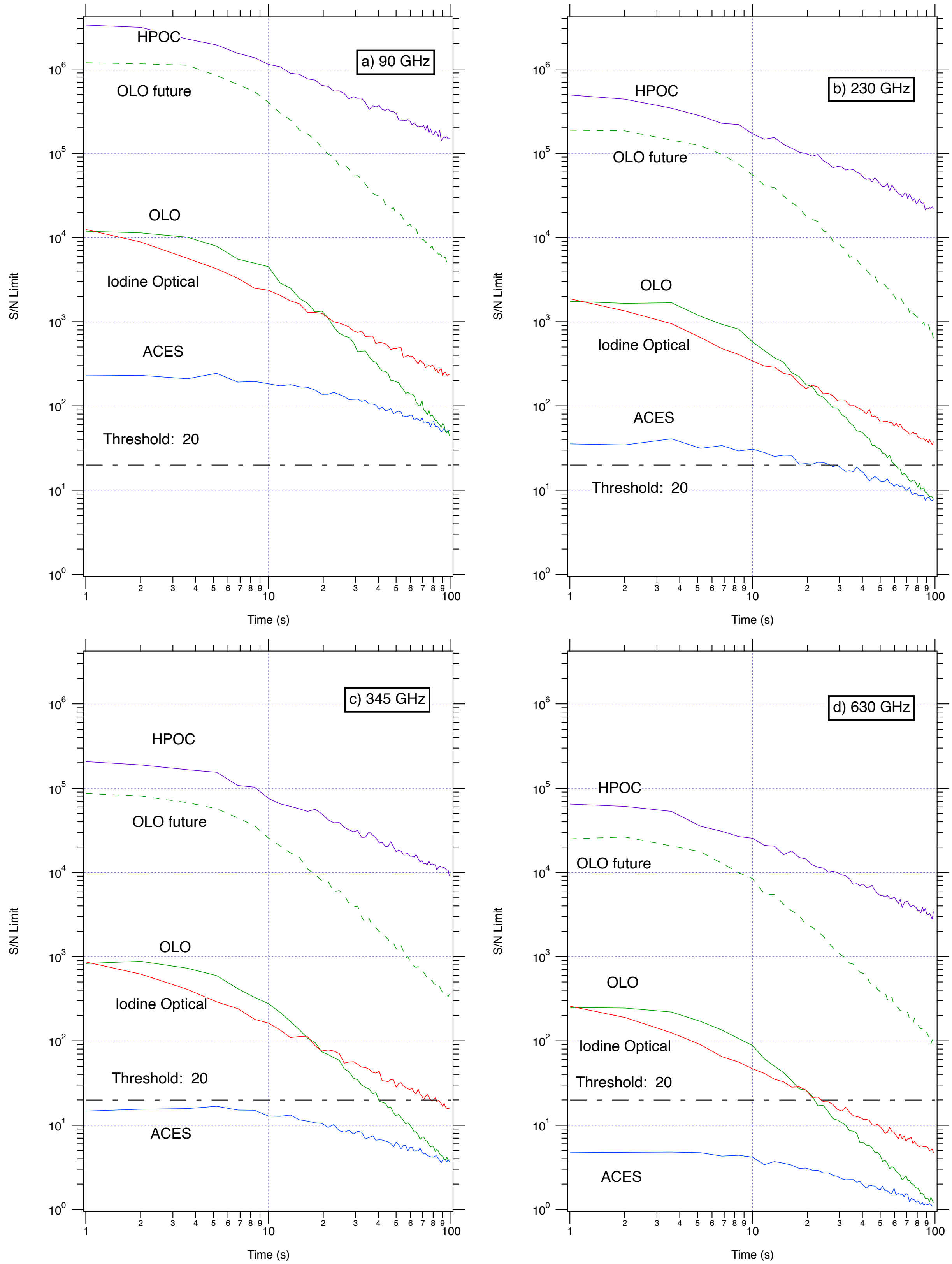
a) 90 GHz
HPOC
OLO future
OLO
Iodine Optical
ACES
Threshold: 20
S/N Limit
Time (s)
b) 230 GHz
HPOC
OLO future
OLO
Iodine Optical
ACES
Threshold: 20
S/N Limit
Time (s)
c) 345 GHz
HPOC
OLO future
OLO
Iodine Optical
Threshold: 20
ACES
S/N Limit
Time (s)
d) 630 GHz
HPOC
OLO future
OLO
Iodine Optical
Threshold: 20
ACES
S/N Limit
Time (s)

**Figure 17.** Clock-induced S/N limits for emerging clocks of interest to space VLBI: ACES (blue), warm cell iodine optical clock (red), space Optical Local Oscillator - OLO (green), future space OLO (dashed green), and a High-Performance Optical Clock - HPOC (purple). The limits are calculated numerically for each of the frequencies 90, 230, 345, and 630 GHz.

## 5 Other Approaches

### 5.1 Comparison of Metrics: the Coherence Function C(T) and "Coherence Efficiency"

We are now in a position to answer the question posed in the introduction on what value of clock coherence is sufficient for high-frequency VLBI measurements. We define the "coherence efficiency" as the value of $C_{RMS}(T)$ for the clock at the time $T$ where the associated clock-induced S/N limit falls below the threshold of 20 for the VLBI frequency of interest. Using equations (10) through (12) for the definition of $C_{RMS}(T)$ and equations (38) through (42) for the visibility S/N, we calculate both quantities for a given clock.

Among clocks that are space-qualified now, the USO is possibly the best option for one node of the visibilities in space. First, we note at what time the S/N limit due to the USO/maser combination falls below 20 for each of the VLBI observation frequencies we have been considering. This is shown in Figure 15 where times 30, 10, and 5 for 90, 230, and 345 GHz respectively (the USO S/N limit is below the threshold at all times for 630 GHz) with fringe fitting enabled. Figure 18 shows $C_{RMS}(T)$ plotted for a visibility consisting of a USO and a ground hydrogen maser for each of these same frequencies. Superimposed on these traces are red dots corresponding to the value of $C_{RMS}(T)$ at the times derived from the S/N results (the coherence efficiency).

The dots show that the coherence efficiency is between 95% and 97% for the frequencies 90, 230, and 345 GHz. This is significantly higher than the recommended value for white phase noise of 85% and 92% for white frequency noise based on 1 radian of phase decoherence. This indicates that, *even with fringe fitting enabled,* $C_{RMS}(T)$ with these recommended values is not a sufficient metric for determining an acceptable integration time in the case of the USO. The situation would be worse without fringe fitting, which removes some of the noise, but this is only true for certain noise types. As pointed out in section 3, with pure white phase noise fringe fitting is not expected to improve the coherence and can actually *add* noise. This is shown in Figure 9 where the S/N limit with fringe fitting enabled is not significantly different than the uncorrected S/N limit for all integration times between 1 and 100 s. While $C_{RMS}(T)$ may be useful in some cases as an approximate metric for frequency standard viability, it can be hard to predict this usefulness when the clock noise has a mixture of pure noise types as is usually the case. Even if a value of 95-97% is chosen, this value is clock-noise-dependent (here the USO/maser combination): a different metric would be needed for each clock type. A more reliable approach is to always use the clock-limited visitibility S/N as derived here.

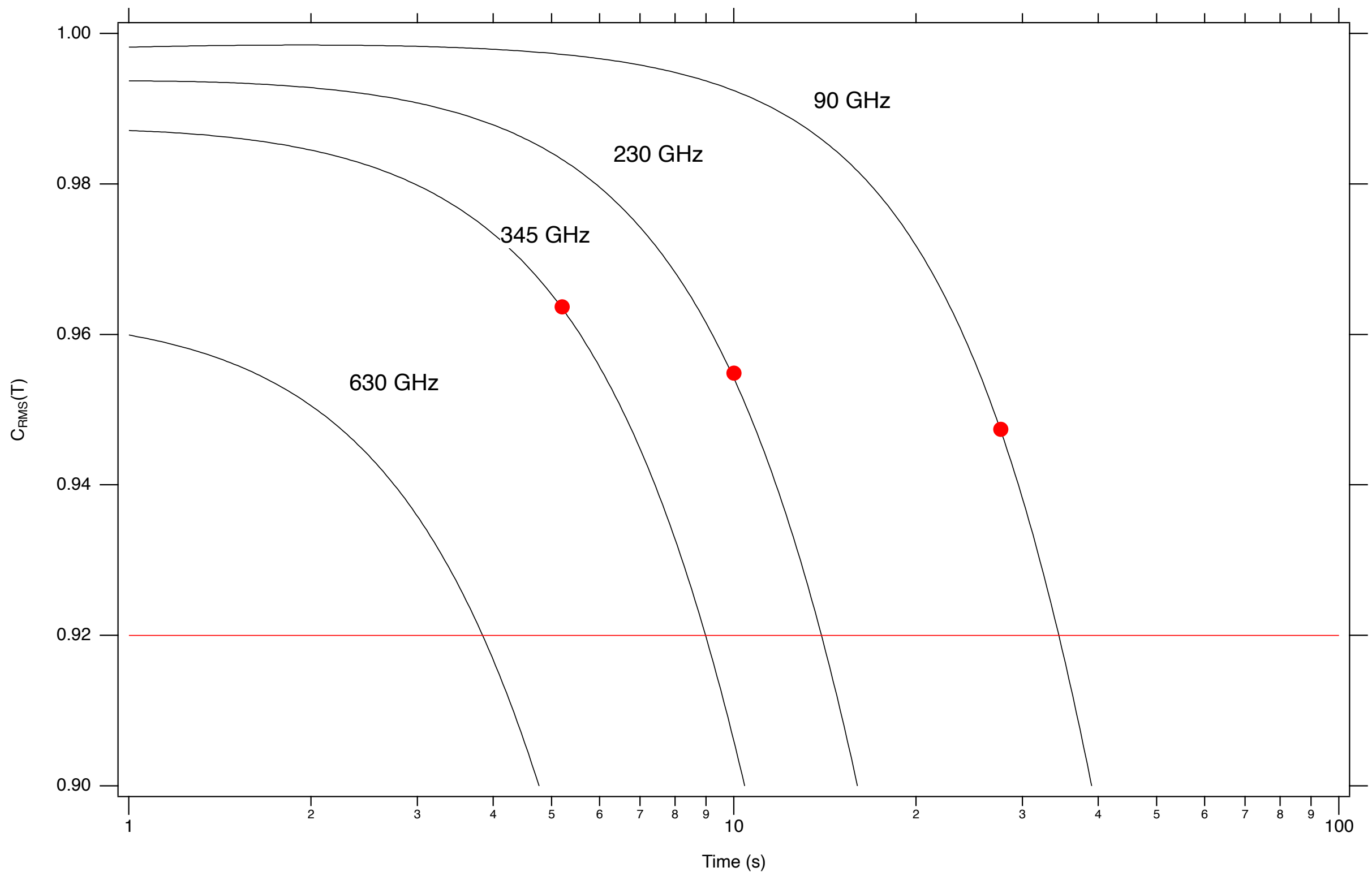

**Figure 18.** $C_{RMS}(T)$ (black) for a visibility with a USO at one node (space) and a hydrogen maser at the other (ground). Each trace corresponds to the coherence function for the indicated VLBI frequency. Superimposed is the coherence efficiency defined in the text for each VLBI frequency and this visibility (red dots). The normally accepted coherence value of 0.92 for white frequency noise is shown as the red line for reference.

### 5.2 A Comment on Frequency Phase Transfer

Since the effect of noise in the clock phase in seconds is multiplied by the observation frequency to obtain the clock phase (and noise) in radians, for low noise and for observation frequencies where phase noise is non-dispersive, it is possible to use a dual-frequency interferometer together with the Frequency Phase Transfer technique to improve the performance of the LO (Zhao et al., 2017). With detection taking place at a high frequency, observations taken at a lower frequency can be used to correct phase errors at the higher frequency. For instance, referring to Figure 15 it might be possible to use a dual-frequency interferometer at 90 and 345 GHz with a USO as the LO: the well-known phase at 90 GHz would be used to correct phase errors at the observing frequency of 345 GHz potentially maintaining a clock-limited S/N above the threshold of 20 for averaging times of over 10 seconds while the single frequency interferometer at 345 GHz using a USO would be limited to about 5 s.

This is a promising approach, but what are the limits of its applicability? For instance, in practice, how effective is the phase correction? Another consideration is that at high frequencies, the frequency standard phase error may be well in excess of 1 radian at short integration times and there might no longer be a simple multiplicative constant relation between the phase at the two frequencies. This would violate the assumption about the effect of the noise and complicate the correction to be made. To address the latter point it is recommended to implement Frequency

Phase Transfer with the multiple frequencies at integer multiples of each other (María J Rioja et al., 2023). When considering a range of potential frequency standards and/or very high frequencies, the phase error can now be many multiples of $2\pi$. Finally, to improve frequency standard coherence using FPT in such a way as to be useful for high frequency VLBI, the improvement must take place at relatively short time scales, on the order of 10 seconds or less. A key aspect of FPT that will determine its usefulness is whether its effectiveness is integration-time-dependent. As we saw in the previous section in the case of the USO, once the coherence drops below about 95% (5 seconds for 345 GHz), the USO noise may begin to limit overall S/N and below 90% (11 seconds for 345 GHz) the fringe won't be unambiguously detectable. Coherence improvements at later times will not help.

A detailed analysis of the uncertainties in the Frequency Phase Transfer technique (Maria J Rioja & Dodson, 2011) applied to a VLBI system that periodically switches between the low and high frequencies concludes that the troposphere is the dominant contributor to phase uncertainty, but this analysis was performed at lower frequencies (eg., 43 and 86 GHz) and appears to have assumed a hydrogen maser for the frequency standard. The total error calculated was approximately 70 degrees in 60 seconds. The phase error for the difference between two hydrogen masers at 86 GHz and 60 seconds would indeed be only about 5 degrees and therefore much less than the tropospheric error. However, a USO is an excellent candidate for the flight oscillator and the phase error at 86 GHz and 60 seconds for a USO combined with a hydrogen maser at the other end would be about 84 degrees and so would dominate other error sources. Further, for an integration time of 60 s, the USO-maser combination would have a phase uncertainty of about 4 radians at 230 GHz, 6 at 345 GHz, and 11 at 630 GHz, all of which would limit VLBI and would not be improved by the phase information at 86 GHz.

Assuming that the baseline with a USO at one end has a hydrogen maser at the other end and that fringe fitting is enabled, the results of the study carried out here (specifically the results shown in Figure 15) suggest that for the VLBI frequency of 90 GHz the integration time should be keep below 40 s. During this time at the higher frequency to be corrected using Frequency Phase Transfer, the USO-maser combination would have a phase uncertainty of about 2.2 radians at 230 GHz, 3.2 at 345 GHz, and 5.9 at 630 GHz. Since these are near to or more than a phase wrap, the higher frequency should be a multiple of 90 GHz, or the lower frequency should be adjusted accordingly. This is promising, but assumes perfect FPT and it remains to be seen what real FPT will be able to do in this context.

## 6 Conclusions

In this paper we have derived an expression for the limitations imposed by the local frequency standards (clocks) on VLBI visibility S/N given the Allan deviation of the clock. This expression, together with the procedure to solve it numerically, gives a method for unambiguously evaluating the viability of a given clock for a specific space VLBI mission architecture. Using the clock limited S/N metric we make the following conclusions:

1. For high-frequency space VLBI, only the USO and the hydrogen maser are viable among currently available frequency standards.
2. USO's are only viable in a narrow range of integration times and frequencies and masers have relatively high SWaP making them less amenable to space applications.

3. The technique of fringe fitting to remove a phase rate improves the S/N limits for some clocks but does not improve it and can actually degrade performance in the case of white phase noise
4. Among emerging new clock technologies we find the warm cell optical clocks and OLOs to be the most promising approaches for future space VLBI applications.
5. We also find that full high-performance optical atomic clocks improve on their OLOs on time scales of less interest to high-frequency VLBI, and that the simpler OLO by itself is likely to be sufficient for the foreseeable future. In addition OLOs are on the verge of being fully space-qualified now in a relatively small-SWaP package, while full high-performance optical clocks would require much higher SWaP and are unlikely to be space-qualified in the near future.
6. The clock coherence time can be useful in estimating clock contributions to VLBI measurements, but that in some cases it can give misleading results and that the clock-limited VLBI visibility S/N is a more reliable metric.

We derived the clock-limited S/N metric by propagating clock noise through the VLBI process, including fringe fitting. The most important places where clock noise contributes are during sampling and down conversion (mixing). We derived a general expression for the VLBI visibility S/N and carried this through to completion for the special case of pure white frequency noise. A comparison between this exact result and a Monte Carlo simulation gives excellent agreement in this case. We then studied contributions to VLBI visibility S/N using simulations of real clocks, which include a mixture of different noise types. We found that even with long-term instability, the hydrogen maser's excellent short-term stability, and in particular its white phase noise characteristic at short integration times, make it almost ideal for VLBI. We then presented results for current space-qualified clocks and found none to be adequate except the USO in a narrow range of VLBI frequencies and integration times and hydrogen masers (here we are referring to hydrogen masers that have operated in space, which have higher instability compared to ground versions).

Using fringe fitting to remove a phase rate improves the situation for noise types whose variance increases with time and we found that the S/N limits imposed by current space clocks was improved using this technique. With fringe fitting enabled, the viable integration time for a USO at one end of the baseline (eg., in space) and a maser at the other end (eg., on the ground) is about 30s at 90 GHz, 10s at 230 GHz and 5s at 345 GHz. We also note that by using the frequency-to-phase transfer technique, it may be possible to extend these times if a second lower frequency is used in the interferometer. However, even with fringe fitting enabled, the same conclusion still holds: among current space clocks, only the USO and the maser are viable. In addition, we also demonstrated that fringe fitting does not increase the viable integration time in the case of clock white phase noise, which has a time-independent variance. In this case fringe fitting can even add noise and degrade overall visibility S/N.

Looking at emerging clock technologies we found that optical clocks would perform extremely well and in particular the warm cell iodine commercial atomic clock has potential to be space-qualified. However, we also find that the Optical LO (OLO) portion of a high-performance optical atomic clock provides most of the performance needs of VLBI and does so with lower SWaP and less complexity than a full optical atomic clock. Thus, for near term space

VLBI applications space qualification of cell-based optical atomic clocks that don't use an ultra-stable cavity such as the iodine clock appear to be promising. For longer-term higher performance space VLBI, such as an all-space network operating at the highest frequencies, further development of space-qualified OLOs may be the best option.

In general, clocks with the following frequency stability characteristics will be most useful for high-frequency VLBI: 1) having a white phase noise characteristic at early averaging times and/or, 2) having a low Allan Deviation value in the short term on the VLBI integration time scale (seconds to 10's of seconds)—a value well below $1 \times 10^{-13}$ at 1 second with clock noise that averages down with time (such as atomic clocks with white phase noie) or a value below $1 \times 10^{-14}$ at 1 second for clock noise that does not improve with averaging time beyond several seconds (such as oscillators). Long term stability beyond the VLBI integration time scale does not significantly benefit VLBI, though as VLBI transitions to an entirely space-based network, the integration time will likely go up from 10s of seconds to 100s of seconds or more since it will no longer be limited by atmospheric decoherence (Doeleman et al., 2011) (in this case $1 \times 10^{-14}$ at 1 second for clock noise that averages down with time or $1 \times 10^{-15}$ at 1 second for clock noise that does not improve with averaging time may be required). This will require a further order of magnitude improvement in the clock Allan deviation, particularly as VLBI moves to the higher frequencies of 345 GHz and above.

For immediate space VLBI applications, the USO coupled with a maser on the ground appears to be adequate with a limited integration time, even up to 345 GHz, as long as fringe fitting is performed to remove a phase rate and the dual-frequency method of frequency to phase transfer (FPT) is used with the caveat that the needed improvements due to FPT have yet to be demonstrated at these frequencies over the short integration times required.

Finally, we note that most conclusions stated here are based on integration times in current use. Longer integration times may be limited by other (non-clock) decoherence effects, while shorter times may be limited by signal strength. VLBI networks with large collection areas that are able to obtain detectable signals in short integration times may be able to do so using only simple USO frequency standards, even up to frequencies of 630 GHz or more. VLBI networks with low collection efficiency such as those in space with necessarily small area antennas that dictate longer collection times to see significant signals, will ultimately need higher performing frequency standards.

## Appendix A: Combining Other Noise Sources

This paper derives the contribution of clock noise to the single baseline visibility S/N. As stated in section 2.1, given that the total single baseline minimum visibility required S/N to unambiguously detect the fringe is usually stated as 6.5, it is conservative to target the clock contribution to be 20. Here we give a more precise target in the case where the S/N due to other sources is known. Defining $S$ as the signal, $\sigma_t^2$, $\sigma_n^2$, and $\sigma_c^2$ as the variance of the total noise, the noise due to other sources, and the noise due to the clock respectively, we then have the S/N for each of these sources as, $\mathcal{R}_t = S/\sigma_t$, $\mathcal{R}_n = S/\sigma_n$, and $\mathcal{R}_c = S/\sigma_c$. Since the clock noise sources and the other noise sources are independent, we can add them in quadrature, that is $\sigma_t^2 = \sigma_n^2 + \sigma_c^2$. With these definitions, we can write,

$$\mathcal{R}_t = \frac{S}{\sqrt{\sigma_n^2 + \sigma_c^2}} \tag{A1}$$

Substituting expressions for $\mathcal{R}_i$ in terms of $\sigma_i$ we have,

$$\mathcal{R}_t = \mathcal{R}_n \frac{1}{\sqrt{1 + \left(\frac{\mathcal{R}_n}{\mathcal{R}_c}\right)^2}} \tag{A2}$$

We can define a S/N reduction factor due to clock contributions as,

$$\eta_{sn} = \frac{1}{\sqrt{1 + \left(\frac{\mathcal{R}_n}{\mathcal{R}_c}\right)^2}} \tag{A3}$$

The value $\eta_{sn}$ varies from 1 to 0 and, when multiplied by $\mathcal{R}_n$, gives the expected total S/N $\mathcal{R}_t$. We can also invert equation (A2) to get a constraint on the clock contribution to the S/N as a function of $\mathcal{R}_n$ and $\mathcal{R}_t$. Solving for $\mathcal{R}_c$ we have,

$$\mathcal{R}_c \geq \mathcal{R}_t \frac{1}{\sqrt{1 - \left(\frac{\mathcal{R}_t}{\mathcal{R}_n}\right)^2}} \tag{A4}$$

Given a constraint on the total S/N, $\mathcal{R}_t$, which is usually taken to be 6.5, and knowledge of the S/N due to all other noise sources, $\mathcal{R}_n$, equation (A4) enables us to place a constraint on the clock contribution to the S/N. The result is graphed in Figure 1 in the text.

**Appendix B: The Expectation of the Cosine of a Stochastic Variable**

The quantity $E\left[e^{i\Phi(t)}\right]$ is central to the calculation of the clock contribution to VLBI visibility S/N. We derive an exact expression for $E\left[e^{i\Phi(t)}\right]$ in terms of the noise characteristics of $\Phi(t)$ where $\Phi(t) = \varphi_2(t) - \varphi_1(t) = 2\pi\nu\big(x_2(t) - x_1(t)\big) = 2\pi\nu\big(\psi_2(t) - \psi_1(t)\big)$ is the total phase difference between the clocks at each node of a visibility. In the derivation we set the clock deterministic offsets to zero and the stochastic clock phase $\psi_i$ of the individual clocks at each VLBI antenna remain. The expectation of the exponential can be calculated from its definition

$$E[e^{i\Phi}] = \int_{-\infty}^{+\infty} p(\Phi)\, e^{i\Phi} d\Phi \tag{A5}$$

We now assume that the noise is Gaussian at each value time $t$ such that the probability density is $p(\Phi) = \left(1/\sigma_\Phi(t)\sqrt{2\pi}\right)e^{-\Phi(t)^2/2\sigma_\Phi(t)^2}$ where $\sigma_\Phi(t)^2$ is the classical variance of $\Phi(t)$. In the following, we generalize the argument given in (Riehle, 2006) to include "weakly" stationary noise types, by which we mean noise with zero mean, but potentially time-dependent variances. Substituting yields

$$E\left[e^{i\Phi(t)}\right] = \int_{-\infty}^{+\infty} \frac{1}{\sigma_{\Phi}(t)\sqrt{2\pi}} e^{-\Phi(t)^2/2\sigma_{\Phi}(t)^2} \cos\Phi(t)\, d\Phi \tag{A6}$$

Note that there is no contribution from the imaginary term since it is odd in $\Phi$ so only the real term is retained. Performing the integration gives

$$E\left[e^{i\Phi(t)}\right] = e^{-\sigma_{\Phi}(t)^2/2} \tag{A7}$$

where $\sigma_{\Phi}(t)^2$ is the standard variance of $\Phi(t)$ at each point in time and each frequency value.

The standard variance of the clock phase is not usually known in advance, but equation (A7) can also be expressed in terms of the Allan Deviation, which is usually well-known. In the main text of this paper, we wish to derive an expression for the variance of the phase in the case of white frequency noise. The phase associated with white frequency noise is characterized by random walk. From Maybeck (1979) we have in general for a random walk process $\varphi_{RW}(t)$ its classical variance taking the form

$$E[\varphi_{RW}(t)^2] = q_{RW}(t - t_0) = q_{RW}t \tag{A8}$$

where $q_{RW}$ is the random walk's diffusion strength and is a constant. Without loss of generality we have also set the initial $t_0$ to zero, thus, turning the time $t$ into an increment. We can now equate the variance in equation (A8) with the variance $\sigma_{\Phi}(t)^2$ in equation (A7) to obtain,

$$E\left[e^{i\Phi(t)}\right] = e^{-q_{RW}t/2} \tag{A9}$$

To evaluate $q_{RW}$, we note that $\varphi(t) = 2\pi\nu x(t)$ where $\nu = \nu_{LO} + \nu_{IF}$ so that

$$E[x(t)^2] = q_{RWx}(t) = \frac{q_{RW}t}{(2\pi\nu)^2} \tag{A10}$$

We can express $q_{RWx}$ (and therefore $q_{RW}$) in terms of the Allan variance $\sigma_y^2(T)$, which is defined by

$$\sigma_y^2(T) \triangleq \frac{1}{2T^2} E\left[\left(x(t+2T) - 2x(t+T) + x(t)\right)^2\right] \tag{A11}$$

where $x(t)$, in the present case, is a random walk process. We define the following random increments

$$\begin{aligned} \Delta_1(T) &\triangleq x(t+T) - x(t) \\ \Delta_2(T) &\triangleq x(t+2T) - x(t+T) \end{aligned} \tag{A12}$$

so that

$$\begin{aligned} x(t+T) &\triangleq x(t) + \Delta_1(T) \\ x(t+2T) &\triangleq x(t) + \Delta_1(T) + \Delta_2(T) \end{aligned} \tag{A13}$$

Substituting equation (A13) into equation (A11)

$$\begin{aligned} \sigma_y^2(T) &= \frac{1}{2T^2} E\left[\left(\Delta_2(T) - \Delta_1(T)\right)^2\right] \\ &= \frac{1}{2T^2}\left(E\left[\left(\Delta_2(T)\right)^2\right] - 2E[\Delta_2(T)]E[\Delta_1(T)] + E\left[\left(\Delta_1(T)\right)^2\right]\right) \end{aligned} \tag{A14}$$

From Maybeck (1979), we note

$$\begin{aligned} &E[\Delta_1(T)] = E[\Delta_2(T)] = 0 \\ &E\left[\left(\Delta_1(T)\right)^2\right] = E\left[\left(\Delta_2(T)\right)^2\right] = q_{RWx}T \end{aligned} \tag{A15}$$

Substituting equation (A15) into equation (A14)

$$\begin{aligned} \sigma_y^2(T) &= \frac{1}{2T^2} E\left[\left(\Delta_2(T) - \Delta_1(T)\right)^2\right] \\ &= \frac{1}{2T^2}\left(q_{RWx}T - 0 + q_{RWx}T\right) \\ &= \frac{q_{RWx}}{T} \\ &= \frac{q_{RW}}{(2\pi\nu)^2 T} \end{aligned} \tag{A16}$$

Therefore, we can express $q_{RW}$ in terms of the Allan variance for *one clock* at 1 second $\sigma_y^2(1)$ as follows

$$q_{RW} = (2\pi\nu)^2\sigma_y^2(1) = (2\pi\nu)^2\sigma_{y1}^2 \tag{A17}$$

where we have defined $\sigma_{y1}^2 \triangleq \sigma_y^2(1)$. Note that the units of $q_{RW}$ are $1/sec$ due to the presence of the variable $T$ evaluated at 1 second. The Allan variance for the difference between *two clocks* will be $2\times$ larger so that $q_{RW}$ becomes

$$q_{RW} = (2\pi\nu)^2 2\sigma_{y1}^2 \tag{A18}$$

$$E\left[e^{i\Phi(t,\nu)}\right] = E\left[e^{i(\varphi_2(t,\nu)-\varphi_1(t,\nu))}\right] = e^{-4\pi^2\nu^2\sigma_{y1}^2 t} \tag{A19}$$

where, in addition to the dependence on time $t$, we have made explicit the dependence on frequency $\nu$ since it will play a role in computing integrals in equation (40).

## Appendix C: Deriving the S/N in the Special Case of White Frequency Noise

We want to derive an expression for the clock contribution to the signal-to-noise ratio $\mathcal{R}_c$ in the case of white frequency noise. To do so requires computing $\overline{\mathcal{V}}_m(t_k)$ in equation (40) and its associated standard deviation $\sigma_{\mathcal{V}_m}$ assuming that all other noise/error sources have been set to zero except for a white frequency noise contribution from the clocks. In that case, we begin by computing $\overline{\mathcal{V}}_m(t_k)$ by substituting equation (A19) in equation (40)

$$\overline{\mathcal{V}}_m(t_k;T) = \frac{\mathcal{V}_0}{T}\int_{t_k}^{t_k+T}\int_{-\infty}^{\infty} e^{-4\pi^2\nu^2\sigma_{y1}^2 s} H_1(\nu_{IF})H_2^*(\nu_{IF})d\nu_{IF}ds \tag{A20}$$

where we note that $\nu = \nu_{LO} + \nu_{IF}$ and $\sigma_{y1}^2$ is the single clock Allan Variance at 1 second of averaging time. In this example we choose ideal filters such that $H_1(\nu_{IF}) = H_2(\nu_{IF}) = H_0$, a constant for $\nu_{IF} \in [\nu_0 - \Delta\nu/2, \nu_0 + \Delta\nu/2]$ and 0 outside of this range. Therefore $E[H_1(\nu_{IF})H_2^*(\nu_{IF})] = H_0^2$. Applying these filters to equation (A20) yields

$$\overline{\mathcal{V}}_m(t_k;T) = \frac{\mathcal{V}_0 H_0^2}{T}\int_{t_k}^{t_k+T}\int_{\nu_0-\Delta\nu/2}^{\nu_0+\Delta\nu/2} e^{-4\pi^2(\nu_{LO}+\nu_{IF})^2\sigma_{y1}^2 s} d\nu_{IF} ds \tag{A21}$$

Note that the result in equation (A21) is real valued. We focus first on the inner integral, this is a Gaussian integral resulting in an error function, but we are primarily interested in high frequency VLBI where the bandwidth $\Delta\nu$ is small compared to $\nu_{LO}+\nu_{IF}$. Under these conditions the exponential is very nearly linear over the band and can be treated as a constant evaluated at the center frequency, $\nu = \nu_{LO}+\nu_0$, such that the integral is closely approximated by,

$$\int_{\nu_0-\Delta\nu/2}^{\nu_0+\Delta\nu/2} e^{-4\pi^2(\nu_{LO}+\nu_{IF})^2\sigma_{y1}^2 s} d\nu_{IF} \cong \Delta\nu e^{-4\pi^2(\nu_{LO}+\nu_0)^2\sigma_{y1}^2 s} \tag{A22}$$

For notational compactness we introduce the following definition

$$\eta \triangleq 4\pi^2\nu^2\sigma_{y1}^2 \approx 4\pi^2(\nu_{LO}+\nu_0)^2\sigma_{y1}^2 \tag{A23}$$

where $\eta = q_{RW}/2$ (from equation (A18)). The units of $\eta$ are $1/sec$ (see equation (A17)). Substituting equation (A22) and equation (A23) into equation (A21) we find

$$\overline{\mathcal{V}}_m(t_k;T) = \Delta\nu\mathcal{V}_0 H_0^2 \frac{1}{T}\int_{t_k}^{t_k+T} e^{-\eta s} ds \tag{A24}$$

Evaluating equation (A24) produces

$$\overline{\mathcal{V}}_m(t_k;T) = \Delta\nu\mathcal{V}_0 H_0^2 e^{-\eta t_k}\frac{1-e^{-\eta T}}{\eta T} \tag{A25}$$

The result in equation (A25) illustrates the nonstationary behavior of the statistics and the need for fringe fitting to take out the effects induced by the presence of $e^{-\eta t_k}$ (fringe fitting will empirically estimate $e^{-\eta t_k}$ and remove this effect). Assuming this has been done over a collection of accumulation periods $T_a$ that have been concatenated $N$ times over a full integration period $T = NT_a$ we would arrive at the following measured estimate of the visibility assuming fringe fitting $\overline{\mathcal{V}}_{mf}(T)$ where, without loss of generality, we have set $t_0 \triangleq 0$

$$\overline{\mathcal{V}}_{mf}(T) \triangleq \sum_{k=0}^{N-1}\overline{\mathcal{V}}_m(t_k)\, e^{\eta t_k} = \Delta\nu\mathcal{V}_0 H_0^2 N^2 \frac{1-e^{-\eta T/N}}{\eta T} \quad \text{(fringe fit result)} \tag{A26}$$

versus a more traditional estimate that evaluates the integral in equation (A25) over the full integration interval $T$ (often the coherence interval $T_c$) to provide the mean value of the measured visibility $\overline{\mathcal{V}}_m(T)$ as follows

$$\overline{\mathcal{V}}_m(T) = \frac{\Delta\nu\mathcal{V}_0 H_0^2}{T}\int_0^T e^{-\eta s} ds = \Delta\nu\mathcal{V}_0 H_0^2 \frac{1-e^{-\eta T}}{\eta T} \tag{A27}$$

Note that since the expression in equation (A27) is now independent of $t_k$ the mean visibility is now only dependent on the integration time.

For the remaining analysis we will proceed with the no fringe fit result in equation (A27). Turning to the second moment calculation (or the mean-square of the visibility) we have

$$
\begin{aligned}
&E[\mathcal{V}_m(T)^2] \\
&= E\left[Re\left\{\frac{\mathcal{V}_0 H_0^2}{T}\int_0^T\int_{\nu_0-\Delta\nu/2}^{\nu_0+\Delta\nu/2} e^{i\Phi(s,\nu)}d\nu_{IF}ds\right\} Re\left\{\frac{\mathcal{V}_0 H_0^2}{T}\int_0^T\int_{\nu_0-\Delta\nu/2}^{\nu_0+\Delta\nu/2} e^{i\Phi(s',\nu)}d\nu_{IF}ds'\right\}\right] \\
&\qquad = \left(\frac{\mathcal{V}_0 H_0^2}{T}\right)^2 \int_0^T\int_{\nu_0-\Delta\nu/2}^{\nu_0+\Delta\nu/2}\int_0^T\int_{\nu_0+\Delta\nu/2}^{\nu_0+\Delta\nu/2} E[\cos\Phi(s)\cos\Phi(s')]d\nu_{IF}dsd\nu'_{IF}ds' \\
&\qquad = \left(\frac{\Delta\nu\mathcal{V}_0 H_0^2}{T}\right)^2 \int_0^T\int_0^T E[\cos\Phi(s)\cos\Phi(s')]dsds'
\end{aligned}
\tag{A28}
$$

where we have used the approximation that across the bandwith $e^{i\Phi(s,\nu)}$ is relatively constant (hence, dropping the dependence on $\nu$ in the present case) and since only the real component participates, replacing $Re\{e^{i\Phi(s)}\} = \cos\Phi(s)$.

To complete the integral in (A28) we need to utilize the following properties of the random walk in phase (or white frequency noise) process (Maybeck, 1979) with independent increments

$$
\begin{aligned}
E[\Phi(s')^2] &= q_{RW}s' = 2\eta s' \\
E[\Phi(s)^2] &= q_{RW}s = 2\eta s \\
E[\Phi(s)\Phi(s')] &= E[\Phi(s)^2] + E[\Phi(s)(\Phi(s') - \Phi(s))] \\
&= E[\Phi(s)^2] + E[\Phi(s)]E[\Phi(s') - \Phi(s)] \\
&= q_{RW}\min(s,s') = 2\eta\min(s,s')
\end{aligned}
\tag{A29}
$$

Turning to $E[\cos\Phi(s)\cos\Phi(s')]$ this can be computed from its definition (for instance, see Equation 3-103 in Maybeck (1979)) using

$$
\begin{aligned}
&E[\cos\Phi(s)\cos\Phi(s')] \\
&= \int_{-\infty}^{+\infty}\int_{-\infty}^{\infty}\frac{1}{2\pi}\left|\begin{bmatrix} 2\eta s & 2\eta\min(s,s') \\ 2\eta\min(s,s') & 2\eta s'\end{bmatrix}\right|^{-1/2} \\
&\times \exp -\frac{1}{2}\left\{\begin{bmatrix}\Phi \\ \Phi'\end{bmatrix}^{\mathrm{T}}\begin{bmatrix} 2\eta s & 2\eta\min(s,s') \\ 2\eta\min(s,s') & 2\eta s'\end{bmatrix}^{-1}\begin{bmatrix}\Phi \\ \Phi'\end{bmatrix}\right\}\cos\Phi(s)\cos\Phi'(s')\,d\Phi' d\Phi \\
&\qquad\qquad = e^{-\eta(s+s')}\cosh(2\eta\min(s,s'))
\end{aligned}
\tag{A30}
$$

Subsituting this result into the double integral in equation (A28) gives

$$
\begin{aligned}
E[\mathcal{V}_m(T)^2] &= \left(\frac{\Delta\nu\mathcal{V}_0 H_0^2}{T}\right)^2 \int_0^T \int_0^T E[\cos\Phi(s)\cos\Phi(s')]dsds' \\
&= \left(\frac{\Delta\nu\mathcal{V}_0 H_0^2}{T}\right)^2 \int_0^T \int_0^T e^{-\eta(s+s')}\cosh(2\eta\min(s,s'))\,dsds' \\
&= \left(\frac{\Delta\nu\mathcal{V}_0 H_0^2}{T}\right)^2 \int_0^T \left[\int_0^s e^{-\eta(s+s')}\cosh(2\eta s')\,ds' \right. \\
&\quad \left. + \int_s^T e^{-\eta(s+s')}\cosh(2\eta s)\,ds'\right] ds \\
&= (\Delta\nu\mathcal{V}_0 H_0^2)^2 \frac{e^{-4\eta T} + 8e^{-\eta T} + 12\eta T - 9}{12(\eta T)^2}
\end{aligned}
\tag{A31}
$$

Finally, subsitituting equation (A27) and equation (A31) into equation (42) (and noting that $t_k = t_0 \equiv 0$ in the present case) yields the following expression for $\mathcal{R}_c$

$$
\begin{aligned}
\mathcal{R}_c(T) &= \frac{|\overline{\mathcal{V}}_m(T)|}{\sqrt{E[\mathcal{V}_m(T)^2] - \overline{\mathcal{V}}_m(T)^2}} \\
&= \frac{1 - e^{-\eta T}}{\sqrt{\frac{e^{-4\eta T} + 8e^{-\eta T} + 12\eta T - 9}{12} - (1 - e^{-\eta T})^2}} \\
&= \frac{\sqrt{12}(1 - e^{-\eta T})}{\sqrt{e^{-4T\eta} - 12e^{-2T\eta} + 32e^{-T\eta} + 12T\eta - 21}}
\end{aligned}
\tag{A32}
$$

Equation (A32) provides a measure of the clock limitation on VLBI visibility S/N in the special case of white frequency noise.

## Appendix D: Numerical Solutions for the Clock-Limited S/N and Time Averaging vs. Ensemble Averaging

We have solved for the quantity $E\left[e^{i\Phi(t,T)}\right]$ exactly in the special case of white frequency noise (equation (A19)). For other noise types and combinations thereof, we use the numerical approach described in section 3.1.6. First we generate a time series for the difference of two uncorrelated clock frequencies and then integrate this to obtain the phase difference, $\Phi(t,T)$. To evaluate the expectation numerically, it would simplify the process if we could apply ergodicity and replace the expectation with a time average. It might seem reasonable to do this because the actual VLBI processing involves time averages to determine the visibility. However, VLBI does not actually take the time average of $e^{i\Phi(t,T)}$. Rather, a real VLBI measurement takes the time-average of the visibility. The visibility can be mathematically expressed as a function of the expectation of this exponential. We can not in general make the ergodic assumption because this only applies to a random variable with a time-invariant variance, which is only the case for white noise. To calculate an expression for the visibility in real VLBI, we must take the expectation of $e^{i\Phi(t,T)}$, *defined* as the ensemble average (Middleton, 1987) not as a time average.

A good example of where non-ergodicity results in $E\left[e^{i\Phi(t,T)}\right] \neq \left\langle e^{i\Phi(t,T)}\right\rangle$ (angle brackets indicate time average) is given by the case of white frequency noise where the frequency is white, but the derived phase is not. Figure A1 shows simulated time series of $e^{i\Phi(t,T)}$ where $\Phi(t,T)$ is derived from 30 trials of white frequency noise. The phase derived from white frequency noise has a random walk characteristic and as expected the variance at each point in time is not a constant. This can be seen in the figure as the increasing spread of traces with time. A simple time average of these traces does not produce the correct expectation because it implicitly assumes that the variance is constant. In the figure, the red trace gives the time average, while the blue trace gives the (correct and different) ensemble average.

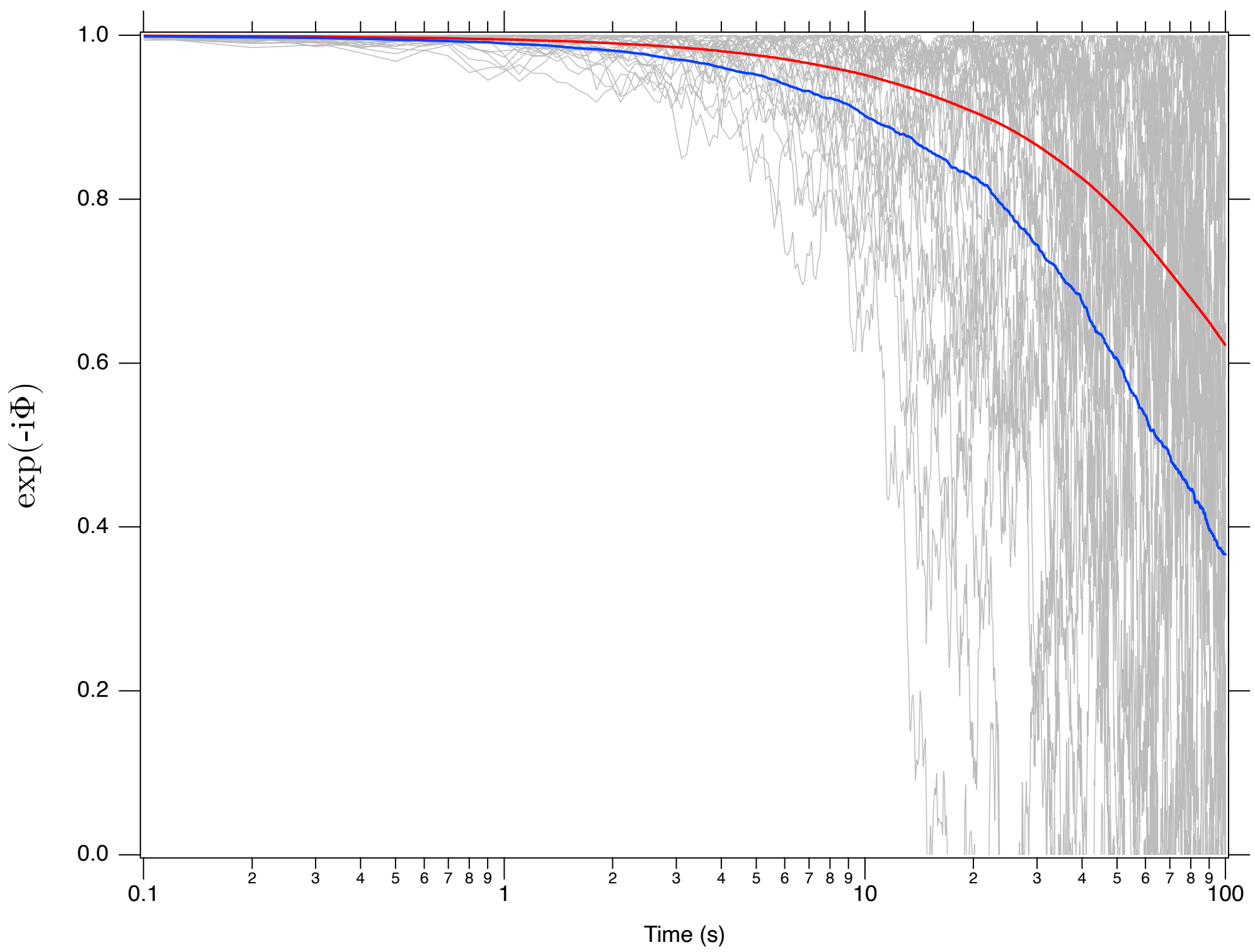

**Figure A1.** An example of 30 runs of the exponential of phase where the phase is derived from white frequency noise time series each having an Allan Deviation of $1 \times 10^{-13}/\sqrt{\tau}$ (gray traces). The red trace shows $\langle e^{i\Phi(t,T)} \rangle = \frac{1}{T}\int_0^T e^{i\Phi(t,T)}\, dt$, while the blue trace shows $E\left[e^{i\Phi(t,T)}\right] = e^{-\sigma_\Phi(t)^2/2}$. The red and blue trace would be the same if $\sigma_\Phi^2$ were constant in time, but for the phase of white frequency noise shown here, this condition is clearly not met.

Another way of viewing this is that for non-stationary noise (e.g., random walk), the expression,

$$\langle e^{i\Phi(t,T)} \rangle = \frac{1}{T}\int_0^T e^{i\Phi(t,T)}\, dt \tag{A33}$$

is itself a random variable with a *different variance at each point in time*—there is no single variance that characterizes the expectation at all times. In order to make statements about it, one is required to derive statistical quantities such as the variance at each point. From equation (A7) in appendix B,

$$E\left[e^{i\Phi(t,T)}\right] = e^{-\sigma_\Phi(t)^2/2} \tag{A34}$$

where,

$$\sigma_\Phi(t)^2 = E[\Phi(t)^2] - E[\Phi(t)]^2 \tag{A35}$$

Here the expectations are ensemble averages (averages over the ensemble at a particular time t calculated separately for all t in a range from 0 to T). The result is a time series of $\sigma_\Phi(t)^2$ that is substituted into equation (A34) to obtain a time series of $E\left[e^{i\Phi(t,T)}\right]$. In a numerical simulation, these expectations (ensemble averages) at each point in time can are obtained by averaging $N_{ave}$ time series of $e^{i\Phi(t,T)}$ where the average is taken at each point $t$ in time.

## Appendix E: Validating the Numerical Approach: Comparing the Exact and Numeric Solutions for $\boldsymbol{E}\left[e^{i\boldsymbol{\Phi}(\boldsymbol{\tau})}\right]$

For most clocks, the phase will not be a pure noise type and we will solve for the expression $E\left[e^{i\Phi(\tau)}\right]$ numerically. Since $e^{i\Phi(t,T)}$ is a function of a stochastic variable we will use a Monte-Carlo simulation to obtain its expectation as a function of $T$. To validate the numerical approach, we perform it in the special case of white frequency noise and compare it to the exact expression for $E\left[e^{i\Phi(\tau)}\right]$ in appendix C equation (A19).

We look at the example where each clock will have an Allan deviation of $\sigma_y(\tau) = 1 \times 10^{-13}/\sqrt{\tau}$ so that their combined noise is $\sigma_y(\tau) = \sqrt{2} \times 10^{-13}/\sqrt{\tau})$ on all time scales (that is, white frequency noise). Here, $\sigma_{y1}^2$ is the Allan variance of a single clock at 1 second - $1 \times 10^{-26}$. In the example of terrestrial VLBI at 230 GHz, the maximum averaging time is usually limited by atmospheric coherence to order of 10 s. However, to see longer-term trends that may be relevant for space-based VLBI we will take the maximum averaging time to be as long as 10,000 s. This corresponds to $f_L = 1 \times 10^{-4}$. We take the upper bound in frequency to be $f_H = 2$ GHz.

Figure A2 shows excellent agreement between the exact and numeric solutions of $E\left[e^{i\Phi(\tau)}\right]$ at the 1% level for 4 observation frequencies of interest to high-performance VLBI: 90, 230 345, and 630 GHz.

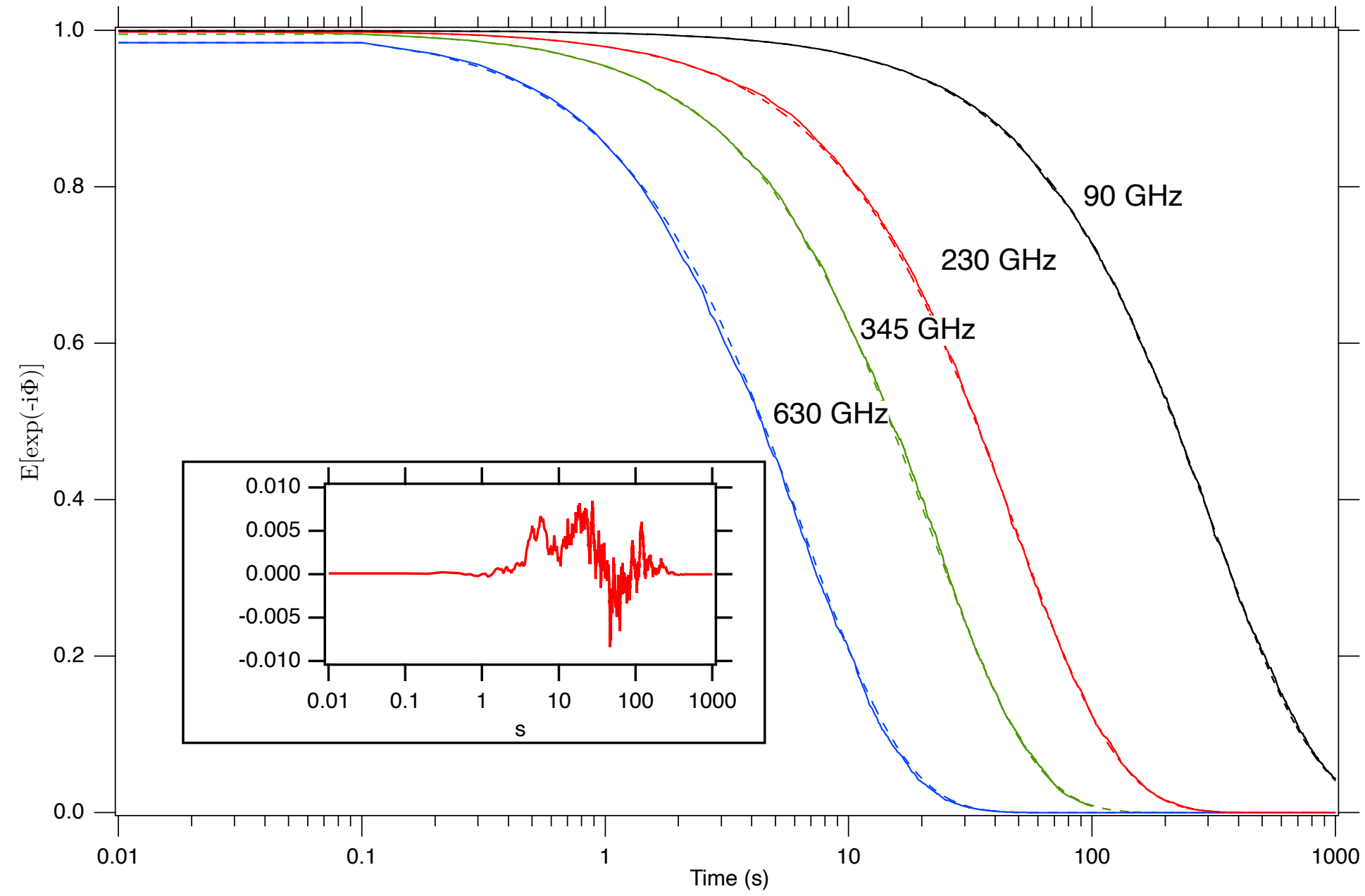

**Figure A2.** A comparison of the exact solution for $E\left[e^{i\Phi(T)}\right]$ in the case of white frequency noise with an Allan deviation for each clock of $1 \times 10^{-13}/\sqrt{\tau}$ (dashed traces) and numerical solution (solid traces) for several frequencies: 90 GHz (black), 230 GHz (red), 345 GHz (green) and 630 GHz (blue). Each numerical solution was obtained by averaging 3000 trials. The inset gives the difference between the numerical and exact solution in the representative case of 230 GHz and shows agreement between the two at the 1% level.

To verify that the numerical simulations of $E\left[e^{i\Phi(T)}\right]$ are relatively insensitive to time step and range used, we compare the exact solution to several numerical solutions taken for different steps in the representative case of 230 GHz. Figure A3 shows agreement at the several percent level.

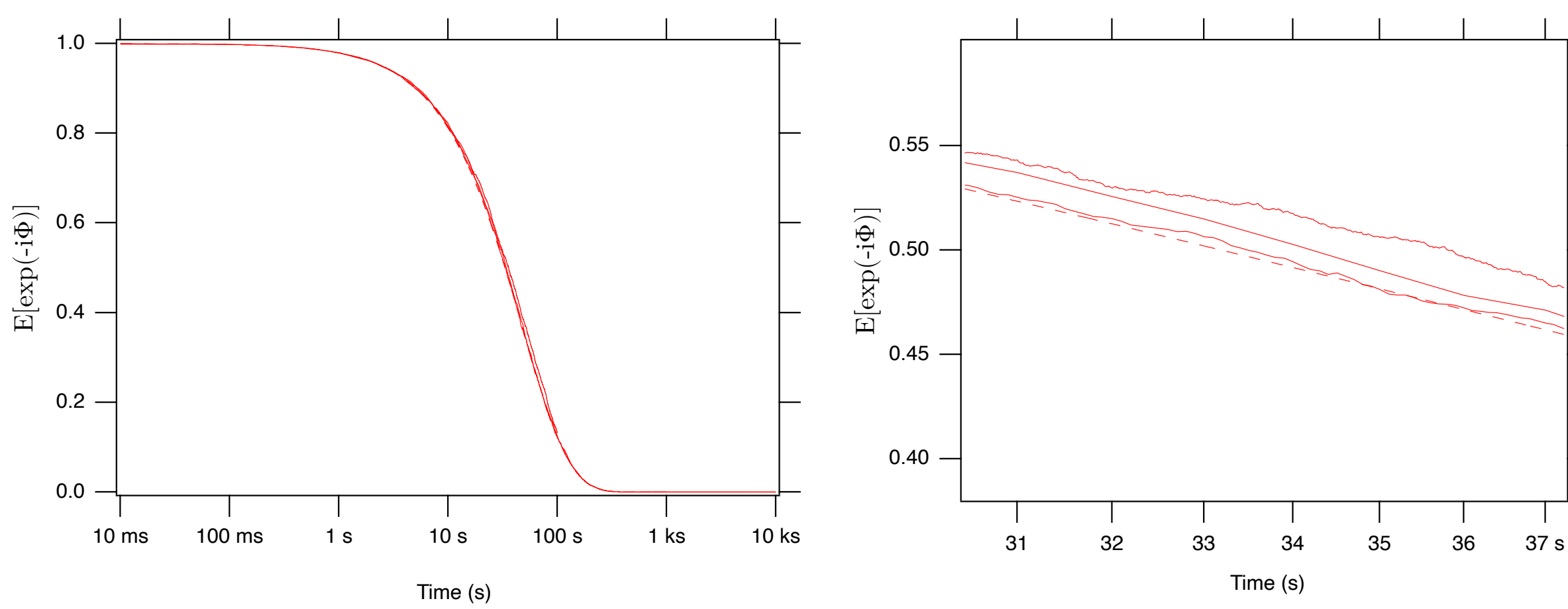

**Figure A3.** A comparison of the exact solution for $E\left[e^{i\Phi(T)}\right]$ (dashed line) with the numerical solution over different ranges: 1–10,000 s, 0.1–1000 s, 0.01–100 s. The graph on the left shows

agreement between all of these cases at the noise level (the four curves are not distinguished). The graph on the right expands in the region of greatest disagreement, which is at the several percent level.

## Appendix F: Estimating Noise Characteristics in Real Clock Allan Deviations

One of the benefits of the Allan deviation metric is that most noise types have different slopes as a function of time (Allan, 1987). This makes it possible to determine which noise processes are governing the behavior of an oscillator or clock at a given averaging time. The Allan deviation for many clocks is often only given at discrete points. From this it is still possible to estimate the noise types involved and to smoothly interpolate between the given points. The Allan deviations given in this paper are derived by estimating the noise types involved for each averaging time and then fine tuning each amplitude such that the resulting Allan deviation passes through the given points. This combination of noise types and amplitudes for each is then used to generate the time series in frequency that are used in the subsequent VLBI simulations.

The Allan deviation for the Radioastron space hydrogen maser serves as a good example of this method. As with most atomic clocks using a USO LO, the Allan deviation at very short times, usually less than 1 second, is likely to be governed by white phase noise. In addition, since the maser is an active device, the white phase noise characteristic continues at even longer times greater than 1 second. It then transitions to white frequency noise and flicker noise, which prevents it from averaging lower ("flicker floor"). Finally, the Allan deviation starts to increase with time due to drift. Figure A4 shows the Allan deviation points provided in Kardashev et al. (2017) superimposed with the derived Allan deviation. Table A1 gives the noise types and amplitudes for all clocks discussed in this paper.

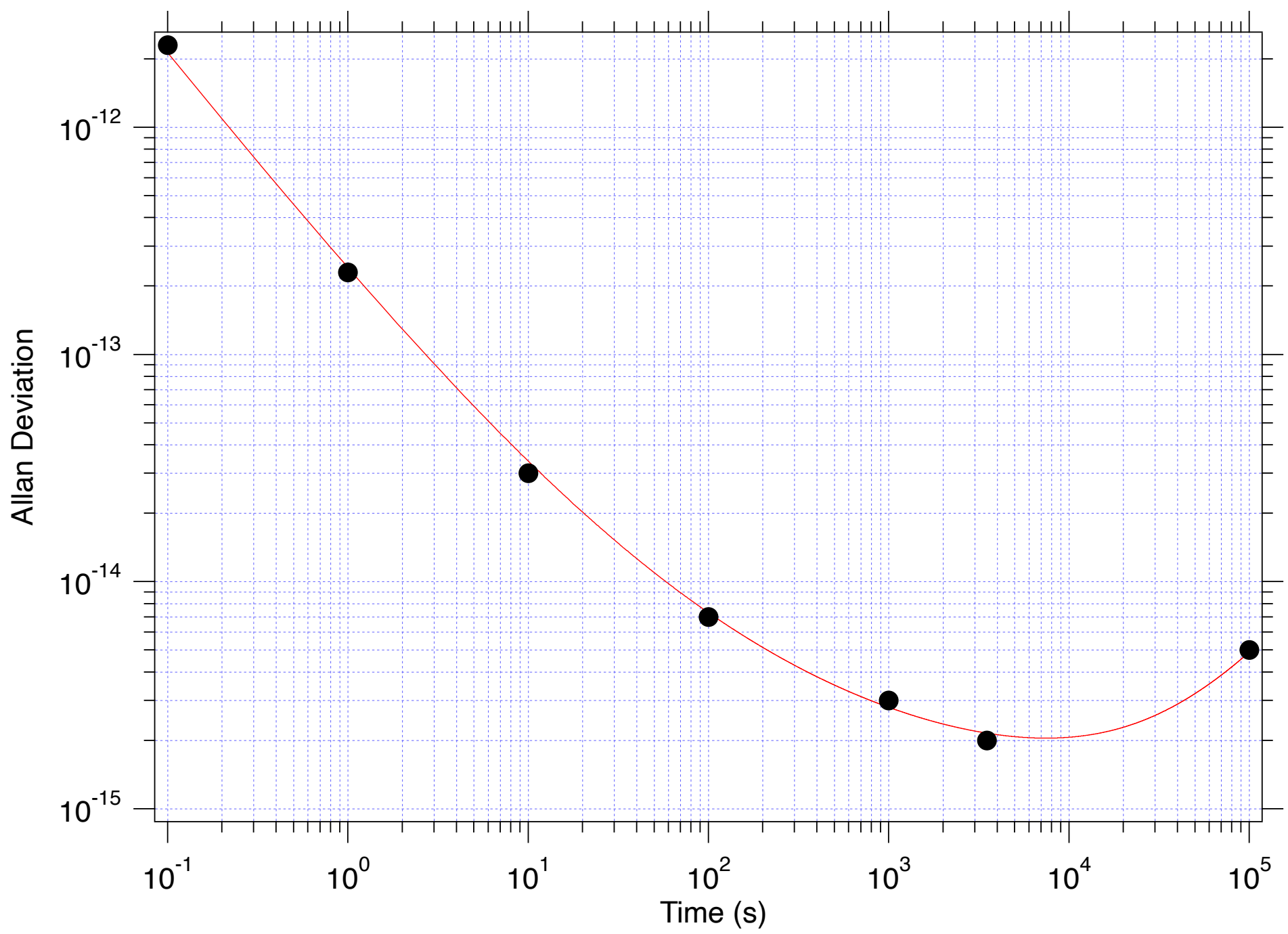

**Figure A4**. Allan deviation points provided in Kardashev et al. (2017) (black dots) superimposed with an Allan deviation derived from a mixture of noise types whose amplitudes are chosen to fit the pointes (red line). The resulting interpolation for this frequency standard is $\sigma_{yRAHM} = 2.2 \times 10^{-13}/\tau + 6 \times 10^{-14}/\sqrt{\tau} + 1.7 \times 10^{-15} + 3 \times 10^{-15} \times (\tau/86400)$ where the terms are white phase, white frequency, flicker frequency, and drift (at one day) respectively.

**Table A1.** Clock noise types, amplitudes, and physical characteristics for clocks considered in this paper. For each noise type, an associated Allan deviation contribution extrapolated to 1 second is given. The noise types are, White Phase Noise (WPN), White Frequency Noise (WFN), Flicker Frequency Noise (FFN), and Random Walk Frequency Noise (RWFN). Some frequency standards exhibit drift on the time scales of interest and this value is given in units of fractional frequency per day. The clocks listed are: the active hydrogen maser (HM) (R. F. Vessot, 1990), the Ultra-Stable Oscillator (USO) ("Ultra Stable Oscillator (USO) for deep space exploration,"), the Radioastron active hydrogen maser (RAHM) (Kardashev et al., 2017), the Rubidium Atomic Frequency Standard (RAFS) used by GPS (Teledyne, n.d.), the Passive Hydrogen Maser (PHM) used by Galileo (Droz et al., 2009), a laser-cooled rubidium clock (cRb) (Liu et al., 2018), a space-qualified active hydrogen maser for the Atomic Clock Ensemble in Space (ACES) mission (Goujon et al., 2010), the Vector Atomics optical iodine clock (Iodine) (Roslund et al., 2023), an Optical Local Oscillator (OLO) consisting of a cavity-stabilized laser and an optical frequency comb for down conversion into the microwave (Zhang & Matsko, 2024), a future version of the OLO (OLO future) with improved performance equal to that now achieved on the ground (Fortier et al., 2011), a future space-qualified high-performance optical clock with performance equal to that now achieved on the ground (HPOC) (Margolis, 2010), and a hypothetical clock with pure white frequency noise (WFN). Note that in the simulations

performed in this paper noise is sometimes not added that would only impact clock performance on times greater than those being simulated (eg., HM drift).

| | WPN | WFN | FFN | RWFN | Drift/day | Volume (liters) | Power (W) |
|---|---|---|---|---|---|---|---|
| **HM** | $8.5 \times 10^{-14}$ | $3.9 \times 10^{-14}$ | - | - | $1 \times 10^{-15}$ | 250 | 75-150 |
| **USO** | $8.5 \times 10^{-14}$ | - | $7.1 \times 10^{-14}$ | $7.8 \times 10^{-15}$ | $1 \times 10^{-11}$ | 1 | 6.5-8 |
| **RAHM** | $2.2 \times 10^{-13}$ | $4 \times 10^{-14}$ | $1.3 \times 10^{-15}$ | - | $3 \times 10^{-15}$ | 120 | * |
| **RAFS** | - | $2 \times 10^{-12}$ | - | - | $1 \times 10^{-14}$ | 4.2 | 39 |
| **PHM** | - | $7 \times 10^{-13}$ | - | - | $1 \times 10^{-14}$ | 22 | 54 |
| **cRb** | - | $3 \times 10^{-13}$ | - | - | - | 200 | 110 |
| **ACES** | $1.4 \times 10^{-13}$ | $5 \times 10^{-14}$ | $1.6 \times 10^{-15}$ | - | $1 \times 10^{-16}$ | 50 | * |
| **Iodine** | - | $2.5 \times 10^{-14}$ | $1 \times 10^{-15}$ | - | $4 \times 10^{-15}$ | 29 | 80 |
| **OLO** | $2 \times 10^{-14}$ | - | $9 \times 10^{-15}$ | $4.5 \times 10^{-16}$ | $1 \times 10^{-12}$ | 100 | * |
| **OLO future** | $2 \times 10^{-15}$ | - | $9 \times 10^{-16}$ | $4.5 \times 10^{-17}$ | $1 \times 10^{-13}$ | < 100 | * |
| **HPOC** | $1 \times 10^{-15}$ | $1 \times 10^{-15}$ | - | - | - | 10 | 33 |
| **WFN** | - | $1 \times 10^{-13}$ | - | - | - | - | - |

**Not published*

## Appendix G. Summary of main results

In Table A2 we summarize the main theoretical results of this work for easier reference and include the results derived for the special case of white frequency noise. In these equations, $\eta \triangleq 4\pi^2\nu^2\sigma_{y1}^2 \approx 4\pi^2(\nu_{LO}+\nu_0)^2\sigma_{y1}^2$ from equation (A23). The general results are from equations (40), (A28), and (42) respectively while the special case results are from equations (A25), (A31), and (A32) respectively.

**Table A2**. Summary of primary theoretical results.

| | **General equation** | **Special result for white frequency noise** |
|---|---|---|
| Expectation of the Mean Visibility | $\bar{\mathcal{V}}_m(t_k;T) = \frac{\mathcal{V}_0}{T}\int_{t_k}^{t_k+T}\int_{-\infty}^{\infty} E[e^{i\Phi(s,\nu)}]H_1(\nu_{IF})H_2^*(\nu_{IF})d\nu_{IF}ds$ | $\bar{\mathcal{V}}_m(t_k;T) = \Delta\nu\mathcal{V}_0H_0^2e^{-\eta t_k}\frac{1-e^{-\eta T}}{\eta T}$ |
| Expectation of the mean visibility squared | $E[\mathcal{V}_m(T)^2] = E\left[Re\left\{\frac{\mathcal{V}_0H_0^2}{T}\int_0^T\int_{\nu_0-\Delta\nu/2}^{\nu_0+\Delta\nu/2}e^{i\Phi(s,\nu)}d\nu_{IF}ds\right\} \times Re\left\{\frac{\mathcal{V}_0H_0^2}{T}\int_0^T\int_{\nu_0-\Delta\nu/2}^{\nu_0+\Delta\nu/2}e^{i\Phi(s',\nu)}d\nu_{IF}ds'\right\}\right]$ | $E[\mathcal{V}_m(T)^2] = (\Delta\nu\mathcal{V}_0H_0^2)^2\frac{e^{-4\eta T}+8e^{-\eta T}+12\eta T -}{12(\eta T)^2}$ |
| Signal-to-Noise Ratio | $\mathcal{R} = \frac{\lvert\bar{\mathcal{V}}_m(t_k;T)\rvert}{\sigma_{\mathcal{V}_m}} = \frac{\lvert\bar{\mathcal{V}}_m(t_k;T)\rvert}{\sqrt{E[\mathcal{V}_m(t_k;T)^2]-\bar{\mathcal{V}}_m(t_k;T)^2}}$ | $\mathcal{R}_c(T) = \frac{\sqrt{12}(1-e^{-\eta T})}{\sqrt{e^{-4T\eta}-12e^{-2T\eta}+32e^{-T\eta}+12T\eta -}}$ |

**Acknowledgments**

The research was partially carried out at the Jet Propulsion Laboratory, California Institute of Technology, under a contract with the National Aeronautics and Space Administration (80NM0018D0004). The authors acknowledge support from NASA Astrophysics Research & Analysis grant 80NSSC22K1747.

The authors would like to thank Shep Doeleman, Michael Johnson, and Lindy Blackburn for many helpful discussions and would like to acknowledges the ideas and advice from the participants in the "Beyond Interstellar: Extracting Science from Black Hole Images" study organized by the W. M. Keck Institute for Space Studies.

**Conflict of Interest**

The authors declare no conflicts of interest relevant to this study.

**Data Availability**

The software used to simulate the clock-limited VLBI visibility S/N in this article is Caltech proprietary software that can be made available to researchers with a license. Please contact the authors for information on obtaining a license to the software. However, the method used is described in section 3.2 and should be sufficient to regenerate the results.